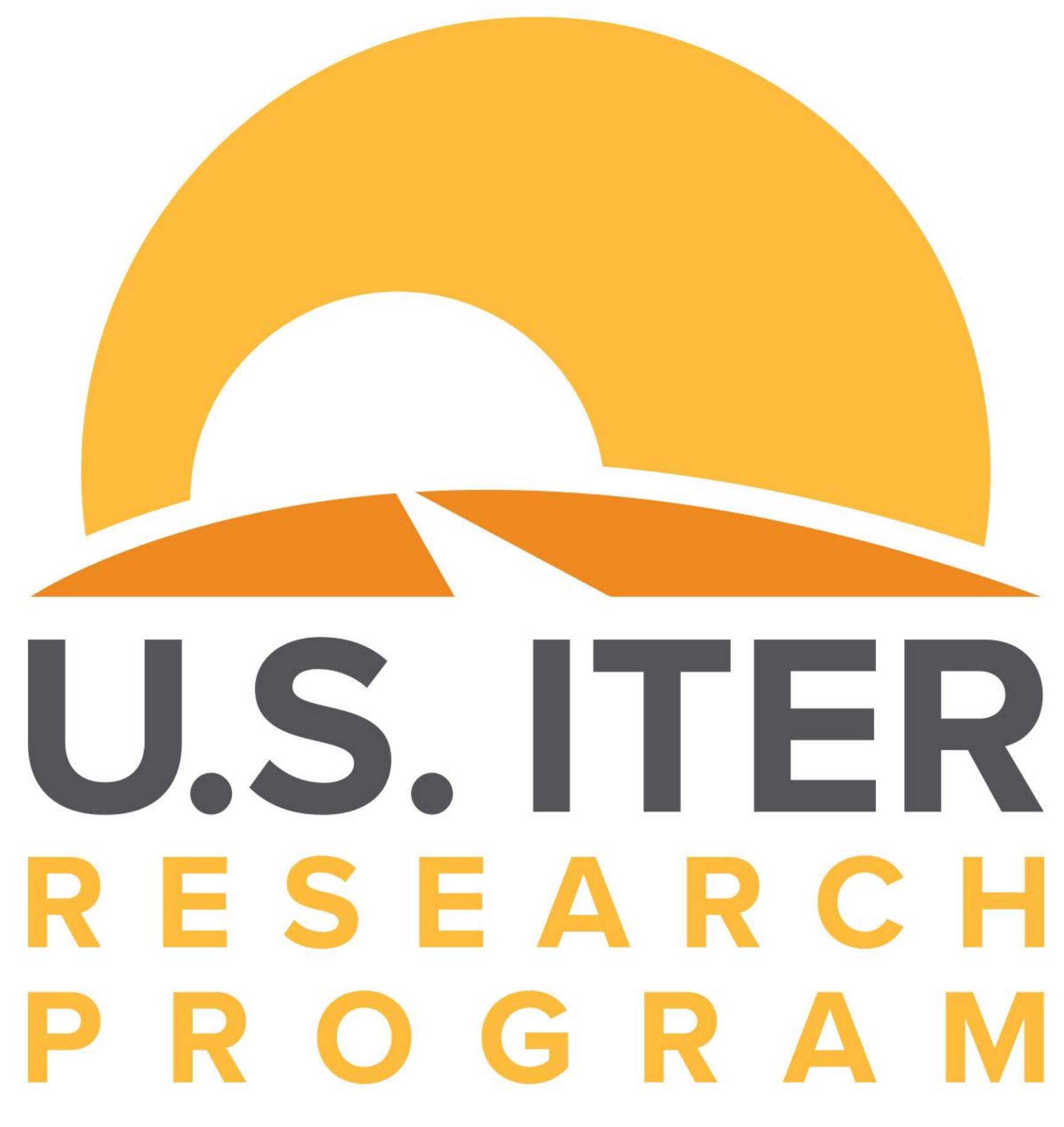

# 2022 Fusion Energy Sciences Research Needs Workshop

**Final Report**
**October 8, 2022**

# Report of the US ITER Research Program Research Needs Workshop

Chair: Charles Greenfield, GA
Co-Chair: Cami Collins, ORNL
DOE FES Liaisons: Matthew Lanctot and Josh King

This report presents the findings of 415 participants from the US fusion community. Despite the challenges of a virtual workshop, participants gave very thoughtful input through discussion meetings, white papers, and presentations, all of which formed the basis for this report. The chairs would also like to express sincere gratitude to all who participated, but in particular the volunteer discussion leaders and scribes, many of whom directly contributed to the text of this report:

Tyler Abrams (GA)
Vittorio Badalassi (ORNL)
Devon Battaglia (CFS)
Matthew Beidler (ORNL)
Theodore Biewer (ORNL)
Rejean Boivin (GA)
Michael Brookman (CFS)
Amelia Campbell (ORNL)
Lane Carasik (VCU)
Livia Casali (Tennessee)
John Caughman (ORNL)
Xi Chen (GA)
Michael Churchill (PPPL)
Jonathan Coburn (SNL)
Diane Demers (Xantho)
Arturo Domínguez (PPPL)
David Donovan (Tennessee)
Ben Dudson (LLNL)
Florian Effenberg (PPPL)
David Eldon (GA)
Darin Ernst (MIT)
Nate Ferraro (PPPL)
Thomas Fuerst (INL)
Yashika Ghai (ORNL)
Jeremy Hanson (Columbia)
Claudell Harvey (US ITER)
Jeffrey Herfindal (ORNL)
Chris Holcomb (LLNL)
David Humphreys (GA)
Paul Humrickhouse (ORNL)
Jacob King (Tech-X)
Florian Laggner (NCSU)
Charles Lasnier (LLNL)
Anthony Leonard (GA)
Robert Lunsford (PPPL)
Rajesh Maingi (PPPL)
George McKee (Wisconsin)
Adam McLean (LLNL)
Saskia Mordijck (W&M)
Hutch Neilson (PPPL)
Andrew Nelson (Columbia)
David Newman (Alaska)
Dmitri Orlov (UCSD)
Carlos Paz-Soldan (Columbia)
Mario Podesta (PPPL)
Francesca Poli (PPPL)
David Rasmussen (ORNL)
Oliver Schmitz (Wisconsin)
Eugenio Schuster (Lehigh)
Michael Segal (CFS)
Sterling Smith (GA)
Philip Snyder (ORNL)
Don Spong (ORNL)
Francesca Turco (Columbia)
Maxim Umansky (LLNL)
Michael Van Zeeland (GA)
Bob Wilcox (ORNL)
Leigh Winfrey (PSU)

The full list of participants is shown in Appendix B.

**NOTE: There are numerous references throughout the report to material posted on the workshop website at https://iterresearch.us. This website will be maintained for at least one year following the release of this report.**

# Preface

The US ITER Research Program Basic Research Needs Workshop, held over the course of several months in 2022, sought to identify steps to be taken to both maximize the return of the US investment in ITER's construction and operation and to ensure US research priorities on ITER strengthen the domestic program aimed at the development of a fusion pilot plant (FPP). Participants in the workshop propose to promptly take three major steps:

***We must establish an equitable, accessible, inclusive knowledge base integrating US and partner developments to support ITER and enable a US FPP.*** The US needs to develop technical expertise in fusion energy by engaging in the full breadth of predictive and analytic modeling, experimental design and measurement, engineering processes and planned operations, beginning with specific research areas identified in the report. US support to ITER is made stronger through development of a knowledge and industrial base for fusion. Operational experience and physics understanding needed to build an FPP will come from timely and robust participation in ITER activities. We should immediately establish the infrastructure and agreements necessary to ensure prompt and equitable access to ITER data by all US institutions, public and private. We must also work towards compatibility of US research software with ITER data standards.

***We must build a structure for maximizing US return on and contribution to ITER advances.*** The report calls for the formation of a funded and diverse US ITER Research Team (USIRT) coordinated by a US ITER Research Coordination Office (USIRCO) that serves as a bridge between the Team, ITER and its Members, and the US domestic fusion program. USIRCO will be charged to maximize the quality of our participation in all ITER operation and research activities, connecting both an on-site US presence with US-based remote participants and the international community. USIRCO should have its own budget and the ability to direct support toward urgent needs or emerging research tasks, and is accountable to the entire US ITER community, including labs, universities, and private companies, through the creation of a US ITER Research Advisory Board (USIRAB).

***We must support education and preparation of the workforce needed to deploy fusion energy.*** The ITER program will span decades, overlapping with the construction of FPPs. Sustained development requires engaged people at every career stage. The US fusion program (both public and private) is experiencing rapid growth, likely accelerating in the coming years, but limited by availability of scientists, engineers, operators, technicians and support staff. Workforce development must remain a priority throughout the ITER program. An inclusive, equitable fusion workforce, prepared by education and experience gained in our schools, universities, labs, and private companies, is needed to contribute to and benefit from ITER. Principles of diversity, equity, and inclusion must be incorporated into our efforts in order to consistently attract and retain the best people for these roles.

***Taking these steps will ensure that the US investment in ITER will return critical knowledge and experience leading toward realization of fusion energy as a clean, reliable energy source.***

# Executive Summary

Thirty-five nations are collaborating to build the world's largest fusion experiment, ITER, in southern France. ITER is designed to prove the feasibility of fusion as a carbon-free and large-scale source of energy that is based on the principle that powers our Sun and stars. ITER represents a massive investment by all of its members (China, the European Union, India, Japan, Korea, Russia, and the United States), and it will produce critical knowledge and experience in science, engineering, and technology that will apply directly to future fusion power plants. Now at an advanced state of construction and assembly, key ITER goals are to:

- Produce 500 MW of fusion power
- Demonstrate the integrated operation of technologies for a fusion power plant
- Achieve a deuterium-tritium plasma in which the reaction is sustained through internal heating
- Test tritium breeding
- Demonstrate the safety characteristics of a fusion device

Since the inception of the ITER Project, the US has adopted a new strategy advocating for a Fusion Pilot Plant (FPP) to be completed as soon as possible, with efforts to include both the public and private sectors. Even with acknowledgment of the aggressive US timelines for fusion energy, the value of ITER has been consistently identified preceding this report, both within community plans and reports from the National Academies of Science, Engineering, and Medicine (NASEM) that have been written since 2018.

With preparations for first plasma over 75% complete and system commissioning already in progress on the ITER site, it is time to establish a research program to ensure that the US is positioned to participate fully in ITER operations. DOE-FES asked the community to report on (a) how US scientists and engineers can best contribute to successful operation of the ITER facility and execution of its Research Program, (b) what science and technology we look to ITER for as we move forward toward fusion electricity, and (c) how best to organize ourselves to carry this out. This report is the product of a Basic Research Needs Workshop including over 420 members of the US fusion community, responding to a charge shown in Appendix A.

This report builds on previous studies and discussions of the scope of the ITER Research Program, and takes as a given the overall ITER Research Plan [ITR-18-003] that was developed jointly by the ITER Organization (IO) and all of its Members. To begin, we considered specific research activities of the US ITER Research Program to both support ITER's success and bring back the results for application to an FPP. Many of these activities can start today (indeed, many are already in progress within the US Fusion Energy Sciences program and within privately funded endeavors) and involve both US-domestic research and on-site work at ITER. All are enabled by two overarching General Initiatives:

GI1: Improve US access to and utilization of ITER information
GI2: Modernize and adapt US codes and data to be IMAS-compatible

These are necessary in order to have an informed US fusion community with the best placement to contribute and with the necessary tools to perform useful research that can be shared with our ITER partners. They need to start immediately, and are important for the success of almost every subsequent activity described here. The Workshop also identified seven cross-cutting Research Missions:

Mission 1: Disruption Prevention and Mitigation
Mission 2: Technology Engagement and Transfer
Mission 3: Materials Evaluation
Mission 4: Heat and Particle Exhaust Handling
Mission 5: Operating Scenarios and Plasma Control
Mission 6: Modeling, Simulation, and Data Handling
Mission 7: Core-Edge Integration

These Missions, in turn, rely on a set of 54 Topical Initiatives (28 to ensure ITER's success and 26 for transferring ITER knowledge and results to an FPP). These initiatives came directly out of discussions by topical breakout groups during the first phase of the Workshop. Of particular note is Mission 2, which includes US participation in ITER system commissioning: Commissioning activities are already in progress, but the current lack of significant US participation limits US impact and return in capabilities, with consequences that grow as it remains unaddressed.

***We must promptly and broadly engage with ITER, to gain knowledge in identified missions and initiatives and to support any needs that may arise in coordination with the US ITER research team and fusion community.***

All of these activities should be taken on in a spirit of collaboration and cooperation with the ITER Organization and our fellow ITER Members. As an effective partner, the US can enable ITER to progress quickly to meet its goals and get the needed results that inform the design, assembly, and operation of a US Fusion Pilot Plant. The Missions and Initiatives address many critical products starting now and continuing through each ITER phase, including physics, engineering, control science, measurement, tritium breeding and handling, operation of a power-plant scale fusion nuclear facility, and licensing. It can be expected that over time we may need to adjust the definition and scope of these Missions to better interface with our partners and to take account of new research results.

***We must establish an equitable, accessible, inclusive knowledge base integrating US and partner developments to support ITER and enable a US FPP.***

While there has been much previous discussion of scientific and technical needs for ITER, details of how the US should organize itself to participate have not previously been discussed in any serious detail. Workshop participants propose a US ITER Research Program composed of three elements (see figure):

1. **The US ITER Research Team (USIRT)** carries out the research program as described in the ITER Research Plan and the Missions and Initiatives described above
2. **The US ITER Research Coordination Office (USIRCO)** provides the necessary coordination for the planning and execution of the US ITER Research Program and is a point of contact with the IO and other ITER Members
3. **The US ITER Research Advisory Board (USIRAB)** provides oversight of the USIRCO and USIRT

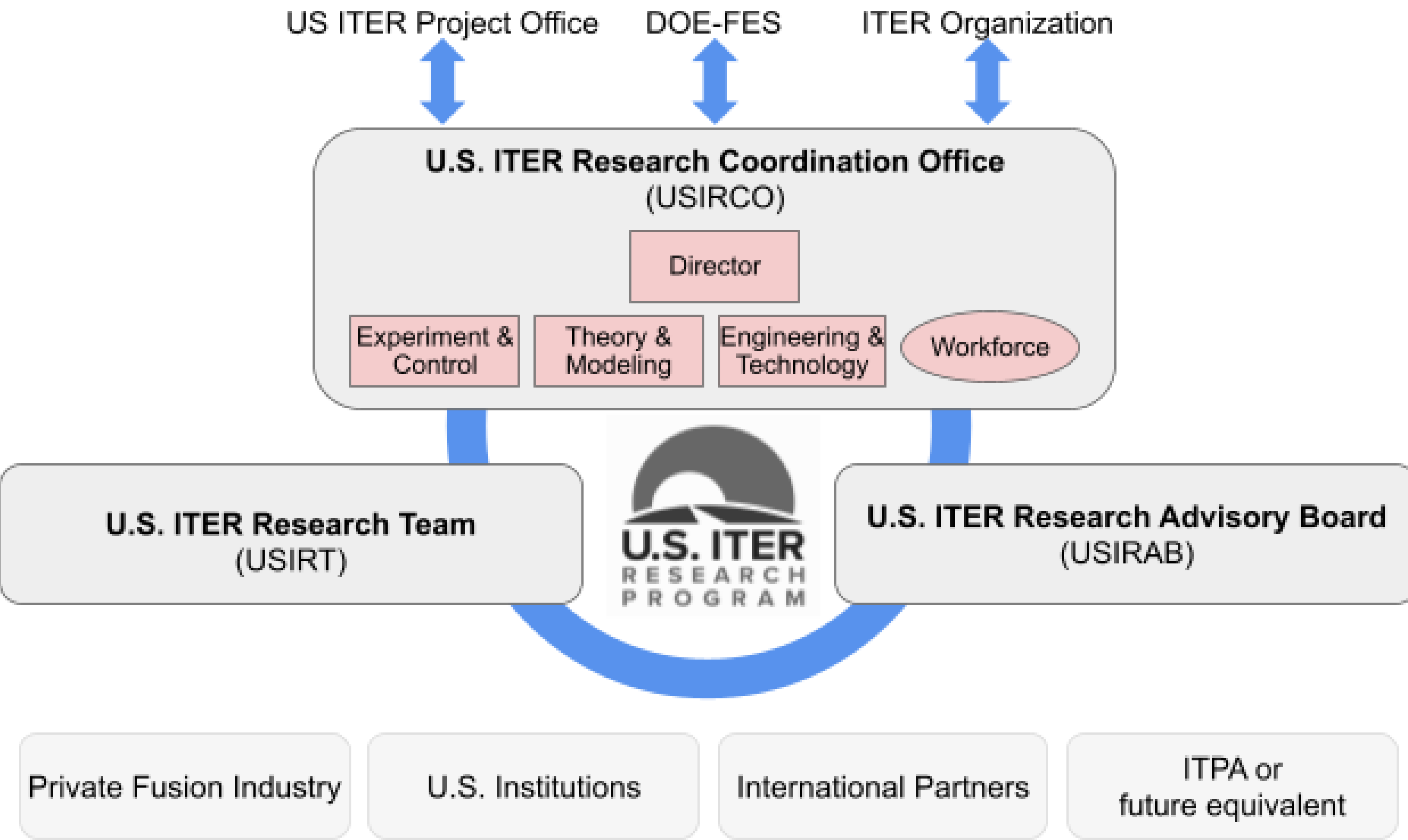

*Organizational structure of the proposed US ITER Research Program*

The program should have points of contact with the US ITER Project Office (USIPO), the IO and other ITER Members, and with DOE-FES. It should also work collaboratively with other elements of the US Fusion program. All segments of the community have roles to play, including laboratories, universities, and privately funded fusion endeavors. In particular, the FES User Facilities have unique capabilities to support both ITER research by acting as “ITER satellites” as well as in training researchers for roles on ITER; the USIRCO should work closely with their leadership to leverage those capabilities. The USIRCO should also look for similar collaborative opportunities with future privately funded facilities.

***We must build a structure for maximizing US return on and contribution to ITER advances.***

ITER Operation is anticipated to continue after first plasma through at least the early 2040s and very likely later. This is a generational program that will span entire careers; many of today’s students and early career scientists and engineers will become senior leaders within the US and international programs during ITER’s lifetime. Mobilization of a workforce is thus a critical element to keep the pipeline of talented researchers stocked at every level and provide opportunities for US researchers to fill critical roles in the US and international programs. Principles of diversity, equity, and inclusion will be incorporated into our efforts in order to consistently attract and retain the best people for these roles.

***We must support education and preparation of the workforce needed to deploy fusion energy***

Participation in a major international project like ITER also raises several logistical challenges. The US ITER Research Program should have personnel working on-site at ITER. Significant support will be needed, both for work and life needs. One challenge may be competition for space at the ITER site. It may be necessary for the US to operate a satellite facility near ITER for US people to be based while spending some of their time on-site. It is likely that most US ITER participants will work remotely from their home institutions for much or even most of their time. This motivates US-based infrastructure, including one or more computational clusters and perhaps one or more remote experimental centers.

***Infrastructure should be provided to fully support USIRT members working both on-site at ITER and off-site in the US***

Laboratories, universities, and private companies working in fusion can all contribute to the success of the ITER research program and can all benefit from the knowledge and experience it produces. Efforts will be needed to minimize obstacles to the free flow of information and valuable contributions of researchers from all segments of the community.

***It should be a goal of the USIRCO and the entire USIRT to make access to ITER participation, data, and research products open to the entire community***

With an aggressive US strategy, ITER will be complemented by a national program of fusion research and technology, technology roadmaps aimed towards commercial viability, active industrial partnerships, technology innovations, and first-phase design/construction of an FPP. Even in the most ambitious scenario, near-term ITER engineering and data will inform FPP component and concept design, and longer-term ITER research tests FPP-specific improvements. In any scenario, ITER provides critical lessons learned that must be incorporated into the FPP and future commercial endeavors. By executing the actions identified in this report, the US can maximize return on ITER investment via a structured, staffed and long-term approach with clearly defined goals, agility, and evaluation of returns from ITER research for the domestic program.

# 1. Introduction

Thirty-five nations including China, the European Union, India, Japan, Korea, Russia, and the United States, are collaborating to build the world's largest fusion experiment, ITER, in southern France. ITER is designed to prove the feasibility of fusion as a carbon-free and large-scale source of energy that is based on the principle that powers our Sun and stars. As ITER is already at an advanced stage of construction and assembly (Fig 1.1) our thoughts are turning toward participation in the operation and experimental exploitation of this unprecedented facility.

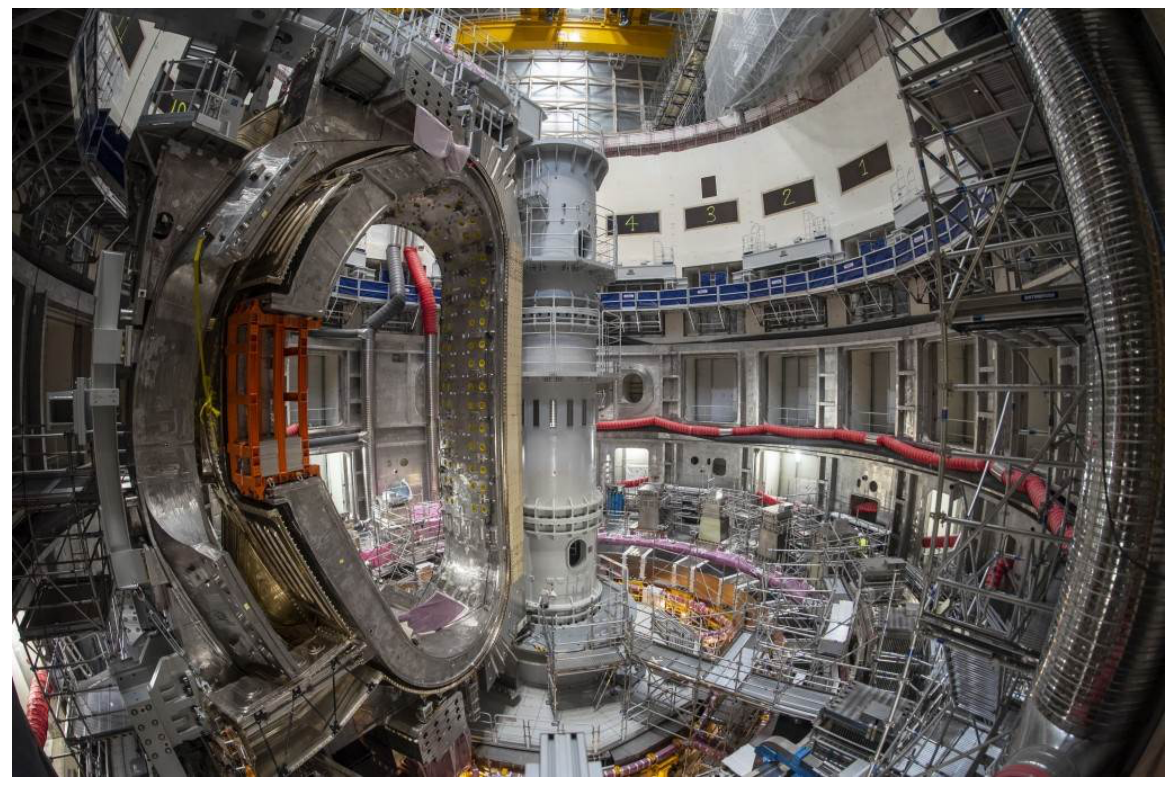

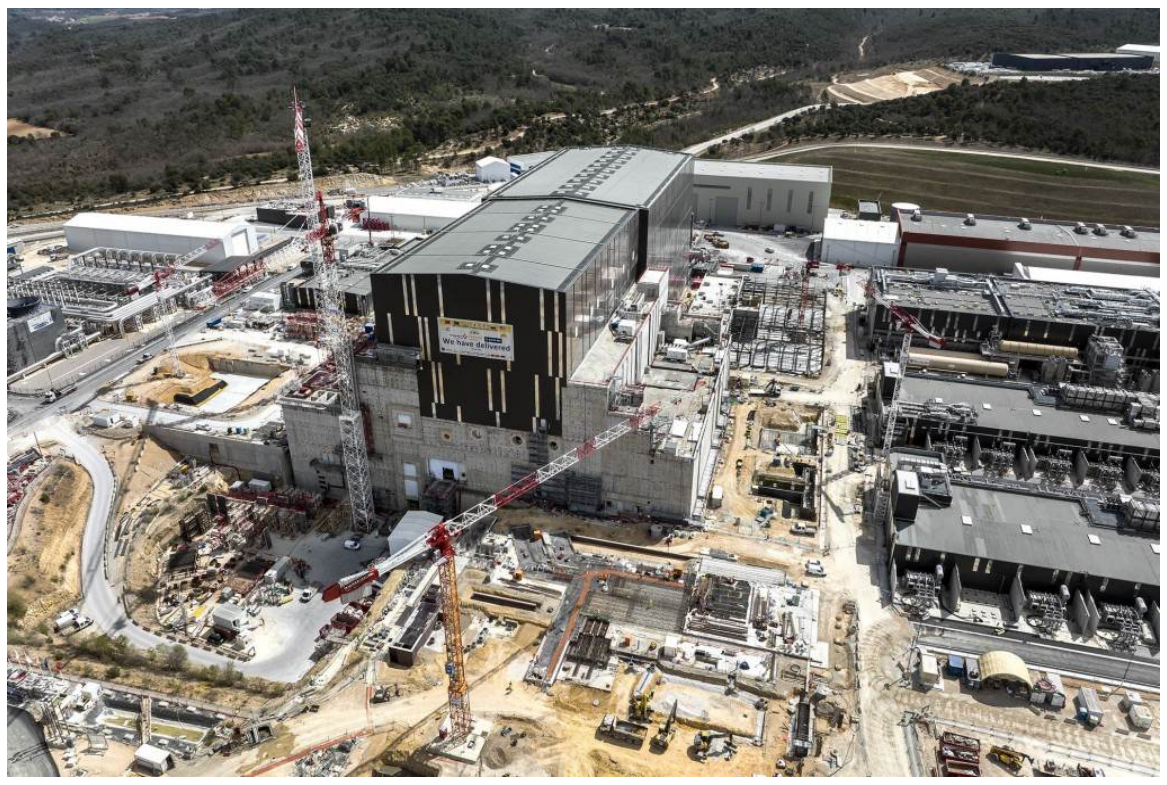

**Figure 1.1. (Left, June 2022) The first vessel segment of the ITER tokamak, complete with two toroidal field coils, is in place in the tokamak pit; (Right, April 2022) aerial view of the ITER site**

ITER will offer opportunities to gain knowledge and experience that can directly contribute to further elements in a strategy aimed at developing fusion as an energy source. In particular, we look to ITER to contribute to the basis for successful operation of a Fusion Pilot Plant (FPP), regardless of its timeline. Even today, ITER can provide essential experience in the design, construction, commissioning, and licensing of a power-plant scale fusion facility. Recognizing the importance of maximizing the return on the US' large investment in ITER and the opportunities presented by ITER, several recent studies including the National Academies Committee on a Strategic Plan for U.S. Burning Plasma Research [NAS2019], the APS-DPP Community Planning Process [DPPCPP2020], and a study by the Fusion Energy Sciences Advisory Committee [FESAC2021] all endorsed the promise of ITER, and made recommendations like this one from FESAC:

> *Ensure full engagement of the US fusion community in ITER by forming an ITER research team that capitalizes on our investment to access a high-gain burning plasma.*

The US Congress, in recent Fusion Energy Science budgets, and DOE-FES itself, have signaled readiness to support this recommendation, starting with a call for this Research Needs Workshop via the charge shown in Appendix A. The charge asks the US fusion community to propose the research scope and organization of a new US ITER Research Program.

Many such workshops like this have been held in the past, but always including major in-person events where participants could work together for several days at a time. With the world still emerging from the COVID-19 pandemic, the workshop could not be carried out in the traditional way. Instead, all meetings were held remotely, via Zoom. To enable high quality discussions, the workshop adopted a set of community agreements, discussion group etiquette, and methods similar to those used in the 2019 APS-DPP Community Planning Process. The need for multiple discussion meetings spanning multiple time zones created some challenges and resulted in the timeline of the workshop stretching to consume most of 2022. However, the remote nature of the workshop also created opportunities. It was possible for participants to collect thoughts individually, and at the same time work simultaneously and collaboratively, with multiple participants editing group notes and the draft report . The remote nature also removed the "travel barrier" for participation, allowing 415 US scientists and engineers to participate (Appendix B). In addition, 28 observers attended the plenary sessions, including eight from the US Department of Energy, six from Europe who are working on their similar process. The observers did not participate in writing this report.

The workshop was carried out in two phases (Appendix C) based on the two parts of the charge (*research* and *organization*). A set of discussion questions, shown in Appendix C and based on the charge, was developed to aid in deliberations aimed at responding to the charge. The phases were held sequentially, with Phase 1 (research) starting with a kickoff meeting on February 9, 2022, and ending with a plenary "Regroup" meeting on March 16. Phase 2, on organization, commenced with a kickoff meeting on March 23 and continued until a "final meeting" on July 13. An early draft of this report was presented to participants at the July 13 meeting.

Following the July 13 meeting, comments were solicited from participants in the form of chits. 49 such chits were submitted . The chairs, with some input from discussion leaders and scribes, responded to all chits. Most of the chits were helpful and incorporated to improve this report. In a few cases, chits were not fully resolved due to extensive complexities and/or limited scope of the workshop. The chits and responses are posted on the workshop website.

Workshop discussions were informed by the plenary talks, 81 white papers submitted by participants (Appendix D), a large library of reference material and previous reports, and the experience and expertise of the 415 participants. All of this material is available on the workshop website at https://www.iterresearch.us (the white papers and presentations will be publicly available, pending permission from their authors, and the website will be maintained for at least one year after this report is submitted). During the workshop, the white papers were all posted anonymously; authors were revealed, with their permission, at the release of this report.

For each phase, a series of breakout discussions were held during the interval between the kickoff and ending meeting, led by volunteers (listed on page 2) selected from among the participants. The groups were organized topically in Phase 1, and at random in Phase 2, with a goal of each group being small enough to give everybody a chance to participate. These leaders and scribes met with the workshop chairs between breakout discussions, and ultimately worked with the workshop chairs to produce the text in this report.

During Phase 1, participants identified a set of 54 Topical Initiatives from 10 topical areas. Of these, 28 are aimed at ensuring ITER's success and 26 look forward to the experience and information that will be gained from ITER for application to an FPP. In addition, two overarching General Initiatives were identified aimed at facilitating information and data flow between the US community and ITER. From these, seven cross-cutting Research Missions are proposed, each of which could - and should - begin immediately. The General and Topical Initiatives and the Research Missions are described in Chapter 2 of this report.

During Phase 2, participants proposed the formation of a US ITER Research Team (USIRT) and described its organization and infrastructure needs. The USIRT, led by a US ITER Research Coordination Office (USIRCO), would be responsible for participation in ITER operation and research, with a charge to ensure ITER's success and that the resulting knowledge and experience can be made available for subsequent US fusion activities, an FPP in particular. This is described in Chapter 3.

In Chapter 4, we discuss mobilization of a US ITER workforce, with a goal of participation at every career level including students and early through late career scientists, engineers, and technicians. It will be critically important to keep the pipeline stocked at every level, affording the US Team opportunities within the greater ITER research program for a variety of technical and leadership roles. Principles of diversity, equity, and inclusion should be incorporated into these efforts in order to consistently attract and retain the best people for these roles.

Related to this is the coordination with other segments of the US fusion community. The USIRT should work collaboratively and inclusively with the university community, the theory and sImulation community, the Technology Program, domestic user facilities and their international counterparts, and the privately funded fusion community to the benefit of all. Interactions with each of these are also discussed in Chapter 4.

This report identifies how the US can gain the most benefit from the ITER experiment, which will provide access to a high-gain, long-pulse burning plasma and provide real-world experience with the operation and control of a power-plant-scale facility. ITER will push plasma science into new regimes, stress-testing models and bringing reality to projected simulations which currently have significant uncertainty. Lessons learned from fusion nuclear technologies in ITER, particularly tritium fuel cycle and neutronics, can significantly impact U.S. FPP design,operations, and safety. The proposed US ITER Research Program calls for the formation of a US Team that can work collaboratively with the ITER Organization and the other ITER Members and execute the critical research missions identified in this report. This program will support the much needed rapid growth of a US fusion workforce for the entire US fusion program. It will raise workforce development and DEI to high levels of importance to attract and retain the best people, gathering talented physicists, engineers, technologists, and technicians from labs, universities, and private companies. ITER operation is on the horizon, and now is the time to act to begin to leverage the large U.S. investment.

# 2. Research Directions for US Engagement in ITER

Given US interests in ITER's success and acceleration of the FPP timeline, enabling ITER to progress quickly to meet its goals is of great value. By applying strengths in all topical areas, the US can be an effective partner in accelerating the development of high-performance, reduced-risk plasma scenarios that allow ITER to reach its goals while avoiding or minimizing the risks of edge-localized modes (ELMs), disruptions, poor confinement, impurity contamination, or excessive divertor heat flux. The US can offer scenario solutions, prepared via integrated simulation framework and coordinated national/international experimental development and demonstration. Finding ways to run with higher safety margins (engineering and physics) will also be important for an FPP.

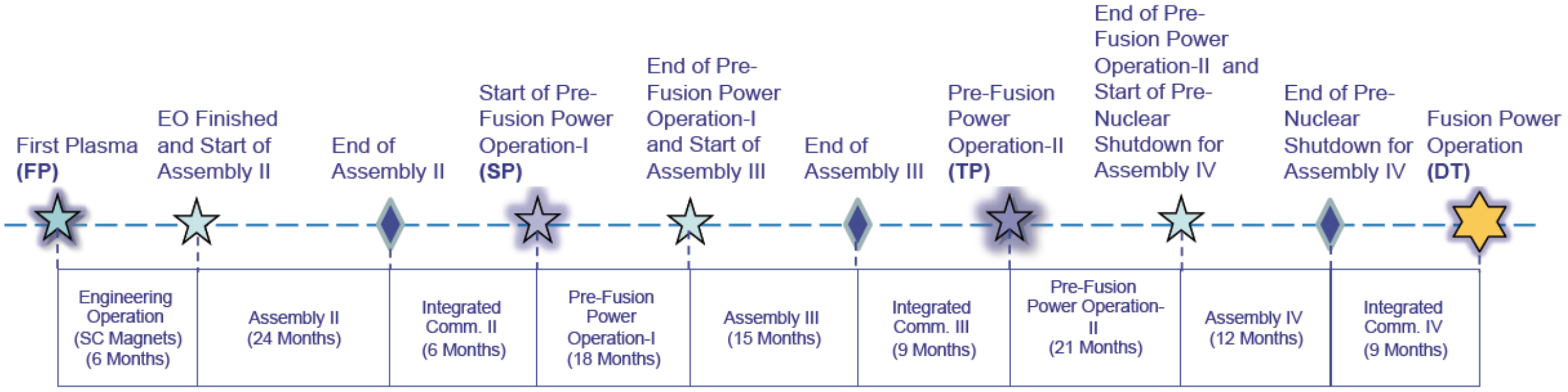

***Figure 2.1. The ITER Research Plan calls for four operational periods interspersed with additional assembly phases on the way to demonstration of high-gain fusion operation.***

The ITER Research Plan [ITR-18-003] divides ITER operation into four operational periods leading toward demonstration of high-gain fusion operation (additional phases are planned but beyond the scope of this report). These will be interspersed with additional Assembly phases during which capabilities (plasma facing components, heating systems, tritium handling capabilities, etc.) will be added. The initial phases will be:

- Assembly I (A1) includes all preparation for First Plasma. At this writing, ITER is in Phase A1.

- First Plasma (FP) will be carried out as part of a sequence of integrated commissioning of the ITER device. The goal of the FP phase is, from a physics point of view, very rudimentary: Demonstrate correct functioning of the tokamak and readiness for operations by producing a plasma current of at least 100kA for at least 100ms. The plasma may not burn through as it will be operated with only temporary plasma facing components in place. From an engineering perspective this will be a critical demonstration. FP will be followed by a period of "Engineering Operation" during which additional plasmas may be run.

- Assembly II (A2) will continue assembly of the tokamak, including the first wall and 20MW of ECI. By the end of A2, ITER will be ready to being experimental operation.

- Pre-Fusion Power Operation I (PFPO-I): This first of two non-activated (hydrogen and/or helium plasmas) experimental campaigns will begin physics studies in the ITER Tokamak. Heating power will be very limited, with 20 MW of ECH (option for 30 MW) available. H-mode operation may require helium plasmas but will be highly desirable. PFPO-I will present an early opportunity to develop operational procedures for all phases of plasma control, ideally including ELM control and commissioning of the disruption mitigation system (DMS). Most diagnostics should be in place, so although PFPO-I will operate only at ⅓ (1.85T) and ½ (2.65T) field, ITER should begin producing significant physics results and allow a beginning of serious model validation at ITER scale. It will be important to validate the DMS' performance here as Assembly Phase III, after PFPO-I, may be the last "easy" opportunity for DMS upgrades.

- During Assembly III (A3), most of ITER's non-nuclear systems will be made ready, most notably the installation of 33MW of NBI and 20MW of ICRI. This is likely to be the last opportunity to improve on SPI as ITER's DMS, if experiments during PFPO-I show it is necessary.

- Pre-Fusion Power Operation II (PFPO-II): The ITER Tokamak will be at near-full capabilities, with 20 (or 30) MW of ECH, 20 MW of ICRH, and 33 MW of NBI available. This will allow most non-nuclear systems to be exercised and characterized to their full capabilities. Full field (5.3T) and current (15 MA) operation will be possible and is included in the Research Plan. Deuterium and tritium will be prohibited in this phase as even deuterium operation could prevent hands-on access during Assembly Phase IV.

- The main task for Assembly IV (A4) will be the installation and commissioning of tritium handling systems. A4 will mark the final opportunity for human access to the tokamak before it becomes irradiated.

- Fusion Power Operation: Nuclear capabilities, especially the infrastructure for handling tritium, will be in place and operation with deuterium and tritium will commence, with the goal of reaching fusion gain of 10. Subsequent FPO phases will increase pulse length and explore other operating scenarios including some aimed at steady-state capabilities.

Each of these phases will be informed by research both past and future. As part of this workshop, participants were asked to evaluate where research is needed to inform and improve the prospects for a successful ITER research program, and what research products from that program can in turn inform the design, assembly, and operation of a Fusion Pilot Plant. This was considered by discussion groups organized by topical areas taking into account:

- Existing US strengths, including modeling and simulation, transient and disruption handling, plasma control, operating scenario development and characterization, etc.
- ITER's identified research needs as documented in ITER Technical Report [ITR-20-008]
- Opportunities for the US fusion program to gain new and unique knowledge and experience that will contribute directly toward an FPP, particularly in areas of engineering, facility maintenance, technology, and diagnostic systems
- Priorities and results that would accelerate the development of an FPP

The participants identified three overarching "General Initiatives" that are described in the next section. Also from these discussions, a total of 54 Topical Initiatives were identified, both for ensuring ITER's success and to contribute to a future FPP. These are listed in Section 2.2 and described fully in Appendix E. Finally, we developed a set of seven Missions (Section 2.3) whose scope encompasses the Topical and General Initiatives. The work described here can begin immediately.

As an active research program, and indeed as part of the larger research program with the IO and other Members, we expect discoveries will motivate changes in our research needs and priorities. The research program described here should not be static and should be revisited periodically in updates to this workshop.

***ACTION: We must promptly and broadly engage with ITER, to gain knowledge in identified missions and initiatives and to support any needs that may arise in coordination with the US ITER research team and fusion community.***

***ACTION: We must establish an equitable, accessible, inclusive knowledge base integrating US and partner developments to support ITER and enable a US FPP***

## 2.1. Overarching Research and Technical Needs

Several general research and technical needs have been identified that cross-cut multiple topical groups. The following two General Initiatives will ultimately impact the success of research within the seven Missions to most effectively contribute to ITER and gather knowledge for a US FPP. Work on these General Initiatives can and should begin immediately. Both are most closely associated with Mission 6 but all USIRT Members should support these where possible and appropriate.

### 2.1.1. GI1: ITER Knowledge

**Objective**: Improve US access to and utilization of ITER information

- Embed members of the US program in existing operation and maintenance teams on site to gather knowledge of technologies and techniques that will be required for an FPP.
- Reduce barriers to access technical information and documents and increase mechanisms to transmit lessons learned.
- Establish methods for long term ITER knowledge management and internalization by the US fusion program, such as up-to-date searchable document and content management systems, databases, data exchange and storage, know-how and training.

Importantly, the products of ITER are more than just scientific results. ITER has established industrial processes and supply chains for many components that will eventually be needed for a power plant scale facility, including high voltage technology, large scale cryo systems, first of a kind components, and peripherals. ITER's nuclear technology, including tritium handling and processing, will contribute new and essential experience and knowledge. The processes involved in assembly, commissioning, and licensing are already happening on an unprecedented scale. It will be key to leverage all of this experience and knowledge in industry as we move towards constructing a pilot plant.

ITER has necessitated the development of solutions for many challenges for the design of a steady-state fusion environment, such as steady state plasma heating, plasma control, radiation resistant diagnostics, managing the tritium fuel cycle, remote maintenance and inspection, plasma facing components, and many others. The US is responsible for 9% of the ITER construction scope, however, it is not clear that this knowledge is being effectively captured to prepare for an FPP. It is important to internalize ITER information to prevent the US from having to repeat the costly lessons already learned from ITER. Beyond successful delivery of US in-kind contributions, it is not a given a priori that the US will have responsibility for operating and exploiting those systems. Such involvement is necessary to gain the understanding of how our systems perform in the field, information that will inform the designs for an FPP; the US must claim operating responsibilities as appropriate. Detailed ITER engineering design solutions for many FPP-relevant engineering, diagnostics, and controls components mitigates design risk and reduces the required FPP design effort. ITER's long-pulse nuclear environment will provide unique information on reliability of components, calibrated diagnostics and controls, and remote maintenance. To fully realize the benefits from this experience, it is essential to have full, low-barrier access to information such as engineering designs, data, codes, performance statistics, maintenance procedures and logs, operational experience, troubleshooting information, etc.

US ITER research must include an on-site presence, enabling vital face-to-face interaction with colleagues from the IO and other parties. Multiple whitepapers called attention to the distinction between "tacit knowledge" and "explicit knowledge": "While explicit knowledge can generally be documented and simply transmitted, tacit knowledge is organically embedded, and can often only be transmitted through human interaction over time." Since there are currently few US personnel "on the ground", we need to begin by gathering documented lessons learned in the assembly phase. Invaluable information gained through on-site participation includes inspection technologies, utilities connect/disconnect, tritium containment, automation of maintenance tasks, remote maintenance, and sensor feedback and human-machine interaction.

A final point on the acquisition and retention of information from ITER is that it is highly desirable for the US to maintain its own domestic copy of all data that ITER produces. Domestic data storage enables USIRCO to efficiently provide data access for US individuals and institutions. Maintaining a domestic copy or data mirror will reduce network traffic across the Atlantic Ocean while ITER is operating, and allow for faster access times for US researchers. In addition it provides a form of backup of the ITER data. Lastly, the data will be in place domestically for when ITER ends operations, and will have accumulated over time, instead of trying to make one mass transfer of a copy at the end of the ITER project.

### 2.1.2. GI2: IMAS Adoption

**Objective**: Modernize and adapt US codes and data to be IMAS-compatible

- Develop an IMAS ecosystem to make adoption of existing codes and workflows easier.
- Develop the infrastructure for converting US experimental data into the IMAS format for inclusion in relevant international databases that can be used for validating ITER operational scenarios.
- Embrace the ITER modeling and analysis tools, and test, validate, and benchmark IMAS workflows at US experimental facilities.

In order for the US to gain and maintain relevance in the ITER research environment, US codes and data urgently need to be adapted to be IMAS-compatible. The Integrated Modeling & Analysis Suite (IMAS) is a comprehensive set of software libraries and codes that use a standardized, machine-generic data model to represent simulated and experimental data with identical structures to support ITER plasma operations and research. US strengths in modeling and simulation of tokamak plasmas could have a transformative impact on the development of IMAS and its capabilities, with direct implications for operational and research capabilities of ITER and for the impact that the US can have on ITER success. Further, the IMAS ontology is ITER's response to the integration challenge in the modeling/engineering space. Any future US FPP development will likely greatly benefit from speaking such a common language, as it will allow different US institutions to contribute to a common design. By leveraging IMAS, each contributor could work on a specific subsystem of the whole plant without each time having to establish new protocols for exchanging information with other contributors.

Adapting existing codes and workflows to IMAS would allow easier and more consistent comparisons between various codes (verification) and between codes and experiment (validation). At a minimum, wrappers should follow the standard approach established by the IO for translation of existing codes to IMAS. The integration of US codes with IMAS should be accelerated, in coordination with the ITER team. Adapting existing US workflows to the IMAS data schema will enable them to be used for ITER, thus ensuring that ITER has the best available tools. Mapping of existing experimental databases to the IMAS data schema should also be enabled. By making all US experimental data available in the unifying IMAS data form, the data of multiple US experiments could more easily be shared and compared by US scientists. Finally, the IMAS workflows should be embraced at US facilities. The complete IMAS software stack (infrastructure, framework, and tools) should then be installed on US-based clusters, and physics modules used and developed to service US device users. Usage at domestic facilities will ensure that the IMAS tools developed are of broad utility and that such knowledge is distributed across US labs, universities, and companies in time for effective engagement in ITER operations.

The US ITER Research Coordinating Office (USIRCO) proposed in this report should work closely with representatives from ITER's Integrated Modeling Expert Group to help organize the implementation of IMAS. Some of the barriers for IMAS adoption are perhaps that reward for implementation does not seem apparent, and access to IMAS can be difficult: IMAS is not open source. However, access is granted through the US ITER agreement, and private or non-government institutions can gain access through collaborative affiliation with an associated government laboratory. The ITER organization outsources the development of IMAS key platforms such as the software infrastructure for data storage, data visualization software, a framework for integrated data analyses, a high-fidelity plasma simulator (HFPS), and a pulse design simulator (PDS). The USIRCO should strategize and facilitate bidding and winning such contracts by US institutions. IMAS compatibility should be considered in hardware requirements and cost analysis of options specific for ITER modeling and data storage needed to facilitate access and utilization of ITER data.

Modeling and simulations will underpin design and operational decisions that have financial and programmatic implications, and in a nuclear environment will be subject to regulations and audit. In general, effort is needed in software engineering across the community to modernize US codes, and research software engineers are needed to help in analysis and in the development of codes. In some cases older models might be suitable, but a plan and support is needed to evaluate, develop/modernize, and maintain these models. In many cases new codes are needed which are scalable on modern hardware and follow modern software standards. Enhanced support is needed to enable career paths specialized in sustainable development of reliable, robust and thoroughly tested software, embedding best practices into developer and user communities, and raising expectations of software quality, testing, and documentation. Delivery of well-documented and usable integrated modeling tools should also include training for users. Two-way communication mechanisms between ITER and US SciDAC (and other) activities are needed to better leverage US efforts and determine the benefit of maintaining US tools vs adopting other international tools from the ITER project.

As a standard, the adoption of the IMAS data schema would benefit greatly from being as widely available as possible. Currently, as a product of the IO, it is restricted to ITER members, and an ITER account is required to access the standard. The US should advocate for the open release of the IMAS data schema. This probably depends on the US delegation to the ITER Council proposing an open release, as the ITER Organization (IO) has indicated it would take an act of the ITER Council to release the schema in a more public way.

## 2.2 Topical Research Initiatives

A set of 28 topical initiatives for ensuring ITER's success and 26 for transferring ITER knowledge and results to a Fusion Pilot Plant was defined by the topical discussion groups and are documented in detail in Appendix E. These are summarized below in Tables 2.1 and 2.2.

**Table 2.1. Topical Initiatives to ensure ITER's success**

| | |
|---|---|
| PMI-1A | Leverage modeling capabilities and lab-scale experiments to predict extreme PMI and material response for ITER PFCs, assess implications, and influence ITER operations |
| PMI-1B | Predict and control the effects of transient events on ITER PFCs |
| PMI-1C | Validate heat flux footprint models and reduce uncertainty in heat flux mitigation requirements |
| PMI-1D | Advance PMI understanding to better assess long term wall material migration |
| PMI-1E | Carry out an advanced materials development program (primarily aimed at FPP, but may apply to future stages of ITER) |
| DIVSOL-1A | Further develop, apply, and experimentally validate plasma edge transport and kinetic neutral transport models for both axisymmetric and non-axisymmetric analysis of SOL and divertor transport |
| DIVSOL-1B | Determine the detachment capability of the ITER divertor in order to meet the requirements for 8 minute long, high-power H-mode discharges |
| DIVSOL-1C | Develop tools for divertor heat flux control |
| DIVSOL-1D | Assess three-dimensional effects on divertor fluxes arising from RMP ELM control |
| SSC-1A | Develop ITER Scenario Solutions |
| SSC-1B | Demonstrate ITER operational pulse planning and control workflow |
| SSC-1C | Manage core and edge instabilities |
| SSC-1D | Qualify techniques to avoid disruptive phenomena in ITER and integrate with plasma control system |
| SSC-1E | Qualify ITER control and exception handling algorithms |
| DISMIT-1A | The US should form a task force to ensure that ITER will not be held back due to disruptions |
| DISMIT-1B | Alternative DMS technologies should be developed in a serious way, especially those that can adequately react to disruptions that rapidly onset due to UFOs |
| ELM-1A | Develop and demonstrate techniques for H-mode access and ELM mitigation/suppression and intrinsically non-ELMing regime access in the conditions anticipated during PFPO-I |
| ELM-1B | Increase focus on modeling ITER scenarios and assess compatibility of the various ELM control approaches and intrinsically non-ELMing alternative scenarios |
| EP-1A | Redirect and coordinate energetic particle research efforts to address ITER-relevant questions. |
| EP-1B | Provide first-principle codes, as well as reduced-physics models that can be integrated into time-dependent integrated modeling codes |
| EP-1C | Develop new synthetic diagnostic algorithms to simulate the various ITER EP diagnostics and enable efficient code validation |
| TC-1A | Expand, develop, and validate full and reduced models to enable proactive evaluation of ITER scenarios, including fundamental questions not addressed in the research plan |
| MODSIM-1A | Develop tools to simulate entire ITER discharges in advance |
| MODSIM-1B | Develop models at all levels of fidelity |
| DIAG-1A | The US should develop a plan and a workforce for participation in the commissioning, operation, and exploitation of the US-contributed ITER diagnostics |
| DIAG-1B | Develop synthetic diagnostics and data interpretation workflows for inclusion in predictive simulations for ITER and for plasma control demonstrations using ITER diagnostic set |
| TECH-1A | Become actively involved in commissioning and startup |
| TECH-1B | Engage US technology experts in addressing near-term ITER needs |

**Table 2.2. Topical Initiatives to prepare for an FPP**

| | |
|---|---|
| PMI-2A | Utilize ITER for experimental and computational studies to advance the high priority CPP FST-SO-A* through understanding of materials interactions, performance, PFC design, and validated predictive modeling capability |
| PMI-2B | Evaluate effects of nuclear environment on materials and the consequent feedback onto PMI and divertor solution |
| PMI-2C | Evaluate build-up and migration of slag and mixed materials and their effects on plasma performance |
| PMI-2D | Inform predictive models on erosion, cracking, melting, and dust formation in ITER conditions to enable projections of overall lifetime of the first wall and PFCs |
| DIVSOL-2A | Develop and test plasma edge and plasma material interaction models, both 2-D and 3-D, that can be validated in a reactor-relevant divertor setting and used in FPP divertor design |
| DIVSOL-2B | Test and validate divertor control schemes for an FPP in ITER |
| DIVSOL-2C | Quantify divertor tolerance to transients |
| SSC-2A | Develop scenario solutions with integrated core-edge approach |
| SSC-2B | Demonstrate operational pulse planning and control workflow, including limited actuators, alpha heating, and ash removal |
| SSC-2C | Manage core and edge instabilities |
| SSC-2D | Qualify disruption prediction, prevention, detection and avoidance and integrate with plasma control system |
| SSC-2E | Test control and exception handling algorithms |
| DISMIT-2A | Demonstrate DMS reliability for proof of viability for tokamak reactors |
| DISMIT-2B | Study runaway electrons seeded by the nuclear environment (tritium decay and Compton scattering) |
| ELM-2A | Develop predictive capability for ELM control in an FPP |
| EP-2A | Use physics results from ITER to validate calculations of EP instabilities, transport rates, and scaling in codes and project alpha heating dynamics for the FPP |
| EP-2B | Use ITER data to validate and make projections of wall loading by energetic particles in a US FPP |
| EP-2C | Use experience with EPs on ITER to inform on the minimum set of measurements to be able to control the reactor, e.g. dedicated measurements for burn control and heat loads caused by EP losses |
| TC-2A | Develop experimentally validated, first-principles predictive capability for burning plasmas, including turbulence and transport, while spanning core to SOL for use in predicting and planning behavior in later phases of ITER and an FPP |
| MODSIM-2A | Carry out validation and further development of physics, engineering, and operational models at reactor scale |
| MODSIM-2B | Enable easy, persistent, and rapid access to ITER data, facilitate remote participation, and rapid turnaround of results |
| MODSIM-2C | Test the full cycle (analysis, predictions, pulse-design, control) |
| DIAG-2A | Utilize ITER to gain experience with the control of a reactor scale, long-pulse burning plasma |
| TECH-2A | Develop relevant technology and integration tools |
| TECH-2B | Gain experience with a tritium fuel cycle |
| TECH-2C | Transfer knowledge of fusion reactor engineering and operations |

## 2.3. Research Missions

Taking the Topical Research Initiatives as input, we have assembled a set of seven initial Research Missions that could be organized within the US ITER Research Program. Tasks within each of these Missions should begin immediately (in most cases, work is already underway) and will evolve through the ITER Research Program. Each Mission will both support achievement of a successful ITER Research Program and provide knowledge and experience that can be applied forward to a Fusion Pilot Plant. Missions cross-cut multiple topical areas, should have clear objectives and seek key results that, when complete, accomplish the objectives, and should be updated periodically.

The Research Missions are:

1. Disruption Prevention and Mitigation
2. Technology Engagement and Transfer
3. Materials Evaluation
4. Heat and Particle Exhaust Handling
5. Operating Scenarios and Plasma Control
6. Modeling, Simulation, and Data Handling
7. Core-Edge Integration

These Missions were identified using input received from the poll following the final workshop meeting, where participants were asked to rank the top initiatives that the US should focus on for ITER's success, and separately, the top initiatives where ITER research would help contribute to a US FPP [see Appendix F]. The US ITER Research Coordinating Office, working with the Research Team, will be responsible for continually evaluating and prioritizing specific research, tasks, and deliverables within these Missions, which will evolve and require a comprehensive view of national and overall ITER Research Program interests.

The organization of the overall ITER research program is as of yet undefined. As that process evolves, US researchers may have opportunities to influence the overall research organization. It is important to note that these Missions will be undertaken as a part of larger efforts involving the IO and all ITER Members. It can be expected that over time we may need to adjust the definition and scope of these Missions to better interface with our partners and to take account of new research results.

### 2.3.1. Mission 1: Disruption Prevention and Mitigation

**Objective**: Ensure that ITER and subsequent tokamaks can operate without damage or loss of operating time due to plasma disruption.

**Background**: Disruptions are considered the largest threat to the ITER Research Program, and effectiveness of a Disruption Mitigation System (DMS) is considered the #1 priority for ITER. ITER’s first DMS is built around Shattered Pellet Injection (SPI). Studies of SPI have expanded from DIII-D to international tokamaks, and many are working together to advance the scientific and technical basis for SPI to the point where we can have confidence in SPI's deployment in ITER. Also, the IO has indicated that there could be an opportunity to upgrade the DMS following PFPO-I, but any alternatives would have to reach a high level of maturity prior to their consideration. Successful mitigation can protect against damage to ITER, but even a mitigated disruption causes loss of operating time and thus limits productivity. Mitigation should be a last resort, and ongoing research into control techniques to operate near stability boundaries and avoid disruption is underway. This line of research will need to continue within the US domestic program and international collaborations, with the results to be deployed in ITER's PCS.

In recognition of the extreme importance ITER, the international community, and the US fusion community place on this area, Topical Initiative DISMIT-1A explicitly recommends "the US should form a task force to ensure that ITER will not be held back due to disruptions;" this should be a guiding principle for the Mission. This Mission will build on long-standing US leadership in Disruption Prediction, Avoidance, and Mitigation (DPAM) and serves to both protect our investment and maintain progress. This Mission will enable close and rapid coordination among domestic efforts and facilities, and should augment the efforts of the existing international ITER DMS Task Force. Table 2.3 indicates the timeline for some of the major elements of WG1. More details can be found in the descriptions of the associated topical initiatives (Table 2.4) in Appendix E.

**Table 2.3. Timeline of Disruption Prevention and Mitigation Key Results**.
For each phase, the timeline is shown for Disruption Mitigation (top) and Prevention (bottom)

|  |  |
|---|---|
| A1/FP | • Working with the international community, complete the scientific and technical basis for the Shattered Pellet Injection (SPI) DMS<br>• Develop alternative DMS technologies for possible installation during A3 |
| | • Validate disruption models on existing tokamaks<br>• Continue to develop and validate control techniques to manage core and edge instabilities; integrate this into ITER PCS<br>• Work to predict impact of mitigated and unmitigated disruptions on plasma-facing components |
| A2/PFPO-I | • Commission ITER SPI system and determine whether modification or alternative technique is needed<br>• If needed, alternative DMS must be brought to maturity in time for installation during A3<br>• Validate DMS disruption simulations on existing tokamaks |
| | • Prepare passively stable operating scenarios for ITER<br>• Continuous process to qualify disruption prediction, detection, and avoidance and integrate with control system |
| A3/PFPO-II | • Any hardware modifications to ITER DMS will be installed during A3<br>• Continue extension of DMS techniques to 15MA in ITER |
| | • Continue to qualify disruption prediction, detection, and avoidance under increasingly challenging conditions and integrate with control system<br>• Test control and exception handling algorithms |
| A4/FPO-1 | • Build confidence in DMS reliability during high-power nuclear operation<br>• Study runaway electrons seeded by nuclear environment |
| | • Build confidence in reliability of control system to maintain high fusion gain while avoiding disruptions |

**Table 2.4. Topical Initiatives Included in Mission 1**

| Ensure ITER's success | | Prepare for FPP | |
|---|---|---|---|
| **A. Disruption Mitigation** | | | |
| DISMIT-1A | Includes "Develop the SPI process and technology to ensure reliability and effectiveness in a reactor environment" | DISMIT-2A | Demonstrate DMS reliability for proof of viability for tokamak reactors |
| DISMIT-1B | Alternative DMS technologies should be developed in a serious way, especially those that can adequately react to disruptions that rapidly onset due to UFOs | DISMIT-2B | Study runaway electrons seeded by the nuclear environment (tritium decay and Compton scattering) |
| **B. Disruption Prevention** | | | |
| SSC-1C | Manage core and edge instabilities | SSC-2D | Qualify disruption prediction, prevention, detection and avoidance and integrate with plasma control system |
| SSC-1D | Qualify techniques to avoid disruptive phenomena in ITER and integrate with plasma control system | SSC-2E | Test control and exception handling algorithms |
| SSC-1E | Qualify ITER control and exception handling algorithms | | |
| PMI-1B | Predict and control the effects of transient events on ITER PFCs | | |

### 2.3.2. Mission 2: Technology Engagement and Transfer

**Objective**: Gain understanding of critical fusion technologies, especially those that form the basis for US hardware contributions. Ensure that US hardware contributions fulfill their roles in the ITER program. Ensure that all knowledge of fusion technology, developed by all parties through all phases of the ITER program, is acquired by the US and made available to support our continuing technology and fusion system studies programs and design of an FPP.

**Background**: Through its contributions to ITER construction, the US is applying knowledge developed through decades of research in its fusion technology programs. ITER commissioning and operation will provide opportunities to learn first-hand how magnet, heating, fueling, diagnostic, and tritium handling technologies perform in a burning plasma system. Responsibilities for their operations are not yet defined, but the US has an interest in being deeply involved in the commissioning and operation of its contributed equipment. In so doing we will be able to ensure, first, that our hardware contributions fulfill their roles in the ITER program and second, that the knowledge gained is coupled back to our continuing technology and fusion system studies programs and design of an FPP. More broadly, involvement in ITER operation can give us critical experience in technologies such as remote handling, neutral beam heating, and plasma facing components, where others have made the major investments, but which will be vital to future US steps in fusion energy development.

**Table 2.5. Timeline of Technology Engagement and Transfer Key Results**

| | |
|---|---|
| A1/FP | ● Actively participate in system commissioning, integrated commissioning and First Plasma, and the Engineering Operation phase<br>● Work with the USIPO to ensure no opportunities are lost for participation in commissioning and operation of US-supplied hardware<br>● Engage US technology experts to identify and address near-term ITER needs |
| A2/PFPO-I | ● Participate in operations of ITER facility |
| A3/PFPO-II | ● Participate in operations of ITER facility<br>● Transfer Knowledge of Fusion Reactor Engineering and Operations |
| A4/FPO-1 | ● Participate in operations of ITER facility<br>● Transfer Knowledge of Fusion Reactor Engineering and Operations<br>● Gain experience with the control of a reactor scale, long-pulse burning plasma<br>● Gain experience with the tritium fuel cycle |

**Table 2.6. Topical Initiatives Included in Mission 2**

| Ensure ITER's success | | Prepare for FPP | |
|---|---|---|---|
| TECH-1A | Become actively involved in commissioning and startup | TECH-2A | Develop Relevant Technology and Integration Tools |
| TECH-1B | Engage US technology experts in addressing near-term ITER needs | TECH-2B | Gain experience with a Tritium Fuel Cycle |
| DIAG-1A | The US should develop a plan and a workforce for participation in the commissioning, operation, and exploitation of the US-contributed ITER diagnostics | TECH-2C | Transfer Knowledge of Fusion Reactor Engineering and Operations |
| DIAG-1B | Develop synthetic diagnostics and data interpretation workflows for inclusion in predictive simulations for ITER and for plasma control demonstrations using ITER diagnostic set | DIAG-2A | Utilize ITER to gain experience with the control of a reactor scale, long-pulse burning plasma |

### 2.3.3. Mission 3: Materials Evaluation

**Objective**: Address the full scope of 'lifetime' for ITER PFCs, observing PFC performance, characterizing how PMI impacts operations, and quantifying issues and limitations imposed by material migration and slag management. Utilize ITER results to predict PFC lifetime and design advanced materials and PFCs suitable for a FPP.

**Background**: The Materials Evaluation Mission is needed to extend our current understanding of PMI to better prepare for ITER and capture ITER's results to inform the choices of materials and mitigation strategies for a pilot plant. Although JET has already carried out studies with its ITER-like wall, ITER's plasma-facing components will enter new regimes that far surpass present-day absolute parameter space. ITER will generate critical data of synergistic effects on neutron irradiation in tungsten and will provide an integrated test of tungsten monoblocks in a divertor at high heat fluxes. The robustness of the actively cooled, monoblock-style ITER divertor will provide critical input on several design choices for the FPP. With respect to materials selection, ITER PFCs will greatly advance our understanding of materials limitations under long-pulse, high flux PMI conditions. ITER's wall materials will almost certainly be inadequate for application in an FPP, motivating further research that must be simultaneously informed by experience in ITER. For example, the resistance of the ITER W monoblocks to various damage mechanisms will provide guidance on the need for new W alloys or composites in the FPP. Thus, a dedicated near-term effort, supported by Mission 2: Technology Engagement and Transfer, is needed to facilitate knowledge transfer of component design, installation, and performance of ITER's plasma-facing components, including plasma physics constraints, thermomechanical requirements, and assembly/remote handling techniques and limitations. Real-time observations of PFC performance at ITER during the start of PFPO-I can immediately inform the FPP strategy during the design phase. The US provided Upper IR/Visible Cameras will serve as a key diagnostic system for divertor surface temperature monitoring and heat load calculation during plasma operation.

The US program will bring expertise in plasma-material science, material testing, high-performance computing, advanced modeling, and simulations to test, analyze and evaluate ITER fusion materials. This Mission will also foster private-public research partnerships to drive fusion material development and component designs for an FPP.

**Table 2.7. Timeline of Materials Evaluation Key Results**

| | |
|---|---|
| A1/FP | ● Leverage modeling capabilities and lab-scale experiments to predict extreme PMI and material response for ITER PFCs and assess implications |
| A2/PFPO-I | ● Utilize ITER to advance understanding of materials interactions, performance, PFC design, and validated predictive modeling capability<br>● Ensure utilization of US supported and other diagnostics to observe PFCs |
| A3/PFPO-II | ● Evaluate PFC heat management and design performance, early material migration, and consequences for plasma operations.<br>● Develop solutions for slag management including in-situ monitoring and removal |
| A4/FPO-1 | ● Utilize ITER to advance understanding of materials interactions, performance, PFC design, and validated predictive modeling capability<br>● Quantify divertor and first wall tolerance to transients<br>● Evaluate effects of nuclear environment on materials |

**Table 2.8. Topical Initiatives Included in Mission 3**

| Ensure ITER's success | | Prepare for FPP | |
|---|---|---|---|
| PMI-1A | Leverage modeling capabilities and lab-scale experiments to predict extreme PMI and material response for ITER PFCs, assess implications, and influence ITER operations | PMI-2A | Utilize ITER for experimental and computational studies to advance the high priority CPP FST-SO-A* through understanding of materials interactions, performance, PFC design, and validated predictive modeling capability |
| PMI-1B | Predict and control the effects of transient events on ITER PFCs | PMI-2B | Evaluate effects of nuclear environment on materials and the consequent feedback onto PMI and divertor solution |
| PMI-1D | Advance PMI understanding to better assess long term wall material migration | PMI-2C | Evaluate build-up and migration of slag and mixed materials and their effects on plasma performance |
| PMI-1E | Carry out an advanced materials development program (primarily aimed at FPP, but may apply to future stages of ITER) | PMI-2D | Inform predictive models on erosion, cracking, melting, and dust formation in ITER conditions to enable projections of overall lifetime of the first wall and PFCs |
| | | DIVSOL-2A | Develop and test plasma edge and plasma material interaction models, both 2-D and 3-D, that can be validated in a reactor-relevant divertor setting and used in FPP divertor design |
| | | DIVSOL-2C | Quantify divertor tolerance to transients |
| | | EP-2B | Use ITER data to validate and make projections of wall loading by energetic particles in a US FPP |

## 2.3.4. Mission 4: Heat and Particle Exhaust Handling

**Objective**: Predict and control both steady and transient heat and particle fluxes flowing to the divertor and main chamber walls, while reducing the transport of eroded PFC's and any injected impurities upstream.

**Background**: In the scrape-off layer (SOL), cross-field and parallel transport connects particles and heat to the divertor targets, and its width determines the required radiative fraction and corresponding impurity seeding levels for acceptable divertor target power loads. The SOL width scaling greatly impacts the ability to design a lower cost, more robust FPP divertor, and measurement and simulation of these processes in ITER is important for determining viable FPP configurations and operational scenarios. Even early data in PFPO-I can begin to validate heat flux width scalings and whether effects (turbulence, ELMs, etc) cause the heat flux width to broaden.

Plasma exhaust, both steady and transient (ELMs), poses a challenge to ITER's divertor regardless of the choice of materials. Measures must be taken to ensure peak energy fluxes striking the divertor do not exceed materials limits. ITER will operate with a partially detached divertor, in which much of the energy will be radiated away before impinging on the divertor surface. Obtaining and controlling this detachment is an ongoing challenge and is a topic of active current research.

Control of ELMs is required for ITER scenarios from the non-active phases of operations to DT burning plasmas for avoidance of uncontrolled erosion of first wall and divertor plasma-facing components (PFCs) and for tungsten impurity control. ITER will have capabilities to use techniques developed in present-day devices to mitigate or suppress the ELMs, but these have not yet been demonstrated in ITER-like conditions, e.g. low torque, high $T_e/T_i$, high $n/n_{GW}$, low collisionality. In addition, confinement degradation typically accompanies ELM suppression by RMPs.

Thus, Mission 4 should work to a) ensure measurements are obtained to validate heat flux width scaling, b) achieve and demonstrate control of divertor detachment, and c) predict and demonstrate control of ELMs, main chamber erosion, and resulting impurity influx and outflux. A coordinated domestic effort in modeling, advanced diagnostic development, and experimental validation on current devices (domestic and international) should be pursued to advance heat and particle exhaust capabilities in preparation for ITER operation. The work must be cognizant of the tightly coupled nature of the problem, where pedestal turbulence and impurity transport, SOL impurity flows and drifts, as well as surface sputtering, erosion, redeposition, and morphology are nonlinearly coupled and depend on boundary temperatures and densities as described in Mission 7: Core-Edge Integration.

**Table 2.9. Timeline of Heat and Particle Exhaust Handling Key Results**.
For each phase, the timeline is shown for divertor detachment (top) and ELM control (bottom).

|  |  |
|---|---|
| A1/FP | • Continue to develop techniques and understanding to access and control of the partially detached divertor conditions anticipated in ITER<br>• Validate heat flux footprint models including the effects of 3D fields used for RMP ELM control and the role of turbulence broadening |
| | • Develop techniques to minimize the power needed to enter H-mode and ensure early H-mode access during PFPO-I<br>• Continue to develop ELM control techniques (RMP, pellet pacing, and naturally ELM-free scenarios) in present-day devices under increasingly ITER-like conditions and use the results to validate predictive models |
| A2/PFPO-I | • Begin to determine the detachment capability of the ITER divertor in order to meet the requirements for 8 minute long, high-power H-mode discharges<br>• Develop techniques to avoid core contamination by tungsten PFC’s as well as any injected impurities |
| | • Access ELMing H-mode, preferably in hydrogen<br>• Apply and begin to optimize ELM control tools<br>• Evaluate and optimize intrinsically non-ELMing scenarios |
| A3/PFPO-II | • Continue developing divertor scenarios as conditions approach those of an ITER burning plasma<br>• Test and validate divertor control schemes<br>• Further develop avoidance strategies to prevent core impurity contamination |
| | ● Continue to optimize ELM control tools and evaluate/optimize naturally ELM-free scenarios |
| A4/FPO-1 | ● Test and validate divertor control schemes for an FPP |
| | ● Develop predictive capabilities for ELM control in an FPP |

**Table 2.10. Topical Initiatives Included in Mission 4**

| **Ensure ITER's success** | | **Prepare for FPP** | |
|---|---|---|---|
| DIVSOL-1B | Determine the detachment capability of the ITER divertor in order to meet the requirements for 8 minute long, high-power H-mode discharges | DIVSOL-2B | Test and validate divertor control schemes for an FPP in ITER |
| DIVSOL-1C | Develop tools for divertor heat flux control | ELM-2A | Develop predictive capability for ELM control in an FPP |
| PMI-1C | Validate heat flux footprint models and reduce uncertainty in heat flux mitigation requirements | | |
| DIVSOL-1D | Assess three-dimensional effects on divertor fluxes arising from RMP ELM control | | |
| PMI-1A | Leverage modeling capabilities and lab-scale experiments to predict extreme PMI and material response for ITER PFCs, assess implications, and influence ITER operations | | |
| PMI-1B | Predict and control the effects of transient events on ITER PFCs | | |
| ELM-1A | Develop and demonstrate techniques for H-mode access and ELM mitigation/suppression and intrinsically non-ELMing regime access in the conditions anticipated during PFPO-I | | |
| ELM-1B | Increase focus on modeling ITER scenarios and assess compatibility of the various ELM control approaches and intrinsically non-ELMing alternative scenarios | | |

### 2.3.5. Mission 5: Operating Scenarios and Plasma Control

**Objective**: Prepare a range of operating scenarios and the tools to access and control them to successfully achieve the goals of each stage of the ITER research program.

**Background**: The US fusion community is a leader in both development of operating scenarios and in plasma control, both of which will be applied to ITER. Plasma control is becoming progressively more precise, leading toward the ability to "dial in" desired scenarios on ITER. Scenario development can be complex, with a need not only to identify desirable operating points for ITER but also stable access and exit paths.

**Table 2.11. Timeline of Operating Scenarios and Plasma Control Key Results**

| | |
|---|---|
| A1/FP | • Continue to develop plasma control techniques for application to ITER, with special focus on managing core and edge instabilities<br>• Develop and demonstrate exception handling algorithms in current tokamaks<br>• Continue to develop controllable operating scenarios that can achieve ITER's objectives for each research phase, with integrated core-edge and from breakdown to shutdown |
| A2/PFPO-I | • Operate the ITER PCS providing a first demonstration; optimize and improve algorithms as appropriate<br>• Demonstrate and optimize operating scenarios appropriate for this early phase<br>• Achieve H-mode and begin first efforts at applying ELM control techniques<br>• Continue scenario development and optimization on domestic tokamak(s) |
| A3/PFPO-II | • Scenario and control development continues as ITER reaches full capabilities |
| A4/FPO-1 | • Full test of exception handling algorithms<br>• Demonstrate operational pulse planning and control workflow, including limited actuators, alpha heating, and ash removal, including<br>○ Demonstrate burn control at Q=10<br>○ Demonstrate steady-state scenarios at Q≈5<br>• Validate predictive capabilities for application toward FPP |

**Table 2.12. Topical Initiatives Included in Mission 5**

| Ensure ITER's success | | Prepare for FPP | |
|---|---|---|---|
| SSC-1A | Develop ITER Scenario Solutions | SSC-2A | Develop scenario solutions with integrated core-edge approach |
| SSC-1B | Demonstrate ITER operational pulse planning and control workflow | SSC-2B | Demonstrate operational pulse planning and control workflow, including limited actuators, alpha heating, and ash removal |
| SSC-1C | Manage core and edge instabilities | SSC-2C | Manage core and edge instabilities |
| SSC-1D | Qualify techniques to avoid disruptive phenomena in ITER and integrate with plasma control system | SSC-2D | Qualify disruption prediction, prevention, detection and avoidance and integrate with plasma control system |
| SSC-1E | Qualify ITER control and exception handling algorithms | SSC-2E | Test control and exception handling algorithms |
| ELM-1A | Develop and demonstrate techniques for H-mode access and ELM mitigation/suppression and intrinsically non-ELMing regime access in the conditions anticipated during PFPO-I | TC-2A | Develop experimentally validated, first-principles predictive capability for burning plasmas, including turbulence and transport, and spanning core to SOL for use in predicting and planning behavior in later phases of ITER and an FPP |
| ELM-1B | Increase focus on modeling ITER scenarios and assess compatibility of the various ELM control approaches and intrinsically non-ELMing alternative scenarios | ELM-2A | Develop predictive capability for ELM control in an FPP |
| EP-1A | Redirect and coordinate energetic particle research efforts to address ITER-relevant questions | EP-2C | Use experience with EPs on ITER to inform on the minimum set of measurements to be able to control the reactor, e.g. dedicated |
| | | | measurements for burn control and heat loads caused by EP losses |
| PMI-1B | Predict and control the effects of transient events on ITER PFCs | DIVSOL-2B | Test and validate divertor control schemes for an FPP in ITER |
| DIAG-1B | Develop synthetic diagnostics and data interpretation workflows for inclusion in predictive simulations for ITER and for plasma control demonstrations using ITER diagnostic set | | |

### 2.3.6. Mission 6: Modeling, Simulation, and Data Handling

**Objective**: Develop and apply Modeling and Simulation tools in preparation for ITER experiments, which in turn provide input for preparing validated models for application to FPP. Integrate IMAS into all modeling, analysis, and simulation codes (**GI2: IMAS Adoption)** and improve US access to and utilization of ITER information (**GI1: ITER Knowledge**).

**Background**: Modeling and Simulation cross-cuts all research areas, and the development and application of validated predictive models is a valuable tool and a critically important product of the entire ITER enterprise. Each stage of ITER operation will improve the predictive modeling tools that can be applied to other elements of the US fusion portfolio up to and including an FPP.

**Table 2.13. Timeline of Modeling, Simulation, and Data Handling Key Results**

| | |
|---|---|
| A1/FP | ● Integrate US codes and tools with IMAS<br>● Provide tools for making ITER data widely available within the US<br>● Develop tools to simulate entire ITER discharges in advance<br>● Develop synthetic diagnostics |
| A2/PFPO-I | ● Apply "Predict First" approach to preparation and execution of ITER experiments and use results to further refine and validate models |
| A3/PFPO-II | ● Apply "Predict First" approach to preparation and execution of ITER experiments and use results to further refine and validate models |
| A4/FPO-1 | ● Use burning plasmas on ITER to validate models spanning the entire device and the entire pulse for application to FPP |

**Table 2.14. Topical Initiatives Included in Mission 6**

| **Ensure ITER's success** | | **Prepare for FPP** | |
|---|---|---|---|
| MODSIM-1A | Develop tools to simulate entire ITER discharges in advance | MODSIM-2A | Carry out validation and further development of physics, engineering, and operational models at reactor scale. |
| MODSIM-1B | Develop models at all levels of fidelity | MODSIM-2B | Enable easy, persistent, and rapid access to ITER data, facilitate remote participation, and rapid turnaround of results |
| DIAG-1B | Develop synthetic diagnostics and data interpretation workflows for inclusion in predictive simulations for ITER and for | MODSIM-2C | Test the full cycle (analysis, predictions, pulse-design, control) |
| | plasma control demonstrations using ITER diagnostic set | | |
| EP-1A | Redirect and coordinate energetic particle research efforts to address ITER-relevant questions | PMI-2A | Utilize ITER for experimental and computational studies to advance the high priority CPP FST-SO-A* through understanding of materials interactions, performance, PFC design, and validated predictive modeling capability |
| EP-1B | Provide first-principle codes, as well as reduced-physics models that can be integrated into time-dependent integrated modeling codes | ELM-2A | Develop predictive capability for ELM control in an FPP |
| EP-1C | Develop new synthetic diagnostic algorithms to simulate the various ITER EP diagnostics and enable efficient code validation | EP-2A | Use physics results from ITER to validate calculations of EP instabilities, transport rates, and scaling in codes and project alpha heating dynamics for the FPP |
| TC-1A | Expand, develop, and validate full and reduced models to enable proactive evaluation of ITER scenarios, including fundamental questions not addressed in the research plan | EP-2B | Use ITER data to validate and make projections of wall loading by energetic particles in a US FPP |
| DIVSOL-1A | Further develop, apply, and experimentally validate plasma edge transport and kinetic neutral transport models for both axisymmetric and non-axisymmetric analysis of SOL and divertor transport | TC-2A | Develop experimentally validated, first-principles predictive capability for burning plasmas, including turbulence and transport, while spanning core to SOL for use in predicting and planning behavior in later phases of ITER and an FPP |

### 2.3.7. Mission 7: Core-Edge Integration

**Objective**: Determine conditions and control requirements for compatibility between the core, pedestal, and boundary regions of the plasma. Coordinate related Topical Research Initiatives to develop effective solutions which prevent erosion of plasma facing components by maintaining heat and particle loads within acceptable limits while avoiding core impurity contamination and consequent performance degradation.

**Background**: Core-edge integration represents a key uncertainty in the design of a high average power FPP. The fundamental challenge arises from the fact that we desire a core plasma that is hot to achieve high fusion gain and an edge plasma that is cold to prevent material erosion from the wall. In presently operating devices, it is not possible to simultaneously achieve a sustained high power density core plasma and a divertor solution with heat and particle fluxes at the same scale as those projected for a pilot plant. Current FPP design scoping relies on extrapolations of models that were validated in regimes at parameters far away from pilot plant conditions. ITER will extend conditions toward burning plasmas starting in PFPO-II, but preparation for ITER will also drive the physics understanding and development of integrated core-edge solutions that could benefit FPP.

For both ITER and FPP success, heat and particle loads on PFCs must be maintained below acceptable limits, impurity sourcing by material erosion must be controlled, and divertor impurity leakage and consequent core contamination must be avoided to maintain fusion performance. In ITER, contamination by tungsten would be especially onerous due to its high radiation efficiency. Suppressing or mitigating ELMs is a requirement to prevent damage to the walls, ideally using intrinsically non-ELMing scenarios, or alternatively via active actuators such as resonant magnetic perturbations or pellet ELM pacing. ELMs are effective in flushing impurities, but ITER will help to address the question of whether impurity content in the core will be reduced due to lower pedestal density gradients, and whether pedestal turbulence or small ELMs are sufficient to minimize core impurity concentrations.

Developing methods and operating scenarios to simultaneously reduce power loads on the divertor target while minimizing core contamination by any injected radiative impurities remains one of the biggest challenges. ITER will provide valuable data in assessing the operational limits and performance with high core radiated power fraction and power through the separatrix close to the L-H power threshold. Control of radiated power fraction must be very precise to avoid transients. At the same time, core particle fueling by pellet injection is expected to be required due to the high neutral opacity in the ITER scrape-off layer, which can increase impurity concentrations as well. There remains uncertainty in the resulting plasma profiles and in scenarios and their access, for example, pellet injection may trigger ELMs in an otherwise ELM-free scenario. All of these challenges are magnified in an FPP.

The highly cross-cutting Core-Edge Integration Mission will connect the work of multiple Topical Research Initiatives and other Missions to address the tightly coupled core-edge integration problem. The Mission should:

- Drive and contribute to the development of integrated simulation capabilities at multiple levels of fidelity that span the core, pedestal, and scrape-off layer, plasma-material interactions, and realistic divertor models.
- Grow and coordinate promotion of operating scenarios and experiments on ITER of US interest for core-edge integration physics and validation
- Leverage existing facilities to prepare for ITER

**Table 2.15. Timeline of Core-Edge Integration Key Results**

| | |
|---|---|
| A1/FP | ● Expand existing model validation efforts using any available experimental data to test extrapolability of core-edge integrated scenarios to ITER<br>● Continue the development of practical codes that address the fundamentally different physics in the pedestal and boundary regions; move beyond using codes which rely on core orderings in the pedestal.<br>● Utilize present facilities to investigate facets of the ITER challenge; for example, freshly boronized conditions to simulate high neutral opacity and high SOL temperatures; test interaction with tungsten PFCs; test methods for maintaining partial detachment without core contamination at low SOL collisionalities.<br>● Apply coupled core-edge models to predict ITER behavior and identify potential challenges to operation, testing models at varying levels of fidelity (though with |
| | kinetic neutrals), including coupled simulations of pedestal/SOL/PMI up to and including fully kinetic models.<br>● Extend intrinsically non-ELMing regimes to improve core-edge integration; leverage turbulence-broadening of heat flux profiles; address key core-edge issues extrapolating these regimes to ITER, such as impurity sourcing and impurity transport, and test pellet fueling. |
| A2/PFPO-I | ● Test and improve models to predict density, temperature and impurity profiles in reactor-like conditions with partially or fully detached divertor targets and scrape-off layers that are opaque to recycled neutral particles<br>● [If H-mode is achievable, otherwise in PFPO-II] Assess access conditions for as many intrinsically non-ELMing or ELM-suppressed scenarios as possible, along with their impact on core confinement and impurity accumulation |
| A3/PFPO-II | ● Use data from ITER integrated plasma scenarios to iterate and improve predictive FPP design models at all levels of fidelity |
| A4/FPO-1 | ● Assess applicability of ITER integrated core-edge scenarios to an FPP device with longer pulse lengths and reactor-relevant hardware and diagnostic limitations |

**Table 2.16. Topical Initiatives Included in Mission 7**

| **Ensure ITER's success** | | **Prepare for FPP** | |
|---|---|---|---|
| PMI-1B | Predict and control the effects of transient events on ITER PFCs | PMI-2A | Utilize ITER for experimental and computational studies to advance the high priority CPP FST-SO-A* through understanding of materials interactions, performance, PFC design, and validated predictive modeling capability |
| PMI-1C | Validate heat flux footprint models and reduce uncertainty in heat flux mitigation requirements | DIVSOL-2A | Develop and test plasma edge and plasma material interaction models, both 2-D and 3-D, that can be validated in a reactor-relevant divertor setting and used in FPP divertor design |
| DIVSOL-1A | Further develop, apply, and experimentally validate plasma edge transport and kinetic neutral transport models for both axisymmetric and non-axisymmetric analysis of SOL and divertor transport | DIVSOL-2C | Quantify divertor tolerance to transients |
| DIVSOL-1D | Assess three-dimensional effects on divertor fluxes arising from RMP ELM control | ELM-2A | Develop predictive capability for ELM control in an FPP |
| SSC-1A | Develop ITER Scenario Solutions | SSC-2A | Develop scenario solutions with integrated core-edge approach |
| ELM-1A | Develop and demonstrate techniques for H-mode access and ELM mitigation/suppression and intrinsically non-ELMing regime access in the conditions anticipated during PFPO-I | TC-2A | Develop experimentally validated, first-principles predictive capability for burning plasmas, including turbulence and transport, while spanning core to SOL for use in predicting and planning behavior in later phases of ITER and an FPP |
| ELM-1B | Increase focus on modeling ITER scenarios and assess compatibility of the various ELM | | |
| | control approaches and intrinsically non-ELMing alternative scenarios | | |
| TC-1A | Expand, develop, and validate full and reduced models to enable proactive evaluation of ITER scenarios, including fundamental questions not addressed in the research plan | | |
| EP-1A | Redirect and coordinate energetic particle research efforts to address ITER-relevant questions | | |

# 3. Organizing for Effective Engagement in ITER

The FESAC report "Powering the Future: Fusion & Plasmas" (2020) calls for the formation of an ITER research team to "make essential contributions to achieving the high gain mission for ITER, exploit access to a burning plasma at reactor scale, and enable US scientists to close the nuclear science and engineering gaps in order to build an FPP" [FESAC 2020, p. 13]. There is broad agreement among participants of this workshop that this team should be structured in such a way as to facilitate the opportunity for participation among all US entities; to speak internationally with a united voice representing their interests; to promote a diverse, equitable, and inclusive workforce; to make decisions quickly, transparently, and with accountability; to prioritize research for the maximum benefit of ITER and the US fusion program; and, fundamentally, to ensure that the US contributes what is necessary for ITER to achieve its mission. In general, there is strong support for creating a centralized organization that can effectively and efficiently serve the mission needs of the US program and ITER, if done in such a way as to promote broad, unfettered, and merit-based access to participation in ITER research by all US institutions, public and private.

## 3.1. Organizational Structure of the US ITER Research Program

Here, we address the workshop charge on the "organization, structure, and modes of operation for flexible, agile, and impactful exploitation of the ITER facility by US participants." We propose a US ITER Research Program with the mission of ensuring ITER's success and returning the experience and knowledge gained to the US fusion program.

***ACTION - We must build a coordinating structure for maximizing US return on and contribution to ITER advances***

The Program should be composed of three elements:

- The **US ITER Research Team (USIRT)**, composed of all individuals contributing to ITER operations and carrying out ITER research for US institutions, charged to propose and execute research and development for ITER and to bring back knowledge and experience to the US program;
- A **US ITER Research Coordination Office (USIRCO)**, responsible for interfacing with DOE and the IO, and for facilitating the broad participation of all US entities in the US ITER Research Team; and
- A **US ITER Research Advisory Board (USIRAB)**, responsible for periodically reviewing and reporting on the effectiveness of the US ITER Research Team organization and leadership.

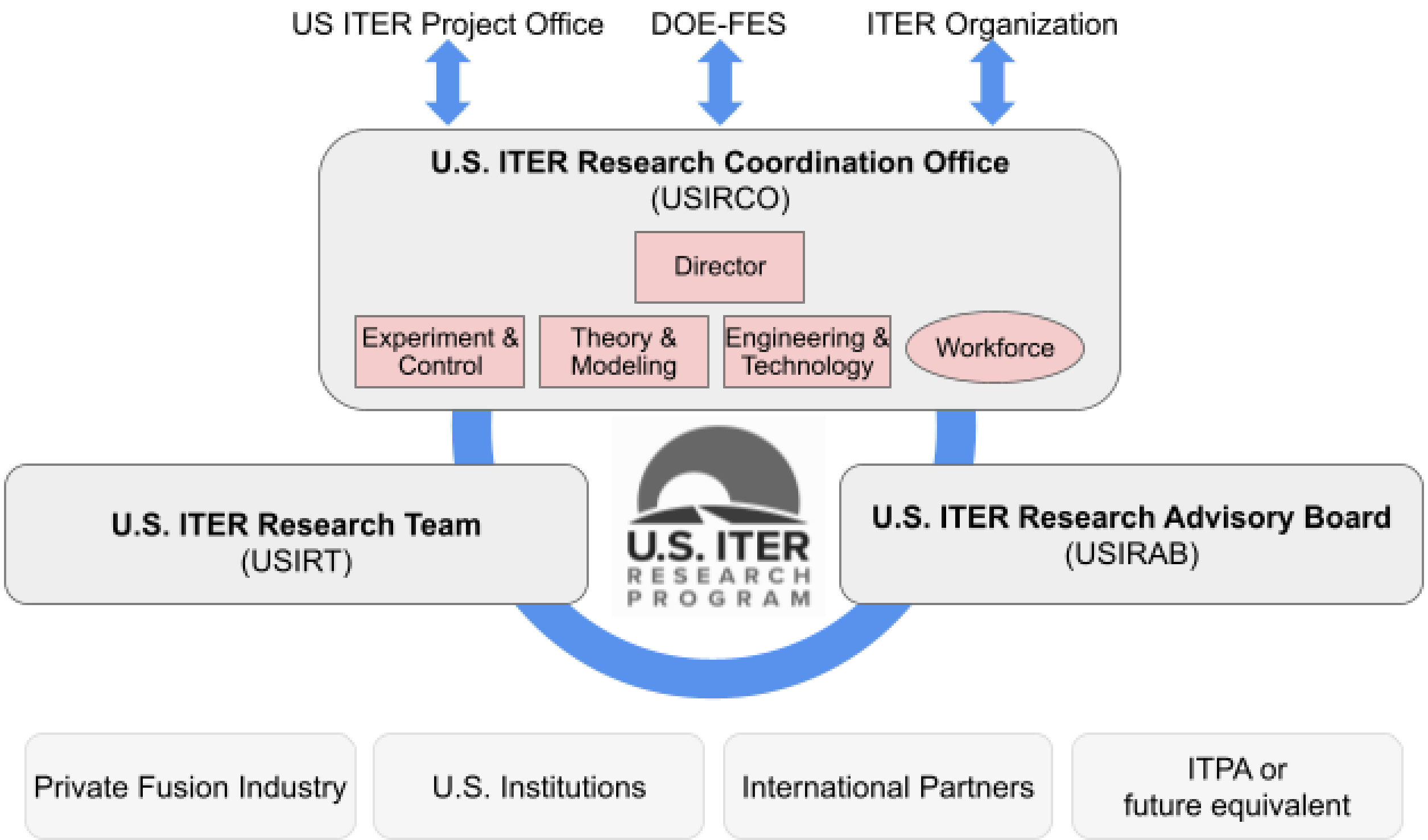

*Figure 3.1. Organizational structure of the US ITER Research Program.*

This structure would be expected to be organized separately but in parallel with the US ITER Project Office (USIPO), which will continue to be responsible for delivering on US contributions to ITER construction, providing hardware through at least the fourth assembly phase prior to Fusion Power Operation. The USIPO, as an organization intended to manage the US in-kind contributions to the construction and assembly of ITER, is sufficiently different in its mission, structure and operation from what we envision for the US ITER Research Program that we do not see significant benefit to creating a unified USIPO/USIRCO organization. Communication and coordination between these organizations will be essential. Moving forward, there may be some areas of overlap in commissioning, operations, and diagnostics, but these can be resolved on a case-by-case basis; it is more important that nothing critical is lost in the gap between the two organizations.

The US shares with the other ITER parties a common interest in ITER successfully achieving its missions. The basic goals of creation, control, and understanding of a high gain, long pulse burning plasma are shared ones. The results needed from ITER for the parties' future steps in magnetic fusion are basically universal, perhaps differing slightly in their order of priorities. Exploiting the ITER facility for scientific benefit will be best accomplished with a One Team approach, in which the parties continue to work in partnership as they have during construction. In a One Team model, multi-party working groups would be formed to attack the scientific and technical challenges at each phase. These would form the "front lines" of ITER research, and be organized by scientific topic or campaign objective. The USIRCO would be responsible for strongly supporting these groups with effective US participation and timely contributions to their work. The US will be a leader to the extent that its expertise and contributions are recognized by all parties as charting a path to overall success.

In the early stages of the US ITER Research Program, we will still be engaged in preparatory research, and the ITPA (International Tokamak Physics Activity, organized under the auspices of the IO) may fill the role of the "One Team" described above. The US program will have to evolve and adapt along with our partners moving into the active ITER research phase.

Broader community interactions with the US ITER Research Program should be explored through community engagement workshops, formal and informal (long form survey) requests for information calls, and external advisory boards (underserved and minority communities, industrial communities, workforce development communities). Interactions and activities should exist through hybrid (mixture of in-person and virtual) events that allow for communities unable to attend or do not thrive in-person to participate in similar manner to in-person communities. For community interactions involving funding opportunities, specific suggestions for different types of engagement are addressed in 4.2 "Working with the broader community."

In defining these elements and their roles, we draw from material provided through white papers as well as presentations at the Phase 2 kickoff on March 23, 2022. Input was received from a number of different community segments, including labs, universities, privately funded fusion entities, and the high energy physics community. Two white papers, "University Participation at the Center of the ITER Research Program" and "A US ITER Coordination Office for Organizing US Participation in ITER Operations," were provided by multi-institutional groups and were particularly impactful in our deliberations.

### 3.1.1. US ITER Research Team

The **US ITER Research Team** (USIRT) is responsible for following through with the research and technical needs outlined in Chapter 2 in order to maximize the return of the US investment in ITER's construction and operation and to ensure US research priorities on ITER strengthen the domestic program to aim at the development of a fusion pilot plant.

The USIRT should provide an equitable, accessible environment where everyone who wants to contribute to the US mission/vision for participation in ITER can participate. Personnel from the US fusion community who are employed by the IO should be welcomed as members with the recognition that, although their job responsibilities may be different than their US-based counterparts, they will be in an excellent position to make valuable contributions toward a US pilot plant. The USIRT should include both on-site (at ITER, but non-IO) and off-site engineers and scientists who engage in ITER operations and research or provide technical support.

The structure of the USIRT should be determined by the USIRCO and is expected to evolve over time. Groups within the USIRT should be organized into **Research Missions**, as described in Chapter 2, that coordinate preparatory R&D and experimental planning starting immediately, and participate in commissioning, operation, and experiments when ITER operates. The USIRT should also carry out specific, time-bound **Projects**, such as modernizing and adapting US codes and data to be IMAS-compatible [General Initiative – IMAS] and improving access to and utilization of ITER information [General Initiative – Knowledge]. Primary responsibility for these crucial initiatives is associated with Mission 6 [Modeling, Simulation, and Data Handling] and Mission 2 [Technology Engagement and Transfer], but contributions should come from across the entire Research Team. USIRT Mission members may initially connect with international activities through existing International Tokamak Physics Activity (ITPA) topical groups. Given the universality and cross-cut nature of ITER issues, we can expect that international working groups would form around the same Mission or Project topics, and that USIRT Members would participate and in some cases lead those groups.

Each of the USIRT Missions or Projects should have a leader, selected as described in Appendix H. These Research Team leaders should work with their groups to identify and prioritize the research tasks that address their objectives and coordinate progress toward their goals. Prioritization should be done considering the context of national fusion energy goals and US FPP progress.

USIRT Leaders should provide regular progress updates and needs to USIRCO. Research could include collaborative activities at other facilities, including domestic (both public and private) and international facilities. For proposed research that would require interfacing with the ITER facility, the USIRCO should provide a Record of Discussion to inform both the reviewers and the applicants of the feasibility, challenges, and mechanisms for interfacing with the facility. The USIRCO should work closely with the IO, DOE-FES, and prospective US participants to ensure that the proposed work does not conflict with IO or ITER Member plans, and can be adequately supported by the program, as is done for proposals for research on US facilities.

Opportunities for funding for USIRT activities should be open to ALL members of US institutions through open DOE calls for proposals. Dedicated funding for routine ITER research (that which does not address emergent issues directly impacting the mission schedule) could be announced through open Funding Opportunity and Lab Announcements from FES, and proposals should be selected by FES for funding through the standard peer review process. Increased interest in USIRT activities could also come through other (future) funding programs, for example, through calls for general burning plasma research or FPP design (including public-private partnership). It should be acknowledged that FOAs put a significant burden on parties involved for submission and more rapid mechanisms are likely to be needed in special cases to address urgent issues. Flexible and rapidly dispersible funding options could involve funded FOAs that have ability to modify or add subcontracts to respond to off-cycle or relatively smaller funding needs. In all cases, the USIRCO should report on priorities for research, hardware, development and training of a diverse fusion workforce, or other topics for consideration in the formulation of funding strategy and selection criteria to ensure US ITER Research Program goals are efficiently accomplished. The USIRCO should have access to some funding that could be distributed to address urgent issues directly.

### 3.1.2. US ITER Research Coordination Office

The **Research Coordination Office** will provide the necessary coordination for the planning and execution of the US ITER Research Program. The USIRCO will ensure that USIRT needs and interests are well supported, issues are rapidly identified and addressed, and US program goals are efficiently accomplished. A primary role of the USIRCO should be to facilitate broad participation of US scientists and engineers in the ITER program. USIRCO should also track the overall status and prioritization of US ITER Research Program activities. The USIRCO should coordinate with the USIPO to ensure that important engineering issues do not get lost in the gap between the two organizations. In order to act efficiently and effectively, the USIRCO will need to have the ability to address emerging research and urgent needs. This will require the USIRCO to have its own budget and authority to direct resources.

The proposed selection process and term lengths of USIRCO appointments are described in Appendix H. These should include an overall Director and Deputy Directors for Research Coordination in major technical areas such as Engineering & Technology, Theory & Modeling, and Experiment and Control, as well as a Deputy Director for Workforce Development (see below). Individuals holding leadership roles or other roles within USIRCO should remain employees of their original institution during their USIRCO term, as with a secondment, and funding from DOE should be made available to support these arrangements.

In order to ensure that workforce development concerns are central to the efforts of the US ITER Research Program, and that the DEI principles are being followed throughout the organization, the USIRCO should include a Workforce Directorate at the leadership level. The directorate would help to recruit, train, retain, and grow the needed workforce. The directorate should be funded and empowered to bring in resources or subject matter experts in DEI that could lend an outside, professional perspective and help develop best practices and strategies for participating research groups, institutions and individuals within the US ITER structure.

The following is a suggested list of roles, responsibility, authority, and accountability of the USIRCO:

- Roles:
  - Interface with the IO
    - Be the primary liaison between the US and the IO for research and operations
    - Serve as the unified voice of the US ITER Research Program when interacting with the IO or other ITER Members
    - Advocate for planning and execution of US interests within ITER research priorities and plans
  - Enable US Research Mission Execution
    - Serve as a liaison between the USIRT and other US entities for rapidly communicating needs
    - Interface with the US ITER Project
    - Track progress and continually reevaluate the ways in which the US can best participate in ITER research and operations.

- Work with domestic facilities and institutions and international facilities to advocate for research supporting US ITER Research Program goals.
        - Connect US researchers to compelling and persistent opportunities to participate in the US ITER Research Program.
    - Disseminate ITER Results
        - Ensure that relevant knowledge gained from ITER participation is dispersed to the domestic program (documentation, workshops, etc).
        - Facilitate coordination, communication, data exchange, and analysis between the on-site and remote participants.
        - Facilitate knowledge transfer of complex requirements for performing successful experiments on ITER, including the extensive pulse and control system modeling requirements.
- Responsibilities:
    - Provide records of discussion to accompany proposals for DOE funding to carry out work that interfaces with ITER facility
    - Ensure and support access to ITER data and software by all US institutions, consistent with ITER agreement
    - Maintain documentation regarding technical processes and information
    - Identify and address urgent scientific and technical ITER needs.
    - Prioritize US research in support of ITER, with input from USIRT, IO, and in cooperation with other elements of the domestic program
    - Work with the IO, DOE, and US entities to expeditiously resolve policies and disputes with regard to intellectual property, review processes, and other rules of engagement with ITER and ITER data
    - Establish relocation and logistical support for personnel from US institutions at the ITER site
    - Determine appropriate balances in personnel and projects among onsite, offsite, and supporting activities
    - Advocate for and connect participants to funding mechanisms that enable broad participation across all institutions, which could include subcontracts from existing grants, or supporting new universities or groups to form proposals or join existing projects.
- Authorities:
    - Strategically assign personnel/projects in order to execute the US ITER Research Program
    - Direct resources to provide support for all USIRT members to access ITER data
    - Direct resources to provide support for making codes compatible with IMAS
    - Direct resources to hold periodic workshops to collect and disseminate information, promote and engage broad participation from all segments of the community, or other topics as necessary
    - Direct resources for workforce development needs including internships, technical or project management training
    - Direct resources to address urgent, emerging ITER needs. This could include support for supplemental run-time at domestic facilities.

- Accountabilities:
    - USIRCO reports to DOE/FES on resources needed, both short- and long-term, to fulfill US commitment to and utilization of ITER research and to enable equitable participation by US stakeholders
    - USIRCO reports to DOE/FES on research priorities and interface costs to inform Funding Opportunity Announcements released by DOE
    - USIRCO will periodically report to USIRAB and seek its advice, including on appointments within USIRCO
    - USIRCO will periodically report to the fusion community on the overall status of Research Missions and prioritization of US ITER Research Program activities

### 3.1.3. US ITER Research Advisory Board

The role of the **Advisory Board** is to provide oversight to ensure that the US ITER Research Team, and in particular the Research Coordination Office, operate effectively and transparently. The primary responsibilities of the Advisory Board are a) to periodically review and report on the effectiveness of the US ITER Team organization and leadership to FES and the US fusion community, and b) to provide feedback on the research priorities in the context of the domestic program. The Advisory Board should publicly report at least once every year, with input from the IO, on the effectiveness, diversity, and inclusiveness of the leadership and organization of the US ITER Team. Meetings of the board should be announced in advance to all members of the ITER Team.

The proposed selection process and term lengths of USIRAB appointments are described in Appendix H.

## 3.2. Infrastructure Needs for the US ITER Research Program

Here, we address the aspect of the charge on the "balance between on-site presence and remote participation as well as coordination between these two modes of operation, and any potential resources that would facilitate cooperation, communication, data exchange, and data analysis." The benefits and resource needs for both on-site and remote participation are discussed.

The US response to the question of balance between on-site and remote participation has ramifications beyond the resources required. It is also of central importance to our aspirations toward leadership and influence in ITER research, as well as our need to gain knowledge and experience for future steps in US fusion energy development. Both on-site and remote participation are necessary, and both must be well supported. In brief, on-site participation is critical if the US team (as distinct from US nationals employed by the IO) is to be a strong and recognized contributor to making ITER succeed in terms of its performance and scientific productivity. At the same time, few researchers will be able to relocate to the project site on a long-term basis. To attract a diverse and highly capable team, we must minimize barriers by making it possible for people to work as full participants from their home institutions or, for that matter, from anywhere.

The optimum balance between on-site and remote will depend on specifics of US involvement not yet determined. An on-site fraction of 10-20% is probably a reasonable assumption for current planning purposes. Participation in ITER operations is certain to be predominantly remote, partly due to limitations on budget and people's ability to relocate, but also because even "on-site" participants will likely be unable to be in the ITER control room itself due to limited space. Fortunately, remote participation in fusion research has an excellent track record already, even as the tools continue to improve and the scope of what can be accomplished remotely continues to expand. What is important is that the US team must function as a coherent entity, fully integrating its on-site and domestically sited components. At the same time, the US team, in collaboration with the IO and other Members' teams, must work as one team toward ITER successfully meeting its objectives.

***ACTION - Provide infrastructure to fully support USIRT members working both on-site at ITER and off-site in the US:***

- Enable on-site participation in ITER by:
    - Establishing US ITER Research Program relocation support to help with policies, economics, and logistics of getting more people on the ground at ITER
    - Establishing a satellite office near ITER and workspace at ITER for on-site participants
- Enable remote participation in ITER by:
    - Utilizing efficient communications technologies between remote and local participants
    - Improving data networking throughput between ITER and the US
    - Establishing data and information repositories in the US
    - Creating one or more remote experimental centers
    - Establishing dedicated computational clusters (new and/or existing) for ITER modeling, simulation, and data analysis

### 3.2.1. On-Site Participation

Starting immediately and continuing through ITER commissioning and research operation there are numerous on-site activities that garner enthusiastic interest from workshop participants. Constraints on participation will have to be agreed within the ITER Council among the Members (including the US) and the IO, taking into account space limitations on the site and limitations on activities of non-IO employees that may be imposed by ASN (the French Nuclear Regulator). It will be to the benefit of the US community, as well as our partners representing other ITER Members, for the environment to be as open as possible. We urge the US representative to the ITER Council to advocate for this openness as on-site activities ramp up.

There may be opportunities to immediately place people on-site via the ITER Project Associate (IPA) scheme, but this is a fairly small program (at this writing there are only two positions posted at the ITER website) and as currently organized will be too limiting. As ITER operation and research ramps up we should expect new mechanisms to create opportunities.

### *On-Site Benefits: Technical Execution and Knowledge Transfer*

Onsite presence is strongly encouraged, beginning now, in order to accomplish the objectives described in Chapter 2, most immediately those of Mission 2 [Technology Engagement and Transfer]. Fulfilling General Initiatives GI1 [ITER Knowledge] and GI2 [IMAS Adoption] will benefit greatly from the interactions made possible by an onsite presence and provide an important service to the entire USIRT, on- and off-site, in setting up mechanisms for data access and qualification. This presence should continue through the ITER program to maintain and support the pipeline for data to pass between the ITER site and the domestic-based Team members.

During the current assembly phase, the US is designing and fabricating equipment necessary for the facility to operate and deliver on its mission. Upon delivery, all contributed equipment will become the property of the ITER Organization. US responsibilities beyond basic support for commissioning are currently undefined, but it is clearly in the US interest to follow up on its contributions by being involved in their operation. This will help to ensure that equipment performs as designed and plays its role in achieving ITER's goals, and will provide essential experience in the operation of a highly integrated nuclear facility to inform the design and planning for next steps in US fusion energy development.

The diagnostics being contributed by the US will play a pivotal role in achieving ITER's research aims. For our diagnostics to be judged a success, we must lead, or at least participate, in their operation including commissioning, maintenance, and any necessary modifications. We must be proactive in ensuring that the data coming from US diagnostics are reliable and fully integrated into the research. Strong on-site US involvement in Diagnostics will be critical for optimizing their operation and their integration into the overall research program.

Participation in system commissioning, which has already begun, as well as the later progression to integrated commissioning leading up to First Plasma, would yield valuable experience to US engineers and potentially be very helpful in bringing up all of ITER's systems. If there are opportunities for on-site participation here they are likely to require long-term assignments of at least one year.

### *On-Site Benefits: Leadership and Influence*

The US can best position itself for leadership and influence by having a strong on-site team of people who, through their technical contributions and their day-to-day interactions with international colleagues, will gain the confidence of their peers and acceptance as valuable and effective collaborators. Research on ITER will be carried out as an international collaboration among the ITER parties. The science and technology goals of ITER, and the associated research topics, are universal. Although the organizational structure of the overall ITER research program is not yet defined, it is reasonable to assume that it will be based on science and technology topics, have clear connections to the Missions defined in Chapter 2, and be a structure reminiscent of how major fusion facilities are operated today. Research groups will be internationally constituted and will need strong leaders who are internationally accepted for their outstanding technical capabilities, their leadership skills, and their full-time commitment to the success of the ITER project.

Access to timely information and the ability to influence decisions are also key benefits of continuous on-site participation. We have experience operating collaborative user facilities and collaborating on other parties' facilities. In all cases, the decision making is predominantly driven by those on-site, including members of the host institution and long-term visitors. Day-to-day decisions, especially those in response to unexpected developments or problems, are necessarily made by a team that together shares the detailed understanding of the physics, the facility and its operation, the available options, and likely outcomes. Our experience working with the IO on ITER design tasks, for example in Diagnostics, is also informative. Despite frequent (multiple per week) meetings between the domestic team and our IO counterparts, our awareness of developments and understanding of issues affecting our work consistently lags that of the on-site IO team. Our ability to influence decisions is constrained. The bandwidth for information exchange afforded by structured one-hour meetings across multiple time zones is simply no match for what is possible among people working and eating lunch together on a day-to-day basis.

### *On-site Resource Needs: Policy, Financial, and Logistics Support*

Significant on-site participation is necessary for the US to achieve its ITER research aims. In order to support staff who will remain members of the US team through their home institutions while assigned to the ITER site, there will have to be mechanisms in place that make it both **economically feasible** and **professionally rewarding** for individuals to accept such assignments. Since some US fusion institutions already assign staff on short- and long-term assignments away from their home institutions, including overseas assignments, models already exist. For a national program such as ITER research, these mechanisms should be standardized to maintain job security, openness and equality of opportunity for such assignments across laboratories, universities, and industry. Additional costs associated with travel, relocation, living expenses, language training, dependent education, extended time off, and regular trips back to the US, etc. need to be planned for as part of the cost of US participation in ITER research. Benefits should be comparable to other federal programs supporting staff and their family members stationed overseas. This should include logistical support for relocation, services such as advice on tax, work visas, benefits, etc. The willingness of individuals to accept on-site assignments will depend on the degree to which the national team and participating US institutions signal the importance of such assignments by facilitating them and avoiding practices that make them unreasonably burdensome or that limit their chances for advancement within their home organizations.

Members of the USIRT performing long-term assignments at the ITER site will need **logistical support for relocating and living** in France. There are several mechanisms already established for early on-site engagement at the ITER Organization which have been highlighted in this workshop. For those opportunities that are managed by the IO, such as staff positions and ITER Project Associate (IPA) assignments, the IO has specific resources in place to assist new hires with their relocation. The ITER Welcome Office remotely arranges the travel logistics including visa applications, shipping of personal goods, and short-term lodging. Once new hires arrive in the country, the Welcome Office contracts with local agencies to tour houses/apartments and set up necessary utilities and services. Such assistance is particularly important in France, where setting up essential services such as bank accounts, phone and internet plans, and even apartment contracts, can take months. The process of house/apartment hunting in France also relies more on local agents and connections, making it difficult to find and book accommodations online prior to travel. Finally, once new hires are established, the Welcome Office arranges for their integration into numerous intercultural and language programs. There is also assistance for family members, including work visas for spouses and assistance with school registration for children at the International School of Manosque. Dedicated resources and local contacts are essential to help staff establish their new mode of living. A smooth life transition to the area, even for shorter-term assignments, is paramount for US staff to realize an easier, more efficient start to their on-site work responsibilities. Workshop participants that have previously worked on-site at ITER attest that such relocation assistance was vital to fulfilling their work obligations. On-site work opportunities that are organized and funded by the USIRCO, rather than the IO, will need to include similar support resources. Ideally, terms could be worked out to enable the ITER Welcome Office to accommodate USIRT researchers that are not explicitly funded by the IO program. In that case, much of its functionality would need to be duplicated for US participants traveling to and from France under a variety of programs (short-term assignments <1 year, Visiting Researchers, Long-term assignment, etc.).

### *On-site Resource Needs: Satellite Office and ITER Site Workspace*

In addition, the US should consider establishing a satellite office in the vicinity of the ITER site. This would be rented office space that would provide staff offices or cubicles, conference space, high-speed internet, data links, access to US domestic computational resources, and communications support for seamless interactions with domestically sited US colleagues as well as with the ITER project site. Such a facility would foster US team coherence while providing the benefits of co-location with the ITER site without burdening the capacity of the site itself. Already there are examples of companies that have established offices in the area for similar reasons. They are often present for face-to-face discussions at the ITER site, and their organizations benefit from those close interactions.

The USIRCO will need to work together with the IO to establish a proper workspace for USIRT researchers at the ITER site. Workspace at the IO headquarters is already very limited, but opportunities may present as the Construction Domain of ITER shrinks and operations begin. Even then, space may be tight after accommodating the needs of the IO's Science and Operation and Engineering Domains. Any excess on-site workspaces will likely need to be shared with visiting researchers from all ITER member nations, not just the US. Advocacy for support and coordination of dedicated workspace for on-site US staff will require some effort from the USIRCO and support from the US representatives to the ITER Council. Additional workers could be accommodated at a separate US facility described above.

In summary, the following resources and USIRCO support will be needed for USIRT members :

- Legal requirements and processes, including tax, employment rights, immigration, etc. that correlate to the specific work assignment (including helping non-US citizens doing work for USIRCO)
- Relocation logistics & expenses
- Support programs for spouse/children for longer-term assignments
- Close coordination with or duplication of the functions of the ITER Welcome Office and Human Resources department
- Contracting with local agencies to assist with establishing living arrangements and utilities
- Arrangement of workspace at the IO headquarters and/or a local, off-site US facility

### 3.2.2. Remote Participation

#### *Remote Participation Benefits: Technical Execution and Knowledge Transfer*

As partners in ITER, the US will have the responsibility and the opportunity to support both the IO central team and on-site members of the US ITER Research Team. Remote participation is essential to support rapid-response function, maintain strong ties with on-the-ground personnel, and maintain information and program continuity. It is anticipated that ITER will operate in a two-shift mode of operation, so remote participants in the US during ITER operations could be working more regular working hours during the night shift in France. Every ITER shot will need to be simulated in advance, and anything more than very minimal between-shot, on-the-fly modification is unlikely. However, the ability to connect during ITER experiments expands the number of scientists who can contribute effectively to monitoring and analyzing ITER data, allowing rapid data dissemination and solution response in the event that problems arise. The benefits of remote participation apply to all the research topics considered in this workshop. Even commissioning and operation activities which require on-site involvement (such as Initiative DIAG-1A, Initiative Tech-1A) will benefit from having strong domestic components. The balance between on-site and remote will vary from topic to topic, but, in all cases, it will be essential for both components to work together as a single unified US team under common leadership.

A key benefit of remote participation is that it provides the broadest possible opportunity for US people to participate in ITER. It enables the program to attract individuals whose talents would be unavailable to ITER if it were a requirement to relocate to the project site or travel extensively. It enables many participants to hold part-time assignments on ITER while maintaining involvement in domestic research activities which may themselves be feeding into ITER needs. Remote participation is a crucial part of a necessary strategy of lowering barriers to participation in order to attract the best from amongst a diverse US talent pool. It also supports the need to maintain strong connections between ITER research and other components of the FES portfolio.

### *Remote Participation Benefits: Leadership and Influence*

Remote, no less than on-site, participation also offers excellent opportunities for leadership and influence within the ITER program. Predictive modeling and simulation of ITER plasma scenarios will have a major role in designing experimental proposals. Analysis of data from ITER experiments will likewise be critical for extracting scientific understanding and informing decisions on research directions. The availability of ITER data will drive rapid advances in the fidelity of models and the capabilities of analysis tools, provided a strong team is in place to capitalize on them. These activities can be performed in the US and indeed the coordination of work amongst a geographically dispersed team of US researchers will benefit if the leadership is also domestically based and well connected to the community.

We anticipate all ITER Members will field domestic teams carrying out similar activities under local leadership. The US Team should focus on applying our strengths to make unique contributions that can best facilitate progress in ITER performance or productivity. The US has prepared for such roles for years, through a variety of computational science programs such as SciDAC and Exascale. We have a capable workforce and powerful tools, and the ability to rapidly take advantage of advances in computational power. Within the larger ITER collaboration, we anticipate this preparation will put the US in a leadership position that will allow us to have major impact on the success of the ITER program.

### *Remote Participation: Resource Needs*

Most of the domestic ITER research team will likely be sited at their home institutions. An infrastructure must be put in place that enables the US team, both on-site and domestic, to function as a coherent entity, charged with the responsibility of delivering on the ITER mission in collaboration with the other ITER Members, and delivering ITER's benefits to the US As in any project, effective communication, including both human and data communication, is key. Clearly many relevant technologies are already in widespread use and new ones are on the horizon (virtual hologram meetings, robotic avatars, virtual reality, etc). While the technological specifics are beyond the scope of this document, there is ample expertise within the DOE Office of Science [DOE2021] community to assess the needs, anticipate likely advances, and recommend appropriate systems. What is important is that priority be given to ensuring that there is excellent communication among all US ITER participants and that the appropriate investments be made.

Networking throughput between ITER and the US will need to be improved to support the data communications that ITER operations will require. The implementation will need to be coordinated between the IO and ESnet with the requirements specified by the USIRT. Although the time scale of data replication is still to be determined, it is requested that the entire ITER database be mirrored within the US [see GI1:ITER Knowledge]. Networking speeds within the US, supplied by ESnet, will always exceed those that are available for transatlantic data transfer. In addition, the lower latency and diverse paths to ITER data within the US will give the USIRT team members faster access to ITER data via a local US data cache for research not tied directly to the day's experiment. Therefore, a substantial ITER data center must be deployed within the US to support the USIRT.

It is anticipated that the proposed infrastructure will allow an individual to fully participate in ITER operations remotely. But extensive remote collaboration experience has shown that gathering groups in a room leads to interactions and conversations that otherwise would not take place. Therefore, for US remote support of ITER experimental operation, it may prove beneficial to have one or several remote experimental centers (REC) that allow USIRT members to gather in one place and work collaboratively. Such an REC would be a substantial space, with large display screens and workstations that mimic the ITER control room. There are already examples of similar capabilities in remote control rooms at PPPL, GA, and MIT used for participation in international experiments and a prototype REC deployed by the EU and Japan as part of their ITER Broader Approach. One can imagine designing the individual pieces of the REC, easing their deployment for multiple locations to fit into existing needs (number of people to support) and available space. The RECs would be natural gathering places for periodic (yearly, bi-yearly, quarterly) hybrid (remote plus in-person) workshops where existing and potential participants can exchange experiences and coordinate future activities, as well as engage in team building activities.

Regarding computational clusters, having ITER's data replicated in the US and available via ESnet means that for a large percentage of use cases, computational power local to the researcher should be sufficient. This therefore would motivate some computational capability collocated with each REE. To support machine learning, it makes sense to have a dedicated cluster next to the replicated ITER data. There are however a series of data analysis use cases where the full power of a large HPC cluster at one of DOE's leadership class computing facilities will be required. In this case, the infrastructure requirement is more aligned with coordinating needs with our DOE/ASCR colleagues than a deployment specific to the US ITER research program. Early discussions with these HPC centers, as with ESnet, will help ensure that the required infrastructure to support the USIRT is in place.

# 4. Mobilizing a US ITER Research Workforce

The US ITER Research Program should begin now, with on-site participation in system commissioning and on- and off-site activities preparing for ITER experiments, and continue through the end of Fusion Power Operation in the 2040s or later. Scientists and engineers supporting areas of technical emphasis and the ITER Research Coordination Office are needed at all career stages, with researchers ready to step into leadership roles almost immediately to meet the technical and programmatic goals outlined in Chapters 2 and 3 of this report. With the generational nature of ITER's multi-decadal research program, today's students and early-career researchers are well positioned to be leaders later in the program and to bring back the experience and knowledge needed for the domestic development of a fusion pilot plant. To meet the enormous scientific and technical challenge of rapidly developing fusion energy, the US fusion program needs to undergo rapid growth, balancing against an expanding domestic research program, both government- and privately funded.

Considering this, it will be critically important to actively grow and retain the pool of talented scientists, technicians, and engineers. Workforce development is an essential component of building up both the US ITER Research Program and the larger domestic program, and we will need to minimize obstacles to participation through consideration of Diversity, Equity, and Inclusion and finding ways to involve communities that have been historically underrepresented in our field.

US ITER researchers will also continue to be members of the larger US fusion community, leveraging connections with domestic experiments, the theory and modeling community, universities, and privately funded fusion to the benefit of both. Barriers need to be identified and addressed, including intellectual property policies that can slow progress.

## 4.1. Workforce Development

Here, we address the charge aspect of "workforce development issues, including transparent mechanisms for broad participation and membership in the ITER research team and opportunities to conduct ITER research guided by the principles of Diversity, Equity, and Inclusion."

As the US fusion community embraces this next phase of fusion development, including the march towards commercialization and the completed construction, commissioning, and operation of ITER, it is clear that the workforce required to tackle the necessary challenges will have to grow. This workforce emcompasses various fields of expertise including research personnel, engineers, technical staff, operations, regulatory personnel, etc. It also involves all career levels, from undergraduate and graduate students, apprentices and postdocs, to early to mid career scientists and engineers to senior personnel. This broad spectrum means that there can't be a one-size-fits-all approach to workforce development and that a variety of strategies and programs are required.

***ACTION - We must support education and preparation of the workforce needed for successful engagement in ITER and FPP:***

- Establish a DEI and workforce directorate (or lead), located within the leadership of the US ITER Research Coordinating Office, to coordinate and follow through with the many aspects of DEI and workforce development [see US ITER Research Coordinating Office].
- Provide workshops and resources for technical training, as well as for leadership and project management development.
- Provide resources for diverse fusion workforce recruitment, including scholarship and fellowship programs to new communities using non-cognitive variable approaches in admission.
- Support both domestic and on-site training and internship programs with an intentional approach towards underrepresented communities. This includes ensuring competitive pay/travel/boarding for ITER-internships (both domestic or IO) by supplementing costs if necessary.
- Enable and support joint training programs with private industry.

### 4.1.1. Principles of Diversity, Equity and Inclusion for the US ITER Research Program

As reflected on various recent community reports, including the APS-DPP Community Planning Process, the Powering the Future report, and the Plasma 2020 report, the US plasma physics and fusion community has begun to recognize the importance of acknowledging and addressing issues of diversity, equity and inclusion in order to attract and retain talent in the field. A framework of DEI should be interwoven into all of the work being done, including our participation in ITER. This framework should include some basic principles that, as a community, we should take as a starting point:

- Diversity (of ages, socio-economic backgrounds, races, ethnicities, genders, gender identities, gender expressions, national origins, religious affiliations, sexual orientations, family education level, disability status, political perspective and other visible and nonvisible differences) leads to innovation, increase in productivity, and an increase in sense of belonging and wellbeing in a community.

- Only through an inclusive and equitable culture within the community can sustained diversity be achieved.

- Bringing members of historically marginalized communities into an environment that is not culturally ready to embrace them leads to harm towards the folks being brought in.

With these principles in place, the US ITER Research Program should embrace best practices in issues of DEI and seek support from subject matter experts in this field, especially as it pertains to issues of workforce development. It is the responsibility of all members of the US ITER Research Program to implement these practices.

## 4.1.2. Ensuring a Ready Workforce for ITER and the Domestic Fusion Program

It is vital that the US participates in ITER to gain experience and expertise from ITER operations, develop our fusion workforce, and apply skills and lessons learned to strengthen our domestic fusion program. Participation in ITER will make invaluable contributions toward the scientific and technical bases for future US fusion operating facilities. The US personnel that will work on ITER should deepen existing US knowledge but should also participate in areas where our experience is limited in order to gain understanding of all ITER systems.

**Recruitment and retention** of our fusion workforce for ITER is, therefore, essential. We must take care to ensure that the US ITER workforce is supported and retained – they will form vital working relationships and knowledge that can be leveraged by the US for future international endeavors and when planning domestic fusion activities. The breadth of workforce needs for ITER operations includes science, technical, and business support workers at all stages of their careers. The US workforce development activities must include engineers, technicians, and diagnosticians who will support installation/commissioning, maintenance, and decommissioning [Initiative DIAG-1A, Initiative TECH-1A]. Proper staff succession and knowledge retention is imperative given that ITER will operate for decades: As leaders and experts in the US fusion research community may pursue new opportunities during ITER operations – mid- and early- career fusion researchers and engineers will need to emerge as the next generations of leaders and senior experts. We must ensure sufficient opportunity for knowledge transfer to both fully support ITER and bring back knowledge gained from ITER to the rapidly evolving domestic program [see General Initiative - Knowledge].

To exercise greater influence on ITER operations, the US needs to earn leadership positions in the ITER research organization. Further information on the current status of US appointments at ITER can be found in Appendix I. Initially, natural leadership may emerge in areas where the US has expertise (i.e. on equipment supplied by the US ITER Project Office) or in areas of significant need (Operations Management). The US ITER Project workforce is already performing ITER activities in the construction and assembly phase; this subset of people needs to be developed and transitioned to be ready to assume subsequent responsibilities for the equipment that we are stewarding. While leaders may be experts in one technical area, all leaders should be informed on US interests at ITER to include desired research agendas and present a consistent message on US involvement and interest. The US ITER Research Program should offer **resources for leadership and project management development** to prepare for both ITER and domestic fusion operations. Effective and efficient leadership can increase overall returns through, for example, increased team motivation, performance, and retention. Workforce development opportunities should be offered to gain valuable lessons in business and administrative aspects, which may include:

- Operations Management
- Environment, Safety, Health, and Quality Assurance management at ITER, to include ITER Regulatory Reporting
- Project Management (ongoing through project construction)
- Planning for Inspection and Maintenance

- The ITER supply chain and procurement management
- International Recruiting for ITER

### 4.1.3. Ensuring Student Participation from Universities

Universities play a pivotal role in forming the future workforce and feeding the pathways towards all STEM research, including the US fusion program. University students are an integral part of the workforce pipeline and engagement through continued and expanded internships, workshops, seminars, and specialized conference activities are needed. The expansion is already necessary when considering existing scientific workforce demands and the growing need for new fields in the workforce with a particular emphasis on engineering, socioeconomics, licensing, and policy. Further, the US ITER workforce should better reflect the demographics of the US itself through intrinsic involvement of historically underrepresented and minority student communities in US ITER workforce development programs. The pipeline of the workforce for ITER activities needs to consider the time scale for ITER operations. Many of the people who will work on ITER are in the very early stages of their education/career. We need to recruit and sustain this future workforce.

The US ITER workforce will require both international (ITER site) and domestic **training and internship programs** that adapt and evolve to changing needs and concerns of students. Existing and future domestic training capabilities (Plasma and Fusion Undergraduate Research Opportunities (PFURO) program, Science Undergraduate Laboratory internships (SULI), Community College Internships (CCI), NSF-Research Experiences for Undergraduates (REUs), currently operational and future domestic and international fusion devices, national hybrid meetings/workshops) can be leveraged to ramp up onsite ITER participation and staff for US commercial fusion entities. The internships need to run with sufficient funding to incentivize continued participation with additional funding support for underserved communities. If possible, on-site ITER participation should occur for students at undergraduate and graduate levels. This should be done with the intention for, both, maintaining the US staffing levels on-site and for students to "bring back the knowledge" in future commercial & national laboratory, and academic environments. Resources should be developed (consolidated) to support the fusion workforce, especially those in underserved/under-resourced groups and for individuals conducting research abroad as part of the US ITER Research Program. This includes a network of support and resources available which prepare them for the work life at ITER. For on-site students, they must have at least two actively involved on-site mentors throughout the duration of their internship. Multiple mentors allows for different perspectives and the flexibility for students to express concerns/needs. Domestic internship programs should be coordinated between US academic institutions, national laboratories, and commercial entities for undergraduate and graduate students from existing and newcomer academic fusion programs. The domestic internships should have similar engagement of multiple mentors including the student's university faculty involved in fusion & plasma physics. This avoids dropping the student into an unfavorable situation where the likelihood of a successful experience is minimized. All efforts should consider financial incentives and generational differences when considering internship salary/pay. For many students, it becomes increasingly less viable to do international and moving domestic internships at wages close to or even double US minimum wage. Funding for moving, travel to and from, local housing, and others will need to be regularly addressed to keep interest strong.

**Workshops and resources to train** and inform US students need to be developed and provided, such as training on IMAS data interpretation workflows or code usage [General Initiative - IMAS], or introductory status and needs of Research Missions. For students, considerations should be made for improved conferencing and technical collaboration tools for hybrid participants [see also Infrastructure Needs]. The format and tools to manage in-person/remote/hybrid conferences and workshops need to be developed and standardized. It is recognized that in-person training is essential to develop many necessary skills needed for fusion energy research, but remote accommodations should be made available when possible for equitable access. A well thought out and managed program for each workshop/training should be put in place. This avoids the common pitfall of science & engineering workshops/training that become a series of unrelated lectures and seminar talks.

**Scholarship** (undergraduate - one or two years) **and fellowship** (graduate - length of masters/doctoral program) opportunities should be explored for fusion related communities similar to the US University Nuclear Leadership Program scholarships and fellowships. The expected merit criteria should be done with the understanding that underserved communities historically do not have the same resources to have similar academic performance to properly served communities (i.e. utilize non-cognitive criteria for admission). The fellowship models should be explored to balance the training needs for all sectors (national labs, industry, and academia) within the US ITER Research Program.

### 4.1.4. Private industry and the US ITER Workforce

The US fusion industry is diverse and includes private fusion companies which are developing their own workforces. There are exciting opportunities to collaborate with private industry in the workforce development space – we can work collaboratively to understand needs and grow our efforts with universities to meet those needs. We should not be in competition with private fusion companies, but aim to recruit enough fusion workers to staff both domestic and ITER workforces.

Private fusion companies are aiming to reach key operational milestones on a different timescale than ITER, which offers opportunities for domestic experience in multiple fusion operations environments prior to ITER operation. The US should prepare **a program to collaborate with private industry** on operational lessons learned in preparation for ITER operations, and share ITER operational experience and data (once available) with private industry to grow the capabilities of our workforce overall [see also 4.2.5].

## 4.2. Working With the Broader Community

The charge to this workshop recognizes the need for ITER research to be well integrated with both components of the US fusion enterprise– the FES-funded community and private fusion entities. In this section we answer the two charge questions that address this need; "coordination across FES programs and activities and the ITER research team, including research on domestic facilities, theory and simulation, technology, and international collaborations outside ITER" and "new opportunities for engagement with the private sector to enable a two-way exchange of scientific research and technological development between ITER and existing US commercial fusion endeavors." We elaborate on the modes of coordination between the ITER research program and selected sectors of the FES-funded program. While not comprehensive, these examples make it clear that the community expects a strong two-way interaction between ITER research and all other elements of the FES portfolio.

This coordination might benefit from creation of a national body with representatives from all stakeholder communities and facilities that can consider research needs and priorities of all segments of the community. This could include, for example, discussion of research within the domestic program that supports the needs of the ITER research program. However, the scope of such a body would probably be broader than that, supporting all aspects of research pointed toward a pilot plant. This might be similar to the Fusion Facilities Coordinating Committee but with broader membership.

***ACTION - It should be a goal of the USIRCO and the entire USIRT to make access to ITER participation, data, and research products open to the entire community*:**

- The USIRCO should facilitate making all data, software, and ITER research products openly available except in the relatively rare cases where proprietary information is involved (4.1.1).
- Establish, with urgency, a process whereby US fusion entities, both public and private, can gain access to ITER-generated information and vice-versa, including Intellectual Property (4.2.1, 4.2.5, 4.2.6).
- Implement a strategy for ensuring that FES-funded theory and model development can be readily extended and applied to ITER where applicable (4.2.2).
- Provide support for making modeling codes IMAS-compatible, and ensure that domestically-developed tools are integrated into IMAS as appropriate [General Initiative - IMAS] (4.2.2).
- Provide financial support for US fusion technology community engagement in ITER by providing base program support for ITER research, subsidizing contract bids in strategically important areas, and facilitating access to ITER design information and data (4.2.3).
- Ensure robust on-site and remote US participation in the commissioning and operation of US-designed and constructed ITER systems (4.2.3).
- Continue research in support of ITER on domestic and overseas tokamaks under the coordination of the USIRCO in cooperation with facility leadership and with participation from USIRT members. Strengthen participation in international ITER information sharing and planning venues, such as the ITPA (4.2.4).
- Develop a strategy for engaging faculty at different Carnegie Commission on Higher Education classifications (R1, R2, PUI, MSI) universities in terms of ITER research leadership, funded research projects, and onboarding of newcomer faculty from underrepresented and underserved communities (4.2.5).
- Conduct a series of 3-4 short 1-2 day hybrid workshops with University faculty, researchers, and research administrators from different Carnegie classifications to evolve engagement and gain a consensus for engagement (4.2.5).
- Expand the existing funding model that provides sufficient resources for existing and new faculty to provide meaningful and needed research contributions. This may include expansion of awards, seed grants, supported/split collaboration grants, supported postdocs/faculty salaries, etc. (4.2.5).
- Implement "Go to where they are" engagement strategies with newcomer faculty at different university classifications through engagement workshops (4.2.5).
- Support the creation of joint appointments between faculty and US ITER Research Program national laboratory relationships (4.2.5).
- Establish a domestic public-private task force to develop a plan for how the two sides will partner in ITER research, operations, preparations, two-way information exchange, and workforce needs (4.2.6).

### 4.2.1. Eliminating Obstacles to Community Involvement

Throughout the workshop, one set of issues that kept coming to the forefront is that of willingness and ability for information in its various forms to be shared between stakeholders.

There is a perception of ITER as having a "culture of secrecy" that results in skepticism and disinterest among many in the US fusion community and can needlessly create obstacles to collaboration. Although lack of sharing is often a default behavior, in most cases there is no rule mandating secrecy. This can be partially alleviated by the USIRCO, with a practice of regularly publicizing information about ITER and providing a gateway for community access to general information, data, software, and research products.

There are more serious concerns about IP, wherein there are conflicts between how it is handled in the ITER Agreement [ITER2007] and the requirements of domestic institutions, in particular universities and privately funded fusion institutes. As these conflicts touch on an international agreement, US government rules and regulations, and a number of different stakeholders, this is beyond the capabilities of the USIRCO to resolve. However, the USIRCO should, with the IO, DOE, and US entities, work to expeditiously resolve policies and disputes with regard to intellectual property, review processes, and other rules of engagement with ITER and ITER data.

### 4.2.2. Coordination with the Theory and Simulation Community

The US Theory and Simulation community is well positioned to contribute to the success of ITER through application of a variety of world-leading modeling capabilities [E.8. Modeling and Simulation]. In turn, ITER will provide unique data that is critical for validating these models in new physics regimes characteristic of burning plasmas in fusion pilot plants. There is significant overlap between the physics priorities of ITER and the priorities of the domestic program regarding model development and validation, especially regarding disruptions, ELMs, edge transport, RF modeling, and fast ion physics.

The US program should act quickly to ensure that theory and model development undertaken for other elements of the FES portfolio can be **readily extended and applied to ITER**, where applicable. Two-way communication mechanisms between ITER and US SciDAC (and other) activities are needed to better leverage US efforts and determine the benefit of maintaining US tools vs adopting other international tools from the ITER project.

Support to **make US modeling codes compatible with the data model used by the ITER Modeling and Analysis Suite (IMAS)** is needed to enable the immediate application of US codes to ITER data (especially for US codes intended to be routinely applied to ITER) and to facilitate the communication of predictive simulation results to other ITER members [see General Initiative - IMAS]. The US presently lacks a coordinated effort around the adoption and development of IMAS, so a centralized activity to support required software modifications across all US institutions would be beneficial. To maximize the return on investment in modeling capabilities, the usability of integrated modeling codes should be prioritized, including adhering to software engineering best practices, porting to HPC hardware, and providing documentation and training for a broad user base.

US scientists and facilities should play a role in helping to develop and validate standardized IMAS workflows that will be used for ITER data analysis. IMAS codes and workflows will continue to evolve before and during ITER operation, so development and validation of IMAS tools will be an ongoing activity. The possibility of using existing data for validation should always be explored before committing additional facility runtime to this type of activity. Presently, IMAS tools are not typically used in experimental analysis in the US; therefore, specific, dedicated funding will be needed to support US researchers' engagement with these tools. In cases where domestically-developed tools essentially duplicate or surpass IMAS capabilities, these domestic tools should be integrated into IMAS, where possible, so that existing domestic expertise and continued investment in these tools benefit ITER research.

### 4.2.3. Coordination with the Technology Program

While it represents a relatively small portion of today's total FES research portfolio, the US nevertheless possesses significant expertise in a variety of fusion energy supporting technologies, and has made important contributions to ITER in these areas. As the US program moves toward an FPP it can be expected that the technology program, and the importance of relevant knowledge and experience from ITER, will increase.

An important takeaway from the workshop was a sense of disconnection from ITER in the technology community that appeared not to exist in many other areas of fusion research. Many fusion technology researchers tend not to engage in ITER-focused work except through specific proposal/tender mechanisms, and then only on very specific tasks, without broader engagement, and terminating with a specific deliverable. Thus, even in areas where the US has made significant contributions to ITER (both analysis and hardware, e.g. neutronics and tritium processing), and which would be natural leadership areas for the US on ITER moving forward, there is currently not presently a clear path forward for such engagement. Action, including targeted funding, is needed to ensure that inclusiveness in ITER extends to the technology community.

**Support for sustained engagement** is critical, both to ensure US researchers obtain the necessary construction, commissioning, and operational experience, lessons learned, and tacit knowledge needed to enable future US facilities [General Initiative - Knowledge], but also to minimize operation and schedule risk for ITER— by providing expertise and leadership where we are best equipped to do so, e.g. on those systems the US has designed and/or built.

Many fusion technology research disciplines are well positioned to provide critical input to ITER, and also stand to benefit significantly from the data it provides, e.g. via exercises like code validation. Contributions in the areas of, disruption mitigation [E.4. Disruption Mitigation], slag management [E.1. Plasma-Material Interactions], diagnostics [E.9. Diagnostics], plasma heating and current drive, nuclear design and integration, tritium management, beryllium handling, magnets, superconductor fatigue, and safety and licensing [B.10. Technology and Integration] were highlighted during the workshop. An effort should be made to **identify ITER needs in these areas and facilitate engagement** of the relevant US experts. Similarly, the USIRCO should ensure access to ITER data for these research communities, and in particular seek to mitigate any obstacles (such as export controls) to such access that may exist.

### 4.2.4. Coordination with the Research on Domestic and International Facilities

The International Tokamak Physics Activity (ITPA) and "voluntary" research in support of ITER are mechanisms for coordination between ITER and research funded on domestic and overseas tokamaks within the current FES portfolio. Present activities, such as the development of hardware innovations, control tools and techniques, and plasma scenarios should continue under the coordination of the USIRCO with participation from the USIRT members. Experimental proposals for ITER gain credibility through testing of ideas and scenarios across multiple devices. Activities should extend beyond present facilities (e.g. DIII-D and NSTX-U) to encompass both upcoming (e.g. MPEX) and potential future facilities (e.g. next-step and fusion technology research).

This arrangement will likely change as ITER begins operation and the FES portfolio evolves to include elements that focus on next-step devices and a roadmap to an FPP. Under this scenario coordination should consider how experiments on ITER can support the FES domestic program. A program to execute experiments on domestic facilities that target model and simulation validation benefits both ITER research and a path to an FPP.

The process of proposing, planning, and executing experiments in domestic facilities will normally follow already established processes in which experimental proposals are considered on their individual merits and support of programmatic goals of the facility. In keeping with the user facility model, we hope opportunities can be extended to the USIRT to propose and carry out experiments that further the US ITER Research Program goals, and in particular be able to prioritize experiments that address urgent issues. This should be done collaboratively with the facility management as we anticipate ITER research goals will be consistent with overall US fusion programmatic goals.

### 4.2.5. Coordination with Universities

Universities play a pivotal role in both research and educational objectives of the US Fusion program through unique research environments and the forming of the future workforce. Universities provide a connection between fundamental and applied research encompassing a large variety of topical fields. Their academic ecosystems provide a wide breadth of expertise and a continuously rejuvenated and dynamic workforce that directly attracts and supplies talent, new skills and fresh perspectives to the US fusion enterprise. The broad expertise in non-fusion topics, as well as in fusion topics like materials and technology that have growing importance, enables universities to tap into and connect with other research areas that provide innovative ideas that can benefit fusion.

Universities incubate innovations that are essential for bringing fusion to commercial reality. As they innovate, they produce diverse talent and they spin-off the private companies that will build the US Fusion Industry. The faculty (junior and senior) and staff researchers envision and build research programs that heavily support the fundamentals and applications within the fusion community. The education & training that Ph.D. students receive goes beyond classroom lectures; an intense mentoring relationship with a faculty member based on involvement in leading-edge research is also required. With ITER, fusion scientists and engineers will encounter challenges and opportunities that are unprecedented in fusion research. In order to fully benefit from ITER's opportunities, we need to grow the number of universities participating in fusion research including ITER, the junior (assistant) & senior (associate and assistant) faculty, staff researchers, and we must invest in student research.

A long term plan, one that leaves room for education and investing in early-career scientists, engineers, and faculty, is needed in order to reap the benefits from ITER today and ensure a sustainable future. Students and faculty must have opportunities to become involved in ITER now; many will rise to leadership responsibilities over the course of the ITER project. Membership in the US ITER research team, including leadership roles, should be fully open to university participants. The budget increases that will be necessary for the US to fully exploit ITER will provide an opportunity to increase the number of universities involved in fusion; proactive measures should be taken to ensure this happens. While many of the larger programs within the US fusion community may be able to begin their participation using existing funds, this is likely not to be the case for smaller, newcomer, underrepresented, and/or underserved groups, universities in particular. It is imperative that funding opportunities be extended to the university community to encourage and welcome their participation and extend leadership opportunities.

Universities also have their own particular needs and parameters that must be addressed in order to fully participate in the US Fusion Research Program and in particular the US ITER Research Team. It is a reality that US universities have competing objectives for their research and educational aspects that are not mirrored within the rest of the US fusion community. This section addresses the expected requirements for an equitable, inclusive, and diverse participation from Universities needed within the US Fusion program. It discusses the factors that differentiate the types of Universities based on their Carnegie Commission on Higher Education classifications and how the US Fusion program needs to treat them in meaningfully similar but unique ways. Faculty at these different types of universities & colleges have large disparities in job responsibilities that impact the ability of individuals to engage with US Fusion. Without engagement of faculty at all experience levels (early-career, mid-career, senior, retired), the workforce pipeline of students and early career scientists will be limited to pre-existing US Fusion relationships.

This section is heavily inspired by and built upon the white papers, "University Participation at the Center of ITER Research Program" and "Coordination of US workforce development efforts for ITER." For information about engagement specific to workforce development around students, please see 4.1.3 "Ensuring student participation from universities."

Regarding the factors that differentiate the types of Universities based on their Carnegie Commission on Higher Education classifications:

- Universities that are classified as "R1: Doctoral Universities – Very high research activity" where faculty at these institutions have a significant research program development focus with a limited teaching focus. Although teaching facing activities are important, engagement will need to be geared at research activities for faculty at these institutions. Particularly, early career faculty in the pre-tenure stage will require funded research engagement or directly risking their livelihood.
- Universities that are classified as "R2: Doctoral Universities – High research activity" where faculty have a more balanced approach between research and teaching. Engagement should be taken under consideration of potentially limited research capabilities as compared to R1 institutions. This more balanced approach can reach the expectations of R1 institutions if the R2 institution has extensive post graduate programs (masters and PhD) that require the same amount of research and funding requirements as R1 institutions.
- Universities and colleges classified as Primary Undergraduate Institutions (PUI) will not have the same research infrastructure found at R1s and R2s. Engagement with faculty should be taken under consideration that teaching and service obligations are the main focus of these faculty. In particular, FOA opportunities will not likely be the best means of engaging these faculty due to limited office of sponsored projects organizations within their institutions.
- Universities and colleges that are classified as Minority Serving Institutions (MSI) span a range of R1, R2, and PUI type of insitutitions. Engagement with MSIs should be done where MSI faculty are treated in an equitable manner as compared to colleagues at non-MSI institutions.

The research program should be focused on both outcomes that directly support US ITER Research Program objectives with the adjacent benefits of student training involving skill development and long term retention. The faculty who are involved will inherently need graduate students, undergraduate students, and high school students to support US ITER research that begins the long term nature of workforce development. The research programs made available for US faculty to apply to should include existing expanded DOE Office of Science FOAs, established training grant programs similar to the National Science Foundation, seed grants for newcomer faculty (early career faculty and newcomer from outside fields), and supplemental joint appointments for faculty between their home institution and relevant US ITER Research Program national laboratories. The strategy for funding USIRP faculty research should be explored through small seed grants (~$250k split over two years of a single project), longer term grants with minimal upfront effort compared to larger grants that individual faculty can be awarded sufficient to support a graduate student full time plus part of their summer, cost of travel to relevant USIRP meetings (>~$600k for 3 years), multi-institutional grants that motivate the inclusion of newcomer programs with early career faculty (>~$1000k for a 4 year period), and supplementing the existing early career program to enable a growth of existing programming funding for different types of science and engineering with FES. The last one could

be done by requesting a larger budget appropriation focused specifically for USIRP intended early career spots. Supplement joint appointments for faculty are intended to enable heavy involvement of faculty in detailed USIRP activities at US National Laboratories where the home institution does not have existing national laboratory relationships. These funding recommendations are based on a rough estimate that a single graduate student costs to external sponsors approximately $60-100k per year that includes tuition, stipends, fees (health insurance). For engineering based fusion programs, it is the expectation of faculty that a PhD or master's students degree is almost if not entirely funded through research programs and consistent funding on the same project prevents issues with extended graduation or inability to complete the degree. Further, the faculty members at most R1 and R2 institutions are required to bring in sufficient funds to cover their salary during the summer months or they are not paid. These funds have to traditionally come from research grants where most grants prevent more than a month of summer salary per year covered. For faculty, every additional effort to develop grant proposals and maintain existing funding levels needed for tenure distracts from contributing to the fusion community in a focused effort. It is suggested that DOE consider grant proposals that enable at least junior (assistant) and mid-career (associate) that provide sufficient funding (~250-350k per year per faculty member depending on the university) of two graduate students and the faculty's summer salary. This level of resources would not only reduce the distractions associated with constant efforts to maintain funding but it would ensure focus on fusion related activities. It should be considered by DOE to provide sufficient funds in future grants to cover at least two months or entirely of engaged faculty to ensure participation on the research at hand as opposed to efforts to make them whole.

The goal for this expanded funding should be intended to establish clusters of fusion and plasma physics faculty at newcomer institutions to expand the community through new university programs. With only one faculty member, the fusion & plasma physics activity will not be sustainable at those institutions with an inability to build a depth of advanced courses that is needed to support ITER. Further, if only one fusion & plasma physics faculty member exists at an institution, then participation in onsite ITER or other activities is unlikely due to the lack of coverage for the faculty technical areas.

In order to sustain programs (educational and research), the faculty members require staff researchers including staff scientists, research engineers, and postdoctoral scientists who have their own incentives and desires for staying at academic institutions. The faculty who hire into these positions have to have enough soft funds (grant/research funds) that well outpaces the existing funding achievable for most junior and senior faculty outside of well-established programs. Well-established programs themselves have significant funding to maintain these positions that can still disappear if not enough grant or research funds materialize. For faculty to hire personnel into these positions, the salaries (funding) have to be sufficiently high for the personnel to take them instead of higher paying jobs. Currently, the US scientific community is encountering issues where recruiting trained and capable scientists/engineers for postdoctoral positions is difficult due to comparatively low salaries (as low as $30k in some cases) to national laboratories or industry (high five figures to low six figures or higher). This is in addition to the temporary nature of postdoctoral positions and most staff research positions (soft money)

where if the university faculty or faculty cluster does not bring in sufficient funds, the position disappears. This prevents a meaningful quality of life from being achieved for the staff scientist and postdoctoral scientists due to constant upheaval and/or uncertainty in their life/career.The engagement of newcomers should be explored through workshops to provide detailed introduction to existing but undersubscribed USIRP research areas at regional (e.g. Southeast, Southwest, Northeast, Midwest, Mountain West, West, Pacific Northwest) and national conferences (APS, ANS, ASME, etc) or ad hoc topical meetings. There will be a significant portion of newcomer faculty that will not have financial support through their institution nor existing research program to attend. It is suggested that strategies for ensuring attendance for these faculty is explored including financial support and fully hybrid meeting strategies.

Participation of university faculty, staff, and students in work with ITER requires a Memorandum of Understanding to be signed between the University and the ITER Organization. Although a specific number was not provided, it was mentioned that the number of MoUs signed between US Universities and the ITER organization is below 10 (details on this statistic can be made available upon request by ITER HR, but the individual MoUs are proprietary). Among the universities with signed MoUs with the IO are the University of Utah, San Diego State University, the University of Michigan, University of Illinois Urbana-Champaign, and Columbia University. Conflicts between ITER's and US Universities' policies on ownership of Intellectual Property generated by the collaboration have been a sticking point preventing wider university participation; indeed the MOUs currently in place have tended toward avoiding rather than resolving IP issues. According to federal statute, that IP belongs to the university, but typically the IO asserts ownership if the work is performed under IO direction, as in the case of an ITER Project Associate. A blanket resolution of this issue is needed; otherwise it will delay or completely block negotiations of individual contracts or agreements between universities and the IO.

In summary, the workshop discussion groups and motivating workshop white papers made clear that the community expects university researchers, including students, to fully participate in ITER research. Besides their obvious role in training and workforce development, involvement of universities is critical for maintaining connections between fusion plasma and engineering research and academic disciplines such as engineering, science, policy focused fields, socioeconomics (sociology & psychology, and economics), licensing, and law that are vital to the long-term development of commercial fusion energy. In the near term, with the US ITER research program in its formative stage, the following actions are needed:

### 4.2.6. Coordination Between ITER and Privately Funded Fusion Activities

The context for coordinating public and private fusion activities is underscored by external events that occurred while the workshop was in progress as well as discussions during the workshop itself. In March, a White House Fusion Summit, attended by representatives of both public and private fusion organizations, was held to accelerate the realization of commercial fusion. Within DOE, a Lead Fusion Coordinator, reporting to the Undersecretary for Science and Innovation, and having coordination responsibilities across the Office of Science, Nuclear

Energy, NNSA, and ARPA-E was appointed. This new cross-department focus on fusion commercialization provides an opportunity for a common context for DOE, through its FES burning plasma program, and US private fusion organizations. Finally, the Administration's FY-2023 budget was submitted to Congress, requesting $32M for public-private fusion partnerships, up from $6M in FY-2022, to fund a new milestone-based cost-share program with private fusion industry, as well as to continue the INFUSE program.

It was noted that there are currently few regular points of contact between ITER and the private fusion community, and this workshop did not significantly ameliorate that problem. Still, important issues were identified, and important recommendations came forward.

One seemingly obvious and immediately available avenue for participation by private entities is the International Tokamak Physics Activity (ITPA), which exists to exchange scientific information and to plan research tasks relevant to ITER. We learned that a policy is in place that provides for private sector participation, though reports of resistance to such participation were heard at the workshop. Going forward the ITPA may evolve and other forums or mechanisms for information exchange and joint planning will surely materialize. Now is the time to **lower any barriers, whether legal or cultural, that may exist inhibiting two-way public-private scientific communication pertaining to ITER**. Developing the necessary processes seems like an achievable first step toward stronger partnerships in pursuit of US fusion goals.

Workshop participants (mostly scientists) generally agreed that eligibility to apply for DOE grants to participate in ITER research should be open to all members of the US fusion enterprise, considered here to include members of private fusion entities as well as publicly funded laboratories, universities, and industry as usual. The level of interest in ITER participation on the part of private fusion companies is not known and likely varies among the various companies. Dialog is needed and the initiative probably resides with DOE.

Regarding the two-way exchange of scientific information and of intellectual property (IP), it is well known that both ITER and private entities have provisions governing protection of IP and sharing of unpublished scientific information. While free information exchange exists in ongoing collaborations between private companies and public organizations (e.g. via INFUSE), there is a need to **broaden access to ITER-generated science and technology information, including data, procedures, manuals, codes, tools, and handbooks.** It was pointed out that the ITER agreement gives the Department of Energy the right to grant public and private US entities access to such information. A process that makes clear how requests for such information are to be handled is needed. Workshop participants, again, are generally in favor of openness in both directions to the extent possible and can only urge that the dialog necessary to minimize any barriers to openness should begin as soon as possible.

Interest in publicly funded research using fusion facilities that have been or are being built by private companies is already widespread in the community and was discussed at the workshop in connection with possible benefits to ITER. Private-side facilities may offer capabilities not available on existing FES facilities that could in principle be used to address ITER R&D issues or to provide experience that can prepare individuals for future ITER responsibilities. A full listing of examples requires broader involvement of private entities than was represented at the workshop. Nonetheless, it is clear that public-private partnerships to exploit such capabilities for ITER benefit could open new opportunities for two-way exchange of scientific and technological information and be a first step toward broader partnership in fusion energy development.

Workshop discussions showed that there are multiple potential avenues for two-way engagement with the private sector but beyond that, further dialogue with the explicit purpose of fleshing out the opportunities is needed. We briefly discussed the role of the Fusion Industry Association (FIA) as a possible central point of contact for helping to instigate such dialog. However, the interests of the various autonomous companies are diverse as are the capabilities they offer, so communication channels acceptable to all stakeholders need to be established as a first step in the dialogue.

# References

DOE2021 **Conference: Lessons from the COVID Era and Visions for the Future Roundtable**, Washington (held virtually), DC (United States), 2-15 Dec 2020 https://doi.org/10.2172/1785683 (2021).

DPPCPP2020 **A Community Plan for Fusion Energy and Discovery Plasma Sciences**, APS-DPP, https://doi.org/10.48550/arXiv.2011.04806 (2020).

FESAC2021 **Powering the Future: Fusion & Plasmas**, Fusion Energy Sciences Advisory Committee, https://drive.google.com/file/d/1YhOZTh7iO5QDPPTC36qojOk4J25JtvkQ/view (2021).

ITER2007 **Agreement on the Establishment of the ITER International Fusion Energy Organization for the Joint Implementation of the ITER Project**, https://drive.google.com/file/d/1griD7KNTK9HqTsn__ZOwJHAGRTL1hady/view (2007).

ITR-18-003 **ITER Research Plan within the Staged Approach (Level III - Provisional Version)**, ITER Technical Report ITR-19-003, https://www.iter.org/technical-reports?id=9 (2018).

ITR-20-008 **Required R&D in Existing Fusion Facilities to Support the ITER Research Plan**, ITER Technical Report ITR-19-003, https://www.iter.org/technical-reports?id=14 (2020).

NAS2019 National Academies of Sciences, Engineering, and Medicine 2019. **Final Report of the Committee on a Strategic Plan for U.S. Burning Plasma Research**. Washington, DC: The National Academies Press. https://doi.org/10.17226/25331.

*Other relevant documents and websites are listed at https://www.iterresearch.us/resource-library.*

# Appendices

## Appendix A: Charge

ITER Research Program - US
Research Needs Workshop Sponsored by FES
1 December 2021
Virtual Meeting via Zoom

**Chair:** Chuck Greenfield, General Atomics
**Co-Chair:** Cami Collins, Oak Ridge National Laboratory

**Purpose of Workshop / Charge:**

As ITER construction nears completion with more than 70% of the machine complete for First Plasma, it is timely for the US to plan its participation during the subsequent operation phase of this high gain burning plasma experiment. The Fusion Energy Sciences (FES) program within the DOE's Office of Science is initiating ITER research planning activities to both maximize the return of the US investment in ITER's construction and operation and to ensure US research priorities on ITER strengthen the domestic program aimed at the development of a fusion pilot plant.

In FY 2022, a Basic Research Needs workshop will be held to engage the US fusion community in the elaboration of a plan describing the formation, organization, and research objectives of a national ITER research effort. Workshop participants should adopt a long-term, comprehensive perspective on the US engagement in ITER research by considering: (1) the present period of ITER hardware commissioning activities up to and including first plasma; (2) the period between ITER First Plasma and the end of Pre-Fusion Power Operation-1 (PFPO-I); (3) the PFPO-II phase; (4) the Fusion Power Operation (FPO) phase, as described in the 2018 ITER Research Plan within the Staged Approach developed by the ITER Organization (IO); and (5) any upgrade program.

This workshop should address issues relating to two major areas:

**Research**

- Areas of research that offer the most opportunities for US leadership in ITER and contribute to its success,while bringing back to the US the necessary experience for accelerating the development of a domestic fusion energy source.
- The essential ITER research products needed to strengthen the domestic program to aim at the development of a fusion pilot plant.
- Capabilities that need to be strengthened and gaps that need to be bridged in the US fusion program to maximize the success rate of experimental proposals submitted by US researchers during the anticipated highly competitive review process for conducting experiments on ITER.
- The role of and opportunities for the US fusion community during the commissioning and operational planning for ITER.

- Areas of facility maintenance where US participation could develop the necessary experience for sustaining fusion systems in a nuclear environment.

**Organization**

- Organization, structure, and modes of operation for flexible, agile, and impactful exploitation of the ITER facility by US participants.
- Balance between on-site presence and remote participation as well as coordination between these two modes of operation, and any potential resources that would facilitate cooperation, communication, data exchange, and data analysis.
- Coordination across FES programs and activities and the ITER research effort, including research on domestic facilities, theory and simulation, technology, and international collaborations outside ITER.
- New opportunities for engagement with the private sector to enable a two-way exchange of scientific research and technological development between ITER and existing US commercial fusion endeavors.
- Workforce development issues, including transparent mechanisms for broad participation in the ITER research effort. Opportunities to conduct ITER research should be guided by the principles of Diversity, Equity, and Inclusion.

In carrying out this study, the committee should take into consideration:

- The 2018 ITER Research Plan within the Staged Approach and associated documents developed by the IO.
- The FESAC Long Range Plan report, “Powering the Future Fusion & Plasmas: A long-range plan to deliver fusion energy and to advance plasma science.”
- Input collected during the two-year process for developing the FESAC Long-Range Plan (e.g., APS-DPP CPP report).
- Reports from community organizations (e.g., US Burning Plasma Organization on the formation of an ITER research team).
- Recent reports by the National Academies of Sciences, Engineering, and Medicine on “Burning Plasma Research” and “Bringing Fusion to the US Grid.”
- ASCR/FES/ESnet Requirements Review Planning ITER Case Study.
- Participation plans developed by the ITER partners.
- Experience and modes of participation from other US research communities that have significant international collaborations.
- The Agreement on the Establishment of the ITER International Fusion Energy Organization for the Joint Implementation of the ITER Project.

The findings of this workshop will be summarized in a report that should be submitted to FES within a month after the meeting.

## Appendix B: Participants

Muhammad Abdelghany (UIUC)
Shota Abe (Princeton)
*Tyler Abrams (GA)
Tsuyoshi Akiyama (GA)
Jean Paul Allain (PSU)
Steve Allen (LLNL)
Robbie Allgood (SRNL)
Himank Anand (GA)
Max Austin (Texas)
David Babineau (SRNL)
*Vittorio Badalassi (ORNL)
Kshitish Barada (UCLA)
Laszlo Bardoczi (GA)
Rhea Barnett (ORNL)
Jayson Barr (GA)
*Devon Battaglia (CFS)
Alexander Battey (Columbia)
Larry Baylor (ORNL)
Torrin Bechtel (ORAU)
*Matthew Beidler (ORNL)
Emily Belli (GA)
Jack Berkery (Columbia)
Nicola Bertelli (PPPL)
Amitava Bhattacharjee (Princeton)
*Theodore Biewer (ORNL)
Tim Bigelow (ORNL)
Serdar Bilgili (WVU)
Douglas Bishop (PPPL)
Tim Bohm (Wisconsin)
*Rejean Boivin (GA)
Phillip Bonofiglo (PPPL)
Alessandro Bortolon (PPPL)
Dan Boyer (PPPL)
Boris Breizman (Texas)
Junge Brian (GA)
*Michael Brookman (CFS)
Jeffrey Brooks (Purdue)
David Brower (UCLA)
Danielle Brown (Stanford)
Galen Burke (LLNL)
Jalaluddin Butt (PPPL)
Richard Buttery (GA)
Igor Bykov (GA)
Feng Cai (PPPL)
Kyle Callahan (UCLA)
*Amelia Campbell (ORNL)
Alejandro Campos (LLNL)
John Canik (ORNL)
*Lane Carasik (VCU)
Jason Cardarelli (Michigan)
Joshua Carlson (Bryn Mawr)
Troy Carter (UCLA)
Ivo Carvalho (GA)
*Livia Casali (Tennessee)
*John Caughman (ORNL)
Mirela Cengher (GA)
Luis Chacon (LANL)
Desmond Chan (Bechtel)
Choongseok (CS) Chang (PPPL)
Brett Chapman (Wisconsin)
Jie Chen (UCLA)
*Xi Chen (GA)
Wilkie Choi (GA)
Hari Choudhury (Columbia)
Satyajit Chowdhury (UCLA)
Colin Chrystal (GA)
*Michael Churchill (PPPL)
Mark Cianciosa (ORNL)
*Jonathan Coburn (SNL)
Ryan Coffee (SLAC)
†Cami Collins (ORNL)
Bruno Coriton (GA)
Alexander Creely (CFS)
Brendan Crowley (GA)
Dionysi Damaskopoulos (Wisconsin)
Julius Damba (UCLA)
Eli Dart (LBNL)
Nicholas Davila (Texas)
Donovan Davino (Brown)
Marc-Andre De Looz (PPPL)
*Diane Demers (Xantho)
Daniel Den Hartog (Wisconsin)
Severin Denk (MIT)
Ram Devanathan (PNNL)
Ahmed Diallo (PPPL)
Patrick Diamond (UCSD)
Erick Diaz (Colorado)
Andris Dimits (LLNL)
Siye Ding (GA)
*Arturo Domínguez (PPPL)
*David Donovan (Tennessee)
John Draper (CGN)
Xiaodi Du (GA)
Robert Duckworth (ORNL)
*Ben Dudson (LLNL)
Andrew Dvorak (ORNL)
*Florian Effenberg (PPPL)
Nicholas Eidietis (GA)
Matthew Eklund (INL)
*David Eldon (GA)
Robert Ellis (PPPL)
*Darin Ernst (MIT)
Ruben Fair (PPPL)
Jiarong Fang (PPPL)
Max Fenstermacher (LLNL)
*Nate Ferraro (PPPL)
Erik Flom (Wisconsin)
Heinke Frerichs (Wisconsin)
Kevin Freudenberg (ORNL)
Alex Friedman (LLNL)
*Thomas Fuerst (INL)
Tripp Fulmer (ORNL)
Enac Gallardo-Diaz (Nevada)
Vincent Galvan (UIC)
Jhovanna Garcia (SDSU)
Gary Garcia Montes (Florida)
Brenda Garcia-Diaz (SRNL)
Andrea Garofalo (GA)
Trey Gebhart (ORNL)
Benedikt Geiger (Wisconsin)
*Yashika Ghai (ORNL)
Arianna Gleason (SLAC)
Siegfried Glenzer (SLAC)
Nikolai Gorelenkov (PPPL)
Richard Goulding (ORNL)
Robert Granetz (MIT)
David Green (ORNL)
†Charles Greenfield (GA)
Brian Grierson (GA)
Robert Grove (ORNL)
Luca Guazzotto (Auburn)
Walter Guttenfelder (PPPL)
Robert Haight (LANL)
Michael Halfmoon (Texas)
Bryant Hall (Princeton)
Chris Hansen (Washington)
*Jeremy Hanson (Columbia)
Michael Hanson (UCSD)
Ben Hardy (ORNL)
Chase Hargrove (PSU)
Jeffrey Harris (ORNL)
*Claudell Harvey (US ITER)
Shaun Haskey (PPPL)
Ehab Hassan (ORNL)
Ahmed Hassanein (Purdue)
Richard Hawryluk (PPPL)
*Jeffrey Herfindal (ORNL)
Kenneth Hill (PPPL)
*Chris Holcomb (LLNL)
Leo Holland (GA)
Rongjie Hong (UCLA)
Suk-Ho Hong (GA)
Ryan Hood (SNL)
Wayne Houlberg (Retired)
Saeid Houshmandyar (Texas)
Eric Howell (Tech-X)
Qiming Hu (PPPL)
Michael Hua (Helion)
*David Humphreys (GA)

*Paul Humrickhouse (ORNL)
Victoria Hypes-Mayfield (LANL)
Valerie Izzo (Fiat Lux)
Stephen Jardin (PPPL)
Trelewicz Jason (SBU)
Sudheer Jawla (MIT)
Tom Jenkins (Tech-X)
Frank Jenko (Texas)
Xiang Jian (UCSD)
Yanzheng Jiangy (GA)
Curtis Johnson (ORNL)
Ilon Joseph (LLNL)
Djamel Kaoumi (NCSU)
Stanley Kaye (PPPL)
Sam Keener (ORNL)
Harry Kelso (TAE)
James Kennedy (Florida)
Charles Kessel (ORNL)
Kyungjin Kim (ORNL)
Sangkyeun Kim (Princeton)
*Jacob King (Tech-X)
Dmitrii Kiramov (Texas)
Alex Klasing (Tennessee)
Andreas Kleiner (PPPL)
E.Christopher Klepper (ORNL)
Veronika Klevarova (Columbia)
Robert Kolasinski (SNL)
Bor Kos (ORNL)
Sean Kosslow (Tennessee)
Gerrit Kramer (PPPL)
Scott Kruger (Tech-X)
Seung-Hoe Ku (PPPL)
Ralph Kube (PPPL)
Atul Kumar (ORNL)
*Florian Laggner (NCSU)
Courage Lahban (NJIT)
Evan Lambert (PSU)
Lang Lao (GA)
Ane Lasa (Tennessee)
*Charles Lasnier (LLNL)
Cornwall Lau (ORNL)
Remi Lehe (LBNL)
*Anthony Leonard (GA)
Jeff Lestz (UCI)
James Leuer (GA)
Jeffrey Levesque (Columbia)
Nami Li (LLNL)
Zeyu Li (ORAU)
Zhihong Lin (UCI)
Zihan Lin (Princeton)
Chang Liu (PPPL)
Deyong Liu (GA)
Yueqiang Liu (GA)
Zefang Liu (GA Tech)
Alberto Loarte (IO)
Lynda Lodestro (LLNL)
Nikolas Logan (LLNL)
John Lohr (GA)
Camila Lopez Perez (PSU)
Jeremy Lore (ORNL)
Kaden Loring (Stanford)
Michael Loughlin (IO)
Maya Lowell (UCLA)
Neville Luhmann (UCD)
Arnold Lumsdaine (ORNL)
*Robert Lunsford (PPPL)
Brendan Lyons (GA)
Xinxing Ma (GA)
Ross MacDonald (GA)
*Rajesh Maingi (PPPL)
Maximillian Major (Wisconsin)
Sam Major (Mines)
Amelia Manhardt (SCSU)
Trevor Marchhart (PSU)
Alessandro Marinoni (MIT)
Michael Mauel (Columbia)
Roberto Maurizio (ORAU)
Kathy McCarthy (ORNL)
Karsten Mccollam (Wisconsin)
*George McKee (Wisconsin)
Kevin McLaughlin (GA)
*Adam McLean (LLNL)
Harry McLean (LLNL)
Michael Meekins (TAE)
Jonathan Menard (PPPL)
Orso Meneghini (GA)
Marco Miller (MIT)
Mianzhen Mo (SLAC)
Steve Molesworth (UCSD)
*Saskia Mordijck (W&M)
Auna Moser (GA)
Rupak Mukherjee (PPPL)
Chris Murphy (GA)
Chris Muscatello (GA)
Alex Nagy (PPPL)
Dominic Napolitano (CFS)
Gerald Navratil (Columbia)
Raffi Nazikian (GA)
*Hutch Neilson (PPPL)
*Andrew Nelson (Columbia)
Federico Nespoli (PPPL)
*David Newman (Alaska)
Murphy Nicholas (Harvard)
Jake Nichols (ORNL)
Martin Nieto-Perez (PSU)
Lauren Nuckols (ORNL)
Tomas Odstrcil (GA)
Shane Olson (Texas A&M)
Masayuki Ono (PPPL)
*Dmitri Orlov (UCSD)
Novimir Pablant (PPPL)
Andres Pajares (GA)
Yogendra Panchal (Tennessee)
Mihir Pandya (Wisconsin)
Tara Pandya (ORNL)
Alexei Pankin (PPPL)
Clemente Parga (Zap)
Jason Parisi (PPPL)
Jm Park (ORNL)
Jong-Kyu Park (PPPL)
Jeffrey Parker (LLNL)
Felix Parra Diaz (PPPL)
Matthew Parsons (PSU)
Sai Tej Paruchuri (Lehigh)
*Carlos Paz-Soldan (Columbia)
Lavanya Periasamy (GA)
Marisa Petrusky (Colorado)
Craig Petty (GA)
Jacob Pierce (UCLA)
*Mario Podesta (PPPL)
*Francesca Poli (PPPL)
Zana Popovic (ORAU)
Douglass Post (CMU)
Kogan Powell (Utah State)
Quinn Pratt (UCLA)
Xijie Qin (Wisconsin)
Tariq Rafiq (Lehigh)
Roger Raman (Washington)
Juergen Rapp (ORNL)
*David Rasmussen (ORNL)
Allan Reiman (PPPL)
Matthew Reinke (CFS)
Ananthi Renanathan (UIUC)
Timothy Renk (SNL)
Terry Rhodes (UCLA)
Valeria Riccardo (CFS)
Nathan Richner (ORAU)
Juan Riquezes (Columbia)
Tom Rognlien (LLNL)
Gilson Ronchi (ORNL)
Graham Rossano (US ITER)
Dmitry Rudakov (UCSD)
Steven Rumbaugh (UIUC)
Steven Sabbagh (Columbia)
Ashrakat Saefan (PSU)
Hanna Schamis (PSU)
David Schissel (GA)
David Schlossberg (LLNL)
Lothar Schmitz (UCLA)
*Oliver Schmitz (Wisconsin)
*Eugenio Schuster (Lehigh)
Filippo Scotti (LLNL)
De Pascuale Sebastian (ORNL)
*Michael Segal (CFS)
Dalena Serena (PRL)

Krasheninnikov Sergei (UCSD)
Morgan Shafer (ORNL)
Amil Sharma (PPPL)
Surja Sharma (Maryland)
Masashi Shimada (INL)
Syun'Ichi Shiraiwa (PPPL)
Daisuke Shiraki (ORNL)
Priyanjana Sinha (PPPL)
George Sips (GA)
Arpan Sircar (ORNL)
Tatyana Sizyuk (ANL)
Valeryi Sizyuk (Purdue)
Tim Slendebroek (GA)
Carli Smith (PSU)
David Smith (Wisconsin)
John Smith (GA)
Michael Smith (ORNL)
*Sterling Smith (GA)
David Smithe (Tech-X)
Sergey Smolentsev (UCLA)
Joseph Snipes (IO)
*Philip Snyder (ORNL)
Wayne Solomon (GA)
Vlad Soukhanovskii (LLNL)
Andrew Sowder (EPRI)
*Don Spong (ORNL)
Bill Stacey (GA Tech)
Gary Staebler (GA)
Peter Stangeby (Toronto)
Charles Stanko (PSU)
Andrey Starikovskiy (Princeton)
Elizabeth Starling (ORAU)
Benjamin Stein-Lubrano (MIT)
Diem Stephanie (Wisconsin)
Ted Strait (GA)
George Sun (Buckner)
Xuan Sun (ORAU)
Shawn Wenjie Tang (UCLA)
Doug Taussig (GA)
Chase Taylor (INL)
Tony Taylor (GA)
Richard Temkin (MIT)
Colin Templeton (LSU)
Dan Thomas (GA)
Arian Timm (Minnesota)
Stefan Tirkas (Colorado)
Peter Titus (PPPL)
Matt Tobin (Columbia)
Dinh Truong (SNL)
*Francesca Turco (Columbia)
*Maxim Umansky (LLNL)
Jonathan Van Blarcum (Wisconsin)
Bart Van Compernolle (GA)
*Michael Van Zeeland (GA)
Venugopal Varma (ORNL)
Jean-Luc Vay (LBNL)
Vladislav Vekselman (TAE)
Brian Victor (LLNL)
Mickey Wade (ORNL)
Francois Waelbroeck (Texas)
William Waggoner (LLNL)
Michael Walker (GA)
Gregory Wallace (MIT)
William Wampler (SNL)
Guiding Wang (UCLA)
Huiqian Wang (GA)
Ling Wang (Stanford)
Weixing Wang (PPPL)
Jonathan Watkins (SNL)
Matt Watkins (GA)
Yumou Wei (Columbia)
*Bob Wilcox (ORNL)
George Wilkie (PPPL)
Theresa Wilks (MIT)
*Leigh Winfrey (PSU)
Andreas Wingen (ORNL)
Brian Wirth (Tennessee)
Zachary Wolfe (ORNL)
Kevin Woller (MIT)
Glen Wurden (LANL)
Xueqiao Xu (LLNL)
Zheng Yan (Wisconsin)
James Yang (PPPL)
Alice Ying (UCLA)
Minami Yoda (GA Tech)
Dennis Youchison (ORNL)
Jonathan Yu (GA)
Leonid Zakharov (LiWFusion)
Dalong Zhang (PNNL)
Menglong Zhao (LLNL)
Ben Zhu (LLNL)

*Workshop Discussion Leader or Scribe
†Workshop chairs

## US Department of Energy Observers

Daniel Clark
Scott Hsu
*Josh King
*Matthew Lanctot
John Mandrekas
Gene Nardella
Guinevere Shaw
James Van Dam

*DOE FES Liaisons

## Observers

Anthony Gerillo (Fusion Energy News)
Nino Pereira (Ecopulse)
Daniel Sioui (GA)
V. Alexander Stefan (Stefan University)
Stan Tomlinson (Quiet Energy)

## EU Guest Observers

Johannes De Haas (European Commission)
Ambrosio Fasoli (EPFL)
Joe Milnes (UKAEA)
Gabriella Saibene (Fusion for Energy)
Robert Wolf (IPP)
Hartmut Zohm (IPP and EUROfusion)

# Appendix C. Organization and Timeline of the Workshop

The workshop was organized in two phases spread out over several months and held virtually. The first phase focused on the scientific and technical content of the US ITER Research Program, and the second on the organizational aspects. Throughout the workshop, participants were asked to answer the charge (shown in Chapter 1), which was recast into a set of six discussion questions:

**Phase 1. Questions to be addressed by breakout groups organized by topic/research area:**

1. How can US research most effectively contribute to the success of ITER?
    - *Where do we have strengths that directly impact research areas identified in the ITER Research Plan?*
    - *What are specific research activities that ITER needs from the US community, and when are results needed to be most impactful?*
2. What essential ITER research products are needed to strengthen the domestic program to address strategic objectives aimed at the development of a fusion pilot plant?
    - *How can physics results accelerate the development of a fusion energy source?*
    - *How can experience and knowledge gained from participating in engineering, facility maintenance, technology, and diagnostic systems on a reactor-scale nuclear facility be applied to the domestic program to develop a fusion energy source?*
    - *How should these goals be prioritized in the mission of a US ITER Research effort?*

**Phase 2. Questions to be addressed by mixed breakout groups organized at random with expertise spread amongst all topical areas:**

3. How should the US organize its activities on the ITER facility?
    - *In forming an ITER Research program, what selection mechanisms could be employed to maintain transparency and ensure broad participation? Program participation should be guided by principles of Diversity, Equity, and Inclusion.*
    - *How can we maximize flexibility, agility, and impact?*
    - *What is the proper balance between on-site and remote participation in ITER research?*
    - *What resources will be needed to facilitate coordination, communication, data exchange, and analysis between the on-site and remote participants?*
4. How can we best position US researchers to capture leadership opportunities and influence within the international ITER research program?
    - *What early opportunities for engagement exist in commissioning and operational planning?*
    - *What preparation, capabilities, new skills, or collaborations are needed to maximize opportunities for US researchers?*
    - *How can we maximize the success rate of experimental proposals submitted by US researchers during the anticipated highly competitive review process for conducting experiments on ITER?*

5. How should ITER research efforts be coordinated with other segments of the FES portfolio, including domestic facilities, theory and simulation, technology, and international collaborations?
6. How can the private sector be incorporated into the ITER research program such that a two-way exchange of scientific research and technological development between ITER and existing US commercial fusion endeavors is enabled?

The questions were discussed in a series of breakout sessions led by volunteers from among the participants. The breakout groups in Phase 1 were organized topically, with experts in each of 10 topical areas (see Appendix A) identifying a series of topical research initiatives responsive to questions 1 and 2. The groups were reconstituted for Phase 2, with the discussion groups organized randomly and a new set of volunteer discussion leaders. From each phase, the discussion leaders and scribes worked with the chairs to produce a draft of this report based largely on their group's deliberations.

The timeline of the workshop is shown in Table G.1.

**TABLE G.1. WORKSHOP TIMELINE (ALL DATES IN 2022)**

|  |  |
|---|---|
| February 9 | Plenary Kickoff Meeting for Phase 1 (Technical scope) |
| February 9-March 16 | Phase 1 breakout groups, organized topically, met to discuss the technical scope of the US ITER Research Program. Discussion leaders, scribes, and chairs worked together to distill the outcomes of these discussions into the draft report. |
| February 9-March 28 | White papers accepted from participants |
| March 16 | Plenary Regroup Meeting to discuss the outcome of Phase 1 regarding technical scope |
| March 23 | Plenary Kickoff Meeting for Phase 2 (Organization) |
| March 23-July 13 | Phase 2 breakout groups, organized randomly, met to discuss how to organize the US ITER Research Program, Discussion leaders, scribes, and chairs worked together to distill the outcomes of these discussions into the draft report. |
| July 13 | Plenary Final meeting: Contents of early draft report presented to all participants. Comments and chits solicited. |
| July 13-September 29 | Discussion leaders, scribes, and chairs worked to produce a final draft. |
| September 29 | Final draft distributed to participants for comment |
| TBD (*very soon*) | Report finalized and presented to DOE |

The agendas for each of the four plenary meetings are shown below. Links to slides and recordings of most (where permission was given by the speaker) plenary presentations can be found at https://www.iterresearch.us/agenda.

## C.1. Kickoff Meeting: February 9, 2022

| Time (EST) | Topic |
|---|---|
| 12:00-12:15 pm | **Introduction and Charge**<br>*Jim Van Dam (Department of Energy)* |
| 12:15-12:45 pm | **Welcome to the US ITER Research Needs Workshop**<br>*Chuck Greenfield (General Atomics)* |
| *How can US research most effectively contribute to the success of ITER?* | |
| 12:45-1:30 pm | **Research Opportunities and Needs to Achieve ITER's Goals**<br>*Alberto Loarte (ITER Organization)* |
| 1:30-1:45 pm | **Assessment on How the US Can Address ITER Research Needs**<br>*Oliver Schmitz (University of Wisconsin-Madison)* |
| 1:45-2:05 pm | **US Contributions to the ITER Integration Challenge**<br>*Raffi Nazikian (General Atomics)* |
| 2:05-2:25 pm | **US ITER Project Overview**<br>*Kathy McCarthy (Oak Ridge National Laboratory)* |
| 2:25-2:45 pm | Break |
| *What essential ITER research products are needed to strengthen the domestic program to address strategic objectives aimed at the development of a fusion pilot plant?* | |
| 2:45-3:15 pm | **ITER Science Contributions to a Pilot Plant**<br>*Mike Mauel (Columbia University)* |
| 3:15-3:45 pm | **ITER Technology Contributions to a Pilot Plant**<br>*Chuck Kessel (Oak Ridge National Laboratory)* |
| 3:45-4:00 pm | **ITER Contributions to Privately Funded Fusion**<br>*Ahmed Diallo (Princeton Plasma Physics Laboratory)* |
| *Wrap-Up* | |
| 4:00-4:10 pm | **Next Steps for Workshop**<br>*Cami Collins (Oak Ridge National Laboratory)* |

## C.2. Regroup Meeting: March 16, 2022

| Time (EDT) | Topic |
|---|---|
| 1:30 pm | **Summary of Phase 1 Discussions**<br>*Chuck Greenfield (General Atomics) and Cami Collins (ORNL)* |

## C.3. Phase 2 Kickoff Meeting: March 23, 2022

| Time (EDT) | Topic |
| --- | --- |
| 12:30-12:45 pm | **Introduction to Phase 2**<br>*Chuck Greenfield (General Atomics)* |
| 12:45-1:45 pm | **Organizing for Success: Meeting Community, Program and ITER Needs**<br>*Mickey Wade (ORNL), Raffi Nazikian (General Atomics), Rajesh Maingi (PPPL)* |
| 1:45-2:15 pm | **Universities at the Center of the ITER Research Team**<br>*Saskia Mordijck (William and Mary)* |
| 2:15-2:45 pm | **Organization and Structure of the US LHC Program**<br>*Frank Würthwein (UCSD)* |
| 2:45-3:00 pm | **BREAK** |
| 3:00-3:20 pm | **Update on EU Preparation for ITER Operation**<br>*Ambrogio Fasoli (Chair of the EUROfusion General Assembly)* |
| 3:20-4:20 pm | **Participation in ITER: Overview of Present Means for Participation and Boundary Conditions to be Considered**<br>*Tim Luce (ITER Organization)* |
| 4:20-4:45 pm | **Private Fusion and ITER**<br>*Michael Segal (Commonwealth Fusion Systems)* |
| 4:45-5:00 pm | **US Workforce Development on ITER**<br>*Arturo Dominguez (PPPL) and Amelia Campbell (ORNL)* |
| 5:00-5:15 pm | **Next Steps for Workshop**<br>*Cami Collins (ORNL)* |

## C.4. Final Meeting: July 13, 2022

| Time (EDT) | Topic |
|---|---|
| 12:30-12:40 pm | **Welcome and Special Surprise**<br>*Cami Collins (co-chair, Oak Ridge National Laboratory)* |
| 12:40-1:00 pm | **Overview of Draft Report**<br>*Chuck Greenfield (chair, General Atomics)* |
| *Chapter 3. The US ITER Research Program* | |
| 1:00-1:35 pm | **Organizational Structure of the US ITER Research Program**<br>*Nate Ferraro, PPPL* |
| 1:35-1:45 pm | **Infrastructure Needs for the US ITER Research Program**<br>*Sterling Smith, GA* |
| 1:45-2:25 pm | **Breakout Group Discussions - Chapter 3**<br>*meeting attendees will be divided into zoom groups* |
| 2:25-2:30 pm | Break |
| 2:30-3:00 pm | Brief Report Back on What Was Heard in Discussion Groups<br>*Discussion group leaders/scribes* |
| *Chapter 4. Mobilizing a US ITER Research Workforce* | |
| 3:00-3:15 pm | **Workforce Development**<br>*Arturo Dominguez, PPPL* |
| 3:15-3:30 pm | **Working with the Broader Community**<br>*Lane Carasik (Virginia Commonwealth University) and Mike Brookman (CFS)* |
| 3:30-4:15 pm | **Breakout Group Discussions - Chapter 4**<br>*meeting attendees will be divided into zoom groups* |
| 4:15-4:30 pm | Break |
| 4:30-5:00 pm | Brief Report Back on What Was Heard in Discussion Groups<br>*Discussion group leaders/scribes* |
| | Wrap Up |

## Appendix D. White Papers Submitted to the Workshop

81 white papers, listed below, were submitted to the workshop by participants for consideration during the discussions. During the workshop, white papers were posted and referred to anonymously to limit any effects of bias. With the release of this final report, the white papers and author lists were made publicly available with permission of their authors. The white papers can be found at https://www.iterresearch.us/community-input.

**TABLE D.1. WHITE PAPERS SUBMITTED TO THE WORKSHOP**

| ID | Authors | Title |
|---|---|---|
| 1 | Professor Elias G. Carayannis; Dr. John Draper | Technopolitical Steering of the U.S. ITER R&D Program |
| 2 | N.C. Logan, J.M. Hanson, E.J. Strait | MHD Stability for, in, and Beyond ITER |
| 3 | Michael Walker | Supporting plasma initiation scenario development for ITER and future devices |
| 4 | Xi Chen, Carlos Paz-Soldan, Raffi Nazikian, Keith Burrell, Qiming Hu, Bob Wilcox | US Preparations for Leadership in ELM control and avoidance on ITER |
| 5 | Sterling Smith, Orso Meneghini, Raffi Nazikian, David Schissel | Increase US adoption of the data language of ITER (the IMAS data schema) |
| 6 | X.D. Du, M.A. Van Zeeland, W.W. Heidbrink | US energetic particle research for ITER success and fusion pilot plant development |
| 7 | D. Humphreys, J. Barr | U.S. Integrated Control and Physics Operations Activities to Support ITER Mission and Maximize Benefits of ITER Collaboration |
| 8 | David P. Schissel, Raffi Nazikian | Enabling Remote Participation in ITER Planning, Experiments, and Analysis |
| 9 | G. Staebler, R. Nazikian, O. Meneghini, D. Humphreys, C. Angioni, C. Bourdelle | Field-testing Pulse Design Simulators for ITER Operations |
| 10 | Lang Lao | Prototyping and Validation of Integrated Data Analysis Tools for Plasma State Determination – EFIT and Beyond |
| 11 | C. Paz-Soldan, N. Eidietis, R. Sweeney, R. Nazikian, J. Herfindal | ITER Engagement for Disruption Mitigation |
| 12 | Xueqiao Xu | Divertor Heat Flux Width Scaling for ITER and Beyond |
| 13 | R. Nazikian, J. Smith and G. Sips | The Value of US Engineering and Technology Contributions to ITER for the Development of US Fusion Energy |
| 14 | Hutch Neilson and PPPL Colleagues | Integrating U.S. Diagnostics into the ITER Research Program |
| 15 | Hutch Neilson and PPPL Colleagues | Building a U.S. Core Fusion Engineering Capability through Participation in ITER Operations |
| 16 | R. Maingi, C.S. Chang, F.P. Diaz | Validating predictive models for the importance of turbulence in setting the heat flux footprint in tokamaks |
| 17 | A.M. Garofalo | ITER Engagement for Baseline Scenario Alternatives |
| 18 | Marc-André de Looz | Cost Reduction: Optimizing Construction Via ITER Experience in Advanced Manufacturing, Logistics and Supply Chain |
| 19 | C. Holcomb, J. Park, A. Garofalo | ITER is a Key Part of Any US Fusion Development Roadmap Aimed at Operating Power Plants Based on Steady-State Tokamak Concepts |
| 20 | Brendan Crowley | Recommended U.S. Neutral Beam Activities to Support ITER Mission and Maximize Benefits of ITER Collaboration |
| 21 | Orso Meneghini | US contribution to IMAS |
| 22 | Anthony Leonard | Recommended U.S. Boundary Plasma Research Activities to Support ITER Mission and Maximize Benefits of ITER Collaboration |
| 23 | T. Abrams, I. Bykov, A. Lasa, P. Stangeby, E. Unterberg | Recommended U.S. PMI Activities to Support ITER Mission and Maximize Benefits of ITER Collaboration |

| ID | Authors | Title |
|---|---|---|
| 24 | Tsuyoshi Akiyama, Michael Van Zeeland, Suk-Ho Hong, and Rejean Boivin | Initiative to contribute to ITER diagnostics and accelerate US FPP diagnostic developments by leveraging ITER experience |
| 25 | James P. Anderson, Michael W. Brookman, Mirela Cengher, Xi Chen, Robert I. Pinsker, Bart Van Compernolle | RF research in support of ITER with a focus on GA's strengths |
| 26 | C.C. Petty, R.J. Buttery, P. Gohil, C.T. Holcomb, R. Nazikian, W.M. Solomon | DIII-D Research Contributions to ITER |
| 27 | Masayuki Ono, Syunichi Shiraiwa, Nicola Bertelli, Roger Raman, Min-Gu Yoo, James Yang, Stephen Jardin, S. Kaye | Provide 2D predictive code for ECH assisted tokamak start-up for ITER |
| 28 | Chase N. Taylor, Masashi Shimada, Thomas F. Fuerst, Pattrick Calderoni | The Need for Test Blanket Module (TBM) Experience |
| 29 | R. Raman, R. Lunsford, C. Clauser, S.C. Jardin, J. Menard, M. Ono | Fast Time Response Disruption Mitigation System for ITER and for an FPP |
| 30 | Tara Pandya, Robert Grove, Bor Kos, Kara Godsey, Michael Loughlin | Neutrons: The Power of Fusion |
| 31 | Tara Pandya, Robert Grove, Bor Kos, Kara Godsey, Michael Loughlin | Neutrons are Powerful and Data Rich |
| 32 | Ralph Kube, Hutch Neilson | Denoising fusion diagnostic measurements in nuclear environments |
| 33 | Robert Ellis, Doug Bishop, Peter Dugan, Hutch Neilson | Why the U.S. Needs to actively engage in ITER engineering commissioning and operations |
| 34 | SangKyeun Kim, Ricardo Shousha, SeongMoo Yang, JongKyu Park, and Egemen Kolemen | Optimized edge confinement and stability via real-time controlled 3D field |
| 35 | Arnold Lumsdaine | Knowledge Transfer from ITER Technology Development Towards an FPP |
| 36 | Dave Babineau, Robert Allgood, Brenda Garcia-Diaz, George Larsen, Robert Sindelar | Fusion Fuel Cycle Technology for ITER and a US Fusion Pilot Plant |
| 37 | Peter Titus, Nicolai Martovetsky | Superconductor Fatigue |
| 38 | Victoria Hypes-Mayfield, Cesar Camejo, David Dogruel, William Kubic, Joseph H. Dumont | LANL Contributions to Fusion Fuel Cycle Technology |
| 39 | Francesca Poli | Coordination of US activities in support of IMAS |
| 40 | M. Podestà, N. Gorelenkov, V. Duarte, M. Ono, N. Bertelli, S. Shiraiwa, G. Wilkie | Energetic particle research contributions to ITER |
| 41 | J.L. Barr, D.A. Humphreys, N.C. Logan, E.J. Strait, N.W. Eidietis, C. Rea, D. Shiraki | Disruption Prevention & Avoidance Research Before and During ITER Operations |
| 42 | D. J. Den Hartog (Univ of Wisconsin-Madison); G. V. Brown, M. E. Eckart, N. Hell (LLNL); C. A. Kilbourne, M. A. Leutenegger, F. S. Porter (NASA Goddard Space Flight Center) | Advancing x-ray diagnosis of burning plasmas with microcalorimetry |
| 43 | S. Shiraiwa, N.Bertelli, P. Bonoli, D. Green, T. Jenkins, J. Myra, M. Ono, D Smithe, and J. Wright | Provide state-of-the-art RF actuator simulation suite and analysis to ITER |
| 44 | Jean-Luc Vay, Eric Sonnendrucker | AMR-PIC simulations for ITER |

| ID | Authors | Title |
|---|---|---|
| 45 | D. J. Den Hartog, V. V. Mirnov, M. A. Thomas (Univ of Wisconsin-Madison); M. Bassan, M. Walsh (ITER Organization); L. Giudicotti (Padova University); R. Scannell (UKAEA-Culham) | Reducing Te and ne measurement uncertainty in high performance ITER operating regimes |
| 46 | Francesca Turco | Core-Edge Integration for ITER scenarios |
| 47 | J. Coburn, R. Hood, R. Kolasinski, R. Nygren, D. Truong, W. Wampler, J. Watkins | Evaluating the installation techniques and performance of ITER plasma-facing components |
| 48 | Michael Segal, Head of Open Innovation, Commonwealth Fusion Systems | The advent of the private fusion ecosystem - Q1,Q2 response to US ITER Workshop |
| 49 | J. Rapp, M. Baldwin, J. Canik, J. Lore, M. Shimada, E. Unterberg | PMI and boundary research on ITER and its role in the US fusion program |
| 50 | Ben Dudson, Alex Friedman, Ilon Joseph | Divertor transport physics validation and prediction |
| 51 | Robert Hager | First-principles modeling of ELMs and ELM-control physics |
| 52 | Michael Churchill, Raffi Nazikian, Ralph Kube, C.S. Chang | U.S. Contributions to ITER in Remote Participation, Networking and Computing |
| 53 | Paul Humrickhouse | Integral-scale validation of tritium transport models with ITER data |
| 54 | Larry Baylor | Fueling and Pumping Research for ITER Success and Beyond |
| 55 | Larry Baylor | Disruption Mitigation Research for ITER Success and Beyond |
| 56 | George Wilkie, Robert Hager, Julien Dominski | Kinetic modeling of reactor-relevant divertor conditions |
| 57 | P.C. Stangeby, E.A. Unterberg, J.W. Davis, T. Abrams, A. Bortolon, I. Bykov, D. Donovan, R. Kolasinski, A.W. Leonard, J.H. Nichols, D.L. Rudakov, G. Sinclair, D.M. Thomas, J.G. Watkins | Developing safe methods to handle PFC slag in ITER |
| 58 | George McKee | Transport Physics for Improved Confinement Projections for ITER |
| 59 | C-S Chang, Felix Para, Rajesh Maingi, Greg Hammett (PPPL); Alex Friedman, Jeff Hittinger (LLNL); Jeff Candy (GA); Scott Parker (U. Colorado), Jim Myra (Lodestar) | Comprehensive extreme-scale modeling of pedestal-SOL-divertor physics for ITER |
| 60 | Jon Menard | ITER Physics Contributions to a Steady-State Tokamak Fusion Pilot Plant |
| 61 | Matthew Parsons | Thoughts on Workforce Development in the context of U.S. Participation in ITER |
| 62 | David Green, Jeff Candy, Francesca Poli, Ben Dudson, Alex Friedman | Accelerate US Integrated Modeling |
| 63 | Arturo Dominguez | Coordination of US workforce development efforts for ITER |
| 64 | Benedikt Geiger | Exploitation of the non-nuclear phase of ITER |
| 65 | D.R. Ernst | Developing Non-ELMing Enhanced Confinement Regimes for ITER |
| 66 | S. A. Sabbagh, et al. | Critical Disruption Prediction and Avoidance Capability and Understanding for ITER |
| 67 | Tatyana Sizyuk | Material mixing and D/T retention, permeation, and tritium inventory |
| 68 | Rick Goulding, John Caughman, Mike Kaufman, Tim Bigelow, and Robert Duckworth | ITER Research Needs in Ion Cyclotron Heating Technology |
| 69 | T.M. Biewer | Establishing fusion reactor control scenarios based on information from a reduced set of nuclear-compatible diagnostics |
| 70 | Severin S Denk, Ralph Kube, Florian Laggner | Integrated data analysis for real-time plasma control and offline applications |

| ID | Authors | Title |
|---|---|---|
| 71 | Hutch Neilson and PPPL Colleagues | Earning Leadership in the ITER Research Program |
| 72 | Oliver Schmitz | Integrated Shared Governance and Funding Approach to US ITER Research |
| 73 | Hutch Neilson and PPPL Colleagues | Q4 U.S. in the Global Context, v1 |
| 74 | Saskia Mordijck, Carlos Paz-Soldan, François Waelbroeck | University Participation at the center of the ITER Research Program |
| 75 | NM Ferraro, BA Grierson, R. Maingi, R. Nazikian (coordinator), WM Solomon, MR Wade | A US ITER Coordination Office for Organizing US Participation in ITER Operations and Research |
| 76 | Marc-Andre de Looz, Hutch Neilson, Douglas Bishop | Q3 - The Value of On-Site ITER Participation |
| 77 | Srabanti Chowdhury, Ahmed Diallo, Neville Luhmann, Jr. | Harsh Environment Microwave Diagnostic Issues for Reactor Plasmas |
| 78 | Nick Murphy | Making open science a core value for US ITER research |
| 79 | Elizabeth Rose Starling | Implementing a Hybrid Conference Model to Promote Equity of Participation |
| 80 | Tara Pandya, Robert Grove, Bor Kos, Kara Godsey, Michael Loughlin | Organizing the Integration of Nuclear |
| 81 | Michael Segal, Head of Open Innovation, Commonwealth Fusion Systems | The definition of US leadership (Q4) |

# Appendix E. Topical Research Initiatives

## E.1. Plasma-Material Interactions

ITER will bring plasma-materials interactions (PMI) science into new regimes of absolute parameter space for scrape-off-layer (SOL) heat flux width, high surface temperature, active cooling, long pulse lengths, high ion fluxes/fluences, 14 MeV neutron damage, and significant tritium inventory. Numerous US community reports stress that solutions to these issues are essential for an FPP. ITER provides an opportunity to advance PMI science by stress-testing existing physics-based models and cross-machine scalings.

Limitations and resulting effects set by PMI are critically integrated and overlap with other areas of this report. Progress will require comprehensive workflows from the edge plasma to the sheath to the wall. Tightly controlled fueling and pumping of radiating impurities in the core and the edge will be needed to induce controlled detachment and reduce peak heat fluxes to acceptable levels. Edge heat fluxes, impurities, and SOL turbulence also impact RF heating and current drive actuators. While SOL heat flux width prediction and control is largely treated in DIVSOL, PMI considerations can provide materials response, effective surface measurement techniques, and divertor design constraints (space, alignment, leading edges) that set requirements for the dissipation given an upstream heat flux.

Materials properties and heat flux removal technology at plasma facing component surfaces presently limit heat flux values to 5-15 $MW/m^2$ perpendicular to the surface and <150-400 $MW/m^2$ parallel to the magnetic field line for solid materials. ITER will provide important information about operating close to this boundary at FPP relevant flux levels. While components in an FPP will likely be gas-cooled instead of water-cooled, understanding and validating models for actively-cooled divertor components is needed, and the success of ITER's carefully designed PFCs will inform FPP design. A dedicated near-term effort is needed to facilitate knowledge transfer of component design, installation, and performance of ITER's plasma-facing components, including plasma physics constraints, thermomechanical requirements, and assembly/remote handling techniques and limitations. Real-time observations of PFC performance at ITER during the start of PFPO-I can immediately inform the FPP strategy during the design phase. The US provides Upper IR/Visible Cameras which serve as a key diagnostic system for divertor surface temperature monitoring and heat load calculation during plasma operation. Therefore, the US must take as much responsibility as possible for the commissioning and operation of this diagnostic system which provides critical data for PMI studies and PFC performance.

Impurities generated by plasma facing components will impact the plasma performance, and wall conditioning has proven essential for achieving high performance. The behavior, impact, and management of net erosion and deposition of solid PFC material is not well understood. Build-up of material from the main wall to the divertor as pulse length increases will be significant. Material that delaminates or re-deposits may need to be removed to prevent excessive dust and UFO disruptions. ITER experience will identify pitfalls associated with maintenance and lifetime estimates for plasma-facing components. An FPP will likely have sacrificial limiters protecting blanket modules. These limiters will erode and transport impurities to the divertor, and thick layers of mixed-material “slag” that can lead to issues with hydrogen (deuterium and tritium fuel) retention. Information on co-deposition and material mixing in ITER will provide validation of "whole-wall" low-Z/high-Z mixed material migration codes which can be used to predict co-deposition and slag production rates and thus inform material selection for FPP. ITER will give information on potential surface morphology changes due to helium (as ash or minority) in reactor relevant conditions of high surface temperature and fluence.

ITER will provide first data of synergistic effects on neutron irradiation in tungsten and will give first information on an integrated test of tungsten monoblocks in a divertor at high heat fluxes. The robustness of the actively cooled, monoblock-style ITER divertor to transients such as ELMs and disruptions, as well as routine thermal cycling, will provide critical input on several design choices for the FPP. The tolerance of high-Z leading edges to melting, and the subsequent level of melt motion, provides information on the feasibility of a monoblock design in the FPP and the level of shaping required to protect these edges. The resistance of the ITER W monoblocks to cracking, recrystallization, neutron damage, and other metallurgical effects at grazing magnetic incidence will provide guidance on the need for new W alloys or composites in the FPP.

**TABLE PMI-1: PMI BEFORE AND DURING EACH IRP PHASE**

| IRP Phase | Configuration | ITER Needs | FPP needs |
|---|---|---|---|
| First Plasma | Temporary steel limiters, no Be first wall tiles or W divertor installed yet, 8 MW EC, glow discharge cleaning | Coordinate with PSI-SciDAC and other US PMI modeling activities to do ITER-centric predictive modeling. | Lessons learned on how to start-up and run discharge without damaging PFCs.<br><br>Knowledge transfer of PFC and divertor design choices, including plasma physics constraints, thermomechanical requirements, and assembly/remote handling techniques and limitations. Observe the installation and alignment procedures in assembly Phase I. |
| PFPO-I | Be wall tiles and W diverter cassettes installed, 20-30 MW EC, | Real-time observation of plasma operations in PFPO-I and PFPO-II with newly installed PFCs.<br><br>Evaluation of any detrimental effects (local PFC misalignments, toroidal asymmetries, non-uniform heat loads, leading edges, etc.) on plasma operations. | Early fuel retention data, thermal fatigue.<br><br>Slag management: monitoring and removal of dust, build-up of mixed materials.<br><br>Observe the installation and alignment procedures in assembly Phase II. |
| PFPO-II | All heating power available in H (or He) plasmas, diverted H mode with ELM control | Determine if first wall power fluxes and ELM mitigation are acceptable.<br><br>Begin to evaluate material migration to divertor and divertor power flux handling to guide operation, determine impact in scenarios, and optimize access to long pulse DT scenarios. | |
| Fusion Power | Full power, integrated DT operation | Preserving integrity of the divertor and the PFCs. | T retention and co-deposits in divertor |

### *Question 1: How can US research most effectively contribute to the success of ITER?*

**Initiative PMI-1A: Leverage modeling capabilities and lab-scale experiments to predict extreme PMI and material response for ITER PFCs, assess implications, and influence ITER operations**

- Apply US strengths of full PMI modeling workflow from edge plasma to sheath to wall.
- Prepare the research program around US-supported diagnostics (including reflectometer, IR).
- Address the full scope of ITER PFC lifetime, including as many PMI issues as possible (recrystallization, fuzz, sputtering, retention, dust and slag management, etc.)
- Conduct experiments on domestic facilities and via international collaboration to evaluate effects and validate models for ITER-like scenarios

**Initiative PMI-1B: Predict and control the effects of transient events on ITER PFCs**

- Organize and apply decades of experience with operation & control of ELMs, disruptions, runaway electrons, etc. to prepare for early ITER operation and protect plasma facing components.
- Identify critical parameters and effective diagnostic integration techniques for limiting main chamber fluxes to control excessive heat and particle fluxes to the ITER first wall.
- Leverage US Upper IR/Visible Cameras for divertor surface temperature and heat flux monitoring and control

**Initiative PMI-1C: Validate heat flux footprint models and reduce uncertainty in heat flux mitigation requirements**

- Increase modeling effort and utilize improved predictions for heat flux width scalings in ITER to resolve uncertainties in heat flux footprint, set requirements for the dissipation, and contribute to decisions on impurity gasses for power exhaust

**Initiative PMI-1D: Advance PMI understanding to better assess long term wall material migration**

- Focus on domestic experiments that can accommodate low-Z PFCs (DIII-D, NSTX-U, university devices) to proxy Be PMI issues, making sure that such results are useful to ITER
- A particular focus on mixed-material deposition is recommended
- Develop solutions for slag management, in-situ removal of dust and divertor armor replenishment

**Initiative PMI-1E: Carry out an advanced materials development program (primarily aimed at FPP, but may apply to future stages of ITER)**

- Novel PFC materials studies could focus on alloys or different crystalline structures of tungsten based components
- Leverage US expertise in “materials by design” (computational, additive manufacturing, joining refractories to other more robust materials) and leading facilities to test such materials under bombardment, including evaluating the impact of neutron damage on performance.

***Question 2: What essential ITER research products are needed to strengthen the domestic program to address strategic objectives aimed at the development of a fusion pilot plant?***

**Initiative PMI-2A: Utilize ITER for experimental and computational studies to advance the high priority CPP FST-SO-A* through understanding of materials interactions, performance, PFC design, and validated predictive modeling capability.** **FST-SO-A: Develop plasma-facing components capable of withstanding reactor-relevant conditions*

- Observe stable plasma operation as well as consequences of disruptions/transients on PFC damage, and operational impact on PMI
- Ensure that crucial data is obtained to validate models predicting magnitude of heat flux footprint
- Observe impact of ITER PFC design/shaping on PMI and plasma performance to inform FPP design requirements.

- Understanding and validate models for actively-cooled divertor components

**Initiative PMI-2B: Evaluate effects of nuclear environment on materials and the consequent feedback onto PMI and divertor solution**

- Quantify tritium retention and breeding
- Quantify irradiation effects on materials

**Initiative PMI-2C: Evaluate build-up and migration of slag and mixed materials and their effects on plasma performance**

- Assess how Low-Z/High-Z material mixing occurs and how the main-wall survives operations.
- Examine how divertor shaping and misalignment issues affect material deposition and operations.
- Assess and quantify fuel retention in slag

**Initiative PMI-2D: Inform predictive models on erosion, cracking, melting, and dust formation in ITER conditions to enable projections of overall lifetime of the first wall and PFCs**

- Validate boundary, wall and system codes used for FPP design

## E.2. Divertor and Scrape-Off Layer

ITER will provide immensely important data and learning experiences for reactor-relevant divertor physics and engineering, where it will operate with a fully water-cooled divertor, with 10MW/m$^2$ heat flux handling capability on a quasi-stationary timescale. ITER will operate in a regime where the particle transport and radiation behavior for neutrals, molecules, ions and impurities will be closer to an FPP than any existing experiments.

In the scrape-off layer (SOL), cross-field and parallel transport connects particles and heat to the divertor targets, and its width determines the required radiative fraction and corresponding impurity seeding levels for acceptable divertor target power loads. A narrow divertor heat flux width is a serious concern to ITER and future reactors, and this area has been identified as a high priority in multiple recent US community reports. Some state-of-the-art modeling of the boundary plasma have encouragingly predicted, however, that turbulence may result in a significantly broader heat flux channel than projected by empirical scaling from existing devices. A coordinated domestic effort in modeling, advanced diagnostic development, and experimental validation on current devices (domestic and international) should be pursued to advance divertor energy and particle transport modeling capabilities in preparation for ITER operation. Improving the fidelity of edge and divertor models, both kinetic and fluid, is an important component of a “predict first” approach, in which robust quantification of the risks to the machine of any proposed experiment will be needed. The SOL width scaling greatly impacts the ability to design a lower cost, more robust FPP divertor, and measurement and simulation of these processes in ITER is important for determining viable FPP configurations and operational scenarios.

Detachment is considered a necessity for ITER operation, where unmitigated peak heat fluxes can substantially exceed limits. Divertor target heat flux and surface temperature will be controlled in ITER with deuterium and impurity gas puffing. Combinations of turbulent SOL broadening and detachment may be needed for acceptable divertor power exhaust while maintaining good core performance. While design of the ITER hardware for heat flux control is mostly complete, the control algorithms are still under development and need to be tested on existing devices. ITER will inform actuator and real time diagnostic requirements, as well as control techniques, for reactors.

ITER will also give first information on an integrated test of tungsten monoblocks in a divertor at high heat fluxes. This topical area combines US strength in plasma materials interactions, dissipative impurity-seeded divertor scenarios, and their coupling to high-performing core plasmas and 3D field application for ELM suppression.

**TABLE DIVSOL-1: DIVSOL BEFORE AND DURING EACH IRP PHASE**

| **IRP Phase** | **Configuration** | **ITER Needs** | **FPP Needs** |
|---|---|---|---|
| First Plasma | Limited heating, no divertor yet | Team in place with divertor control algorithms and diagnostic interface | Ready for rapid assimilation of ITER divertor spectroscopy, particle and heat flux data to test and validate against models. |
| PFPO-I | Operation at low power levels (20-30 MW EC) and low density, divertor installed but radiative divertor operation not yet required | Apply edge transport modeling in pulse simulator to baseline ITER divertor operation and possible scenario optimization, ensure experiments will obtain key validation data | Determine heat flux width scalings and whether effects (turbulence, ELMs, etc) cause the heat flux width to broaden. |
| PFPO-II | All heating power available in H (or He) plasmas, diverted H mode with ELM control | Demonstrate radiative divertor control and compatibility with plasma fueling, ELM control, and core performance for applicability and optimization in DT FPO | Acquire lessons learned from divertor power flux measurement, control, and mitigation techniques in PFPO-I&2 for FPP divertor design |
| Fusion Power | Full power, integrated DT operation | Demonstrate controlled divertor power flux at full | Determine how erosion and redeposition with DT |
| | | power with impurity and ELM control | plasma affects divertor performance, lifetime, maintenance, apply to FPP divertor operation and scenario |

*Question 1: How can US research most effectively contribute to the success of ITER?*

**Initiative DIVSOL-1A: Further develop, apply, and experimentally validate plasma edge transport and kinetic neutral transport models for both axisymmetric and non-axisymmetric analysis of SOL and divertor transport**

- Address the challenge of understanding the SOL heat flux width along with the interaction of the SOL plasma with the material interface in the divertor and the main chamber numerically using state-of-the-art code developments
- Utilize US facilities and international collaboration for validation, including DIII-D, NSTX-U, and upcoming SPARC and iDTT devices, as well as US exascale computing facilities
- Utilize and enhance diagnostic development (e.g., thermographic, imaging, spectroscopic, and turbulence measurements) in existing facilities with the goal of code validation, including high fidelity capacity and high temporal+spatial resolution, for synthetic diagnostics and couple to US diagnostic efforts at ITER

**Initiative DIVSOL-1B: Determine the detachment capability of the ITER divertor in order to meet the requirements for 8 minute long, high-power H-mode discharges**

- Address explicit requests in the recent urgent research needs document associated with the ITER Research Plan
- Establish partially detached, dissipative divertor through impurity seeding and demonstrate stable exhaust of main species and impurities including helium ash

**Initiative DIVSOL-1C: Develop tools for divertor heat flux control**

- Develop and validate control algorithms
- Develop reduced models for control scenarios
- Interpret the divertor state, including target heat flux, with limited diagnostic sets.

**Initiative DIVSOL-1D: Assess three-dimensional effects on divertor fluxes arising from RMP ELM control**

- Include access to detachment in impurity seeded conditions and sustainment of the particle exhaust properties
- Utilize existing domestic and international facilities to address tasks in the ITER research plan to determine impact on the baseline ITER divertor and possible optimization in PFPO for FPO

***Question 2: What essential ITER research products are needed to strengthen the domestic program to address strategic objectives aimed at the development of a fusion pilot plant?***

**Initiative DIVSOL-2A: Develop and test plasma edge and plasma material interaction models, both 2-D and 3-D, that can be validated in a reactor-relevant divertor setting and used in FPP divertor design**

- Prioritize and utilize ITER scans and data for determining SOL radial transport and divertor heat flux width
- Determine implications of far SOL radial transport for FPP main chamber erosion and overall lifetime, and determine implications for FPP divertor volume, core density and tolerable impurity densities.

**Initiative DIVSOL-2B: Test and validate divertor control schemes for an FPP in ITER**

- Resolve divertor heat flux control challenges beyond those in existing experiments, including limited diagnostics, longer timescales for actuators such as gas puffing, and quantifying the helium exhaust capacity.
- Integrate diagnostics, divertor actuators, 3D fields and other constraints into workable control schemes.

**Initiative DIVSOL-2C: Quantify divertor tolerance to transients**

- Confirm maximum tolerable transients in a burning plasma device.

## E.3. Scenarios, Stability, and Control

ITER represents a massive step beyond present-day tokamak facilities in terms not only of performance, but in its scale and complexity. Its plasma control system (PCS) will need to be able to accurately steer the plasma conditions, maintaining high performance while controlling instabilities that can degrade the plasma or even cause a disruption. This is more than just improving on what we do today; entirely new challenges await us for equilibrium control, stability control, burn control, helium ash control, tritium inventory control, and investment protection.

Stability control is of particular importance. High performance operation will naturally challenge stability limits whose crossing could lead to degradation or even disruption. The closer we can reliably operate to those limits, the better the performance. This will be even more important in a power plant, where that performance translates to return on investment. It will be of crucial importance that we understand where those limits are and have the ability to predict and steer away from such unstable situations.

ITER will have flexible capabilities to explore a variety of operating scenarios, and will be tasked with developing optimized scenarios to meet its goals of pulsed Q=10 and steady-state Q=5 performance. ITER will benefit from efforts in the US domestic program to develop, demonstrate, and extrapolate promising operating scenarios via a combination of experiment, theory, and simulation.

The Fusion Power Operation phase of ITER will bring new challenges. At fusion gain Q=10, ⅔ of the plasma's heating power will come from the fusion reactions themselves. Our ability to control the plasma through heating sources is then reduced. ITER will provide an opportunity to develop burn control in such circumstances in preparation for even higher gain (lower auxiliary heating power) operation in a power plant. The challenge of burn control can and should be simulated in present-day tokamaks, but the experience of operating at high gain will be a much more stringent test.

Preparatory research for ITER operation should target:

- Development, experimental demonstration, and qualification of operating scenarios for First Plasma and PFPO plasmas, including robust initiation and H-mode operation
- Preparation, demonstration, and application of operating scenarios for both the Q=10 and Q=5 steady-state ITER missions
- Maintaining performance while avoiding core and edge instabilities that could reduce performance, lead to early termination of a discharge, or even damage the device
- Preparation of plasma control system algorithms applying techniques that include stability control and exception handling
- Development and experimental demonstration of a control architecture, including supervisory system and actuator management system, enabling integration of multiple control objectives while resolving in real time the tradeoff between performance and stability.
- Developing and qualifying the end-to-end workflow needed for ITER pulse preparation and validation, control trajectory and algorithm application, and as well as between-pulse assessment and monitoring for ongoing operations

These areas are key to successful operation of any reactor-grade fusion facility. Domestic research has consistently recognized this with major present-day research lines via experiment, theory, and simulation. Translating the US leadership in these preparatory activities into participation in ITER operations will provide necessary experience for designing and operating next-step devices.

Together, these areas are already major emphases in the US Fusion research program, and represent potential major contributions to the success of both ITER and an FPP. ITER's FPO phase represents a major, possibly first, opportunity to face the challenge of burn and He ash control where the majority of the heating power is provided by fusion alphas.

**TABLE SSC-1: SCENARIOS, STABILITY, AND CONTROL BEFORE AND DURING EACH IRP PHASE**

<table>
<tr><th>IRP Phase</th><th>Configuration</th><th>ITER needs</th><th>FPP needs</th></tr>
<tr><td>First Plasma</td><td>Limited simple scenario, little or no heating, ≤1MA, mostly bare walls</td><td>Research on domestic and international tokamaks, coordinated with theory and modeling, to identify a range of operating scenarios for later phases, and tools to obtain and control them, should be at an advanced stage of development. First plasma scenario is very simple and is not required to burn through.</td><td>Robust startup in a large volume, superconducting device that is achieved with limited toroidal electric field, small prefill gas pressure and passive stability with substantial induced currents in passive structures.</td></tr>
<tr><td>PFPO-I</td><td>20-30MW of heating power into H and/or He plasmas → limited capabilities to commission control tools and explore scenarios</td><td rowspan="2">All control tools should be at a mature level and have already been subsumed into the ITER Plasma Control System. There may be continuing opportunities for improved control algorithms. The workflows developed should be contributing to ITER operation. ITER will operate with progressively more challenging scenarios with experience and additional heating power by PFPO-II. Opportunities for domestic efforts (experiment, theory, simulation) to qualify scenarios and control algorithms prior to implementation on ITER.</td><td>Demonstrate capabilities of an advanced plant-scale control system, albeit with relatively simple plasmas. Limited demonstration of scenario solutions.</td></tr>
<tr><td>PFPO-II</td><td>Full operating capabilities but limited to H (or He) plasmas</td><td>Reactor grade plasma provides relevant experience in advanced plasma control and scenario development.</td></tr>
</table>

| FPO | DT fuel ➞ burning plasma | Burn and ash (He) control will be new challenges in ITER. Opportunities to simulate in domestic facilities will be critical for minimizing disruptivity and need for online optimization. | Full demonstration of candidate scenarios and complex control solutions for an FPP. |
|---|---|---|---|

***Question 1: How can US research most effectively contribute to the success of ITER?***

**Initiative SSC-1A: Develop ITER Scenario Solutions**

- Develop plasma initiation solutions with ECH pre-ionization and heating via experiment and theory
- Advance the predictive capability required to design integrated scenarios consistent with reactor-level boundary and divertor operation
- Present a menu of high-performance scenarios to ITER
- Engage with international partners and ITER to advance the scenario simulation framework necessary for pulse planning and scenario optimization

**Initiative SSC-1B: Demonstrate ITER operational pulse planning and control workflow**

- Field-test and qualify the ITER experimental planning procedure and Pulse Design Simulator (PDS) in domestic facilities and international collaborations.
- Emulate ITER control through operation of ITER Plasma Control System and algorithms on domestic tokamaks

**Initiative SSC-1C: Manage core and edge instabilities**

- Develop and optimize passively stable scenarios for ITER by leveraging modeling and experimental testing while at the same time develop control and actuation approaches towards active stabilization of key MHD instabilities such as EF, RWM, NTM
- Apply experiment and theory to advance the capability for integrated RMP ELM control with error field correction techniques

**Initiative SSC-1D: Qualify techniques to avoid disruptive phenomena in ITER and integrate with plasma control system**

- Characterize and identify disruptive event prediction, detection, and avoidance.
- Characterize control actions to prevent and avoid disruptive phenomena.

**Initiative SSC-1E: Qualify ITER control and exception handling algorithms**

- Advance the control-oriented predictive capability required to develop control algorithms, observers/estimators to complement or curate diagnostic measurements, control-integration schemes, actuator sharing strategies, and real-time stability/performance tradeoff solvers.
- Demonstrate continuous control solutions for long-pulse, self-heated plasma. Lead in implementing and testing ITER-specific algorithms.

- Demonstrate effective ITER-specific exception handling and disruptive event response.

***Question 2: What essential ITER research products are needed to strengthen the domestic program to address strategic objectives aimed at the development of a fusion pilot plant?***

**Initiative SSC-2A: Develop scenario solutions with integrated core-edge approach**

- Demonstrate predict-first and scenario optimization on ITER with reactor-relevant actuators
- Quantify robustness of scenarios and control algorithms to perturbations in ITER and how it may impact FPP design and operation
- Pursue alternative scenario development on ITER including those with higher safety margins, lower current, improved confinement, and/or incorporation of AT physics
- Engage in efforts to achieve steady-state Q=5 goal including consideration of needed upgrades to the ITER H&CD systems

**Initiative SSC-2B: Demonstrate operational pulse planning and control workflow, including limited actuators, alpha heating, and ash removal**

- Develop techniques to control dominantly self-heated burning plasmas
- Test pulse planning by participation in ITER and other experiments using the ITER PDS
- Qualify high-reliability control systems, procedures and algorithms that directly translate to an FPP

**Initiative SSC-2C: Manage core and edge instabilities**

- Utilize high performance operation in PFPO-II to predict and demonstrate passively and actively stable operation at high performance
- Operating in this unexplored domain will provide important lessons on passive-stabilization physics and active-stabilization approaches
- Exploit Fusion Power phase to extract early alpha particle physics and implication for FPP operation, including effects of alpha particles on MHD stability, impact of EP mix with alphas on linear and non-linear stability thresholds.

**Initiative SSC-2D: Qualify disruption prediction, prevention, detection and avoidance and integrate with plasma control system**

- Actively participating in the implementation phase(s) in ITER (not only design) as well as in the commissioning phase(s) and joint experimental planning to adopt solutions and extract lessons learned in characterizing and identifying disruptive event prediction and detection
- Demonstrate disruptive event prediction and detection in the unique burning plasma regime
- Characterize control actions to prevent and avoid disruptive phenomena, including conditions transitioning from external to alpha heating

**Initiative SSC-2E: Test control and exception handling algorithms**

- Develop continuous control solutions for long-pulse, incorporating lessons learned from ITER, including real-time control and estimation, exception handling and disruptive event response, to inform FPP on solutions for licensing of reactors
- Diagnostics, divertor actuators, 3D fields and other constraints must be integrated into workable control schemes. ITER's environment will much more closely match an FPP's situation in this regard than existing devices. ITER will provide a lot of practical experience in successfully integrating these elements.

## E.4. Disruption Mitigation

Although strategies are being put into place to avoid situations where a disruption might occur, the occasional disruption is inevitable, whether deliberately triggered as part of ITER's research program, because of plasma instabilities, or because of a hardware failure. As ITER's operational parameters are ramped up to full 15MA DT scenarios, developing disruption mitigation solutions as a last line of defense will be essential.

The US has had a leading role in preparing ITER's chosen baseline disruption mitigation strategy, Shattered Pellet Injection (SPI). SPI was first demonstrated on the DIII-D tokamak and later, in collaboration with US researchers, at JET, KSTAR, and others. Without the confidence to reliably mitigate disruptions and runaway electrons on ITER, plasma performance may be significantly limited due to imposed limitations on the allowable maximum current.

The baseline system is already in advanced stages of design. Any further qualification of SPI must be completed in the near term with a thorough understanding of and extrapolation needed for the various SPI utilization schemes required before the system is used in ITER. Although the DMS will not be available for First Plasma, manufacture and assembly in port plugs will already be in progress by that time. First operation and validation of the DMS will occur during PFPO-I.

In case the performance of SPI is inadequate during PFPO-I, there may be an acute need to prepare an alternative technique in time for installation during Assembly Phase 3 (between PFPO-I and PFPO-II). Such an alternative must be developed to maturity before this time period to allow for design and construction towards deployment. While work in this area has already begun, an optimistic lead time of 5 years necessitates additional priority before FP.

**TABLE DISMIT-1: DMS BEFORE AND DURING EACH IRP PHASE**

| IRP Phase | Configuration | ITER Needs | FPP Needs |
|---|---|---|---|
| First Plasma | Completed and validated design; manufacture and assembly of SPI system underway | Complete research required to freeze SPI design | |

| PFPO-I | Fully installed and operational system | Commission SPI system and determine whether modification (e.g. installing injectors at port 11) or alternative technique is needed | Model validation experiments carried out to project forward to pilot plant. Reliability and longevity assessment begins. |
|---|---|---|---|
| PFPO-II | Identified modifications complete; fully installed and operational alternative DMS (if needed) | Operation and validation continues with increasing current and stored energy | Expand studies to upgraded or alternative technique. |
| FPO | Fully operational baseline or alternate DMS that can be used with confidence | Confidence in DMS essential in all operating scenarios | Evaluate impact of RE seeded by nuclear environment. |

### *Question 1: How can US research most effectively contribute to the success of ITER?*

**Initiative DISMIT-1A: The US should form a task force to ensure that ITER will not be held back due to disruptions**

- Develop the SPI process and technology to ensure reliability and effectiveness in a reactor environment
- Coordinated efforts should use existing domestic and international devices as model validation platforms, use ITER engagement to qualify the baseline SPI approach to disruption mitigation
- Validate simulation capabilities with present experiments and develop predictive modeling
    - Predictive capability is extremely important since present tokamaks cannot replicate the conditions in a high-current discharge in ITER
    - Efforts of associated SciDACs should be focused on the goal of developing real predictive capabilities by increasing inter-SciDAC and experimental engagement
- Identify aspects that could potentially benefit from machine learning, e.g. SPI shard modeling
    - Perform cross-validation on existing devices to develop general models that are applicable to ITER
- Utilize any and all opportunities to study the DMS in the high current RE avalanche regime
    - End-of-life campaigns in existing devices such as JET
    - Engagement with high current private sector devices such as SPARC

**Initiative DISMIT-1B: Alternative DMS technologies should be developed in a serious way, especially those that can adequately react to disruptions that rapidly onset due to UFOs**

- Development must start immediately to be sufficiently mature to deploy on ITER before PFPO-II
- Explore development strategies that are less dependent on copious experimental time on major facilities, i.e., carry out basic R&D before installing on a tokamak

***Question 2: What essential ITER research products are needed to strengthen the domestic program to address strategic objectives aimed at the development of a fusion pilot plant?***

**Initiative DISMIT-2A: Demonstrate DMS reliability for proof of viability for tokamak reactors**

- Conduct dedicated model validation experiments on ITER to determine DMS performance, scalability, and implications for the operational window of an FPP
- Assess reliability and longevity of DMS systems during long pulse ITER ops
- Develop the SPI process and technology to ensure reliability and effectiveness in an FPP environment (if SPI successful in ITER)

**Initiative DISMIT-2B: Study runaway electrons seeded by the nuclear environment (tritium decay and Compton scattering)**

- Further develop runaway mitigation techniques unique to burning plasmas during FPO
- Validate and calibrate models used to project to a pilot plant

## E.5. Edge Localized Mode Control

Control of ELMs is required for ITER scenarios from the non-active phases of operations to DT burning plasmas for avoidance of uncontrolled erosion of first wall and diverter plasma-facing components (PFCs) and for tungsten impurity control. ELM control is a relative US strength with historic expertise, and we have been and should continue dedicating significant resources toward this topic in support of early ITER operation.

ITER will rely most heavily on 3D fields (Resonant Magnetic Perturbations generated by 27 ELM coils) for ELM suppression, but will be capable of utilizing other control (e.g. pellet pacing) or avoidance (naturally ELM-free operating scenarios) techniques.

**TABLE ELM-1: ELM CONTROL BEFORE AND DURING EACH IRP PHASE**

| **IRP Phase** | **Configuration** | **ITER Needs** | **FPP Needs** |
|---|---|---|---|
| First Plasma | All necessary captive hardware systems (e.g. ELM coils) are complete | Complete basis for ELM control in ITER scenarios via experiment and simulation | |
| PFPO-I | Power supplies and ancillary hardware installed and ready to begin experiments; ≥20MW ECH | Obtain robust H-mode (at least in He, preferably in H) and begin commissioning of ELM control tools | Collect data for model validation and lessons learned to develop predictive capability to project to FPP |
| PFPO-II | All heating power (≥73MW) available | Complete commissioning in hydrogen H-modes, provides PFC protection at high power | Continue at higher power and performance |
| FPO | Full capabilities, DT nuclear environment | Confidence in ELM control essential in all operating scenarios | Study ELM controlled and ELM-free scenarios in self-heated conditions |

### *Question 1: How can US research most effectively contribute to the success of ITER?*

**Initiative ELM-1A: Develop and demonstrate techniques for H-mode access and ELM mitigation/suppression and intrinsically non-ELMing regime access in the conditions anticipated during PFPO-I**

- Demonstrate ELM control in H-modes with dominant ECH heating and in helium plasmas
- Improve prediction of H-mode threshold in PFPO-I conditions: Is a 10MW ECH upgrade (to 30MW) necessary and sufficient for robust H-mode and regular Type-I ELMs so that it can explore ELM control at 1/3 field?
- Commissioning of ELM control tools in ITER should start as early as possible but will require a robust ELMing H-mode as a test platform. The ability to do this during PFPO-I will depend on availability of adequate heating power..
    - Full complement of heating systems expected in PFPO-II (≥20MW ECH, 20MW ICRH, 33MW NBI) should be more than sufficient to obtain robust H-modes
    - Techniques for reducing the L-H power threshold should be investigated
- Development and demonstration of intrinsically non-ELMing, high performance naturally ELM-free operating scenarios should be included in the ITER experimental program and in advance planning

**Initiative ELM-1B: Increase focus on modeling ITER scenarios and assess compatibility of the various ELM control approaches and intrinsically non-ELMing alternative scenarios**

- Increased modeling efforts are needed, especially for intrinsically non-ELMing regimes (to complement existing important activities in RMP-ELM modeling)
- Carry out careful model validation from experiments on multiple (existing and future, domestic and international) devices to extrapolate both intrinsically non-ELMing regimes and active ELM control actuators (RMPs, pellets) to ITER
- Collaborative activities should be expanded

### *Question 2: What essential ITER research products are needed to strengthen the domestic program to address strategic objectives aimed at the development of a fusion pilot plant?*

**Initiative ELM-2A: Develop predictive capability for ELM control in an FPP**

- Understand the successes and failures of ELM control in each operational phase of ITER
    - ITER can provide extremely valuable information on how ELM control and intrinsically non-ELMing regimes extrapolate from present experiments to

reactor conditions, for example: high density and low collisionality, low $\rho^*$, low rotation, exhaust power near the L-H threshold power

- Validate models for projecting ELM-controlled/non-ELMing scenarios to FPP, and identify FPP hardware requirements
    - Order of preference for FPP: (1) Intrinsically non-ELMing scenarios, (2) RMP ELM suppression, (3) ELM pacing
- Collect lessons learned from integrating ELM suppression into control systems
    - Examples: RMP ELM control with error field control, ELM control in self-heated conditions, etc.

## E.6. Energetic Particles

ITER is a particularly important step for energetic particle (EP) research, the mission of which is to demonstrate adequate confinement of alphas for high gain and safe operation of self-sustained burning plasmas. ITER will demonstrate long-pulse, controlled alpha particle heating at reactor-relevant parameters. Heat loads on plasma facing materials induced by the loss of alpha particles and other fusion products in burning plasmas will need to be minimized. ITER will also be able to explore helium ash removal from the plasma core, something essential for maintaining high hydrogenic purity (to avoid fuel dilution) at reactor scale. ITER will be used to validate modeling of energetic particle behavior in burning plasmas and inform on a minimum set of diagnostics and actuators that are needed to monitor and control fusion performance in future fusion reactors.

One of the strengths of US contribution to ITER is the development and validation of EP transport models for time-dependent simulations. These models are needed to simulate scenarios,evaluate impact of EP transport on fusion performance, and quantify EP losses and resulting heat loads to the wall. Model development is needed to include multiple energetic particle species and interactions (including NBI+RF,alphas, and interactions with thermal plasma populations) and the impact of different types of instabilities, not limited to Alfvén Eigenmodes.

Current EP research can most effectively prepare for ITER by ensuring that operational regimes with effective alpha heating and minimal losses can be confidently predicted and achieved with practical control techniques available in ITER. Tools such as TRANSP and OMFIT could be further leveraged to perform predictive modeling for ITER scenarios. Present devices (both in the US and abroad) should continue to be utilized for validation work of initial value and reduced models and codes required for ITER success, including collaborations with JET and possibly SPARC to grow our understanding of alpha-particle physics before and during the phases of ITER operation. Both EP diagnostics and energetic particle driven instabilities in ITER will be very different from what is encountered in present US devices. Efforts should increase to ensure that a sufficient coverage of EP dynamics will be provided by the systems presently planned for ITER and a plan is in place to ensure proper interpretation and maximum extraction of ITER DT physics. This implies a greater involvement with both the ITPA EP and the ITPA Diagnostics groups.

**TABLE EP-1: EP BEFORE AND DURING EACH IRP PHASE**

| IRP Phase | Configuration | ITER Needs | FPP Needs |
|---|---|---|---|
| First Plasma | Limited heating, no significant EP population | Completed integrated EP modeling workflow and application to ITER scenarios, including pulse design | Projection of EP stability and core transport in integrated models. |
| PFPO-I | 20-30 MW EC, no significant EP population | Completed plan for monitoring EP population, burn control, wall loads, and utilizing synthetic diagnostics. Validation of EP physics models and applicable regimes is at an advanced stage. | Pulse design with EP physics, including wall loading. |
| PFPO-II | 33 MW NBI + 20 MW IC in H (or He) plasmas, MSE available to constrain magnetic equilibrium reconstruction | Integrated prediction and demonstrated control of EP confinement in H/He/D and DT plasmas with ICRF+NBI EP populations and available actuators, project to prepare/optimize access to long pulse DT scenarios | Minimum set of required EP diagnostics and actuators for a reactor. |
| Fusion Power | Alpha heating, alpha driven EP instabilities, control of non-linear heating response | Demonstrate predictive control of alpha stability and heating, wall loading, He ash removal | Burn control techniques, effects of alpha-driven instabilities. |

### *Question 1: How can US research most effectively contribute to the success of ITER?*

**Initiative EP-1A: Redirect and coordinate energetic particle research efforts to address ITER-relevant questions**

- Provide improved, validated RF models for time-dependent integrated simulations with NBI+RF and alphas through coordination of EP+RF communities.
- Provide practical tools for EP losses, heat loads, including multiple EP species and instabilities.
- Apply tools to assess the impact of energetic particle transport in various scenarios and phases in the discharge. Validate under conditions of relevance for ITER and codes within whole device modeling suites required for ITER success.

**Initiative EP-1B: Provide first-principle codes, as well as reduced-physics models that can be integrated into time-dependent integrated modeling codes**

**Initiative EP-1C: Develop new synthetic diagnostic algorithms to simulate the various ITER EP diagnostics and enable efficient code validation**

- Ensure ITER systems provide sufficient coverage of EP dynamics.

***Question 2: What essential ITER research products are needed to strengthen the domestic program to address strategic objectives aimed at the development of a fusion pilot plant?***

**Initiative EP-2A: Use physics results from ITER to validate calculations of EP instabilities, transport rates, and scaling in codes and project alpha heating dynamics for the FPP**

- Determine if alpha heating can be effectively used in plasma start-up phase to minimize external heating, and determine what levels of control and external heating sources will be needed to maintain the steady-state phase of the reactor operation.

**Initiative EP-2B: Use ITER data to validate and make projections of wall loading by energetic particles in a US FPP**

- Extend EP studies and codes to beyond the last-closed-flux-surface to include heat loads by lost energetic particles in order to assess potential risk for an FPP.

**Initiative EP-2C: Use experience with EPs on ITER to inform on the minimum set of measurements to be able to control the reactor, e.g. dedicated measurements for burn control and heat loads caused by EP losses**

## E.7. Transport and Confinement

Particle, momentum, energy, energetic particle, and impurity transport in the core, pedestal, and SOL play a critical role in achieving ITER goals. In a burning plasma, the three regions are strongly coupled, all transport channels are coupled, and fusion self-heating further magnifies this nonlinear interdependence.

Continued advancement in this area leverages the work we can do with our present-day capabilities (devices, diagnostics, modeling tools,…), and new tools must be continuously developed to address the different parameter regimes in ITER relative to present machines. This effort leads toward validated understanding of transport and confinement behavior for use in predictive simulations that are already essential for preparation and optimization of operating scenarios for ITER (see 2.5).

ITER will move into new regions of parameter space, and experience tells us there are likely to be surprises (good and bad) . But ITER will not be the final step on the quest for fusion energy; it will be followed or accompanied by devices (such as an FPP) that move even further into unexplored regions of parameter space. Data from ITER and its extensive suite of diagnostics will be essential for validating the models that can be used to design and optimize those next steps.

**TABLE TC-1: TRANSPORT AND CONFINEMENT BEFORE AND DURING EACH IRP PHASE**

| IRP Phase | Configuration | ITER needs | FPP needs |
|---|---|---|---|
| First Plasma | All necessary captive hardware systems (e.g. ELM coils) are complete | Continue to leverage existing facilities to produce validated high fidelity models that can extrapolate to predict behavior in ITER | |
| PFPO-I | Power supplies and ancillary hardware installed and ready to begin experiments; ≥20MW ECH | Utilize these models for application to ITER in developing scenarios and carry out experiments in a "predict-first" mode. Continually work to improve and validate models. | Successively improved validated predictive capability as ITER ramps up toward its full performance can be applied to a pilot plant design |
| PFPO-II | All heating power (≥73 MW) available | | |
| FPO | Full capabilities, DT nuclear environment | Preparation for FPO will entail improved predictive understanding of isotope effects and presence of energetic and thermalized alpha particles | Add burning plasma characteristics at low $\rho_*$ to the set of validated models |

### *Question 1: How can US research most effectively contribute to the success of ITER?*

**Initiative TC-1A: Expand, develop, and validate full and reduced models to enable proactive evaluation of ITER scenarios, including fundamental questions not addressed in the research plan**

- Develop practical predictive capability for pedestal-SOL turbulence coupled with collisional transport and other important physics (radiation, neutrals, impurities, finite orbit widths, orbit loss, etc.), including the role of turbulence in heat flux width.
    - Build on US strengths in the theory and first principles simulation of fusion plasma turbulence to develop practical and validated simulations of pedestal and SOL turbulence and transport.
    - These simulations must be accurate, but also practical, to be useful for ITER and FPP. The new capability must address the relevant physics of the pedestal and SOL and not simply consist of running codes based on fundamentally invalid core orderings. New codes must be developed.
    - Such new predictive capability should include intrinsically non-ELMing regimes.
- Develop predict-first capability for ITER experiments
    - New capabilities are needed to address turbulent and collisional transport simultaneously in SOL, pedestal, and core, including physics important in burning

plasmas such as isotope effects, energetic particle interactions with turbulence, strong electron heating, and impurity transport, all at small ρ*.
    - Evaluate whether ITER diagnostics will be sufficient for validation, and sensitivities of fusion performance to modeling errors.
- Leverage existing facilities for validation, and to understand beneficial confinement in alternative regimes, including non-ELMing regimes. Useful validation exploiting turbulence measurements and comprehensive simulations requires a more focused effort.

***Question 2: What essential ITER research products are needed to strengthen the domestic program to address strategic objectives aimed at the development of a fusion pilot plant?***

**Initiative TC-2A: Develop experimentally validated, first-principles predictive capability for burning plasmas, including turbulence and transport, while spanning core to SOL for use in predicting and planning behavior in later phases of ITER and an FPP**

- Careful measurements of all transport related quantities, including fluctuations, in ITER will add critical regions of parameter space to the existing database for validation
- These measurements will be needed at all stages of ITER operation, including both non-nuclear and burning plasmas
- Validate against measurements of all transport quantities, including fluctuations, in ITER discharges in a variety of conditions

## E.8. Modeling and Simulation

An important goal of US fusion modeling and simulation is to accelerate the availability, usage, and support of the integrated modeling capabilities required for both ITER and domestic fusion efforts. Modeling can identify better methods of operation or identify where new problems may arise in regimes different from existing devices. Extensive modeling of discharge scenarios to explore the robustness of the plasma control system and proximity to operational boundaries will likely be required prior to approval of high-power ITER discharges. Modeling is essential, for example, in disruption avoidance and mitigation, because disruption stresses and their effects cannot be safely validated on ITER at full power, yet have implications for safety and licensing purposes. Robust software engineering of control systems is critical for a future power plant which will need to be automated as much as possible.

Development of multiple levels of fidelity hierarchy are needed across all topical areas. The US should prioritize development of modeling capabilities that can be applied to systems and plasma regimes to which ITER provides unique access, both to position the US to benefit maximally from ITER data, and potentially to accelerate the achievement of ITER's mission. Access to as much information on ITER workflows as possible (including development notes, source code, usage, and capabilities) aids FPP model development efforts.

Experience on ITER will provide valuable insights into the challenges of control algorithms for an FPP, where limited measurements are expected to be available due to the harsh neutronic and radiation environment. Plasma control will require accurate determination of the plasma state including its equilibrium, confinement, and stability properties from the limited diagnostic capability, and physics models will need to evolve to include system response and engineering aspects to meet regulatory and safety requirements. Real-time plasma control will need to be hardened against the failure of single sensors. It is especially important to integrate disruption prediction, avoidance, and mitigation into the plasma control systems. The US can leverage and apply extensive experience in plasma control to improve the necessary analysis tools and physics models for a burning plasma environment in ITER, prototype the workflows, and where beneficial, validate them against present tokamaks in preparation and parallel to ITER.

In particular, emphasis should be placed on simulation capabilities for the boundary and divertor, including reduced models and improved predictive modeling for heat flux width and divertor transport, PMI, RF interactions, and coupling plasma models to constraints on bulk materials, structural support, and the blanket test modules. The self-consistent understanding and prediction through modeling of the inter-related turbulent transport and pedestal-SOL-divertor plasma behaviors in the low-$\rho^*$ (ratio of gyroradius to equilibrium scale) machines such as ITER can have significant impacts on timelines for demonstrating high power discharges with power exhaust management. The US has significant advantages with available extreme-scale computers for open-science research. This uniquely enables first-principles-based boundary plasma modeling that solves non-equilibrium multi-physics in complex geometries that include the magnetic separatrix and the material wall. Multi-physics includes neoclassical physics, microturbulence and blob-filaments, edge localized modes, neutral particles, atomic physics, and impurities. These phenomena are typically scale-inseparable in both space and time, interact nonlinearly, and make analytic or reduced modeling difficult. There are multiple groups in the US who have been developing such codes and attempting to use the understanding they offer to improve reduced-model simulations of both the edge and the whole device.

Lessons learned from modeling for fusion nuclear technologies in ITER, particularly tritium fuel cycle and neutronics, can significantly impact US FPP design, cost, operations, and safety. Modeling of tritium transport in ITER is needed to predict tritium flows and inventories, quantify uncertainties, and perform safety analysis for both ITER and to guide FPP-relevant design and research. US involvement in the ITER fuel cycle control software and simulator enables development of a fuel cycle for a US FPP by providing a modeling framework and implementation of best practices. Recommended activities involve ITER pre-operation development of detailed computational models of the facility and its subsystems, enabled by full access to ITER tritium plant and facility design details. While the principal research product of interest here (complete code validation and measured tritium distributions in ITER) cannot be obtained until the corresponding phase of ITER operation begins in 2036, relevant hydrogen and deuterium data might be obtained in earlier phases. Nuclear analysis will also have implications for remaining design, operations, and safety of ITER as well as US FPP, and development, documentation, and validation of necessary tools and software will be highly important for success. Current strengths of the US nuclear design, analysis, and integration (NDAI) that could benefit ITER include: efficient utilization of high performance computing platforms;

development and application of HPC Radiation Transport tools; conversion of computer-aided design (CAD) models to transport models; and improved rubric for iterative design between the development of CAD models, the implementation of these models into neutronics calculations, and the application of neutronics results in multiphysics calculations.

Further discussion of resources that are needed to enable IMAS code compatibility can be found in General Initiative - IMAS, as well as coordination of efforts across the FES portfolio [Coordination with Theory and Simulation Community].

**TABLE MODSIM -1: MODSIM BEFORE AND DURING EACH IRP PHASE**

| IRP Phase | Configuration | ITER Needs | FPP Needs |
|---|---|---|---|
| First Plasma | Limited heating and diagnostics | IMAS compatible codes, data workflows are well established and validated with existing devices. Pulse simulator implemented, utilizing reduced models. Modeling and integrated commissioning ensures breakdown/startup success. | Integrated core-edge models for design, computational model development for fuel cycle and nuclear analysis |
| PFPO-I | Operation at low power levels (20-30 MW EC) and low density | Advanced integrated, predictive modeling and demonstrated scenario control with exception handling. Modeling to enable early H-mode access, heat-flux width validation, ELM control, success against disruptions, etc. | |
| PFPO-II | All heating power available in H (or He) plasmas, diverted H mode with ELM control | Scenario prediction and validation including RF models, extensive test of plasma control system, actuators, and proximity to operational boundaries. | Insights into the challenges of control algorithms with limited diagnostics for an FPP |
| Fusion Power | Full power, integrated D and DT operation, full | Complex, long pulse scenario control, | Full demonstration of control algorithms with limited |
| | diagnostic set for validation | validated high fidelity plasma simulator | diagnostics and burn control, complete tritium fuel cycle and neutronics code validation |

***Question 1: How can US research most effectively contribute to the success of ITER?***

**Initiative MODSIM-1A: Develop tools to simulate entire ITER discharges in advance**

- Integrate reduced models into full discharge simulation for application to predict-first experiment design, control, and results interpretation.
- Focus should include boundary tools for divertor heat load, ELM suppression, as well as tools for RF and disruption physics
- Ensure control in highly demanding ITER nuclear environment, including safety constraints and limited diagnostic capability

**Initiative MODSIM-1B: Develop models at all levels of fidelity**

- Accelerate development of practical reduced (faster) models needed for pulse design and optimization
- Ensure predict-first works through systematic, large-scale validation of numerical models on existing and planned experiments, extending international collaborations

***Question 2: What essential ITER research products are needed to strengthen the domestic program to address strategic objectives aimed at the development of a fusion pilot plant?***

**Initiative MODSIM-2A: Carry out validation and further development of physics, engineering, and operational models at reactor scale**

- Assess scientific and technical readiness (using Technology Readiness Levels or scientific equivalent) in order to track the status of model validation, critical gaps, and put resources where needed, especially where there are implications for safety and licensing.
- Validate pre-simulated ITER scenarios in order to bring that knowledge to FPP design and operation and assess where important differences may exist
- Qualify automated tools and integrated real-time controls to guide requirements for reactor-compatible diagnostics and sensors

**Initiative MODSIM-2B: Enable easy, persistent, and rapid access to ITER data, facilitate remote participation, and rapid turnaround of results**

- Enable transport of large amounts of data and practice analysis workflows for rapid turnaround to catalyze application of early results to FPP design
- Increase resources to modernize codes, to provide documentation and support, and to utilize modern computer architectures effectively

**Initiative MODSIM-2C: Test the full cycle (analysis, predictions, pulse-design, control)**

- Gain experience, best practices, and lessons learned for integrated tools and workflow on a large-scale fusion device (e.g. burning plasma + tritium handling + other hardware/software), including operating in a mode of "Fusion Pilot Plant simulator" with reduced actuators and diagnostics

## E.9. Diagnostics

The top US goals for diagnostics on ITER are a) to ensure the success of US ITER diagnostics systems as measured by compliance with design requirements and utility of the instrumentation to enable achievement of ITER's performance and scientific missions, and b) to maximize the knowledge gained from US experience with ITER diagnostic systems in order to accelerate the development of diagnostic and control systems for a US FPP. Both goals require that US researchers be active participants in the commissioning, operation, and exploitation of (at least) the US-contributed ITER diagnostics. Such involvement is necessary if the US is to capture the diagnostic operational experience and performance results in a burning plasma environment. Through the design and operation of ITER diagnostics, a US talent pool including scientists, technicians and engineers will be developed that is capable of applying this knowledge to diagnostic systems for a US FPP.

The US is responsible for supplying 7 diagnostic systems, which includes design, fabrication, and delivery (see Appendix A Table 1 -US Diagnostics). These diagnostics are critical for ITER, and successful delivery is a high priority. The designs of ITER diagnostics are fairly established at this point, though some remaining R&D/prototyping is needed for each specific diagnostic. While some US ITER diagnostic concepts have already been tested on existing US facilities, continued R&D would improve ITER implementation and potentially demonstrate new techniques and hardware that could be applied to ITER in the future. Currently, US institutions not part of the 7 system design teams have limited/no access to ITER Diagnostics. A mechanism should be put in place to ensure US fusion scientists and engineers (both public and private) have access and avenues for engagement with ITER Diagnostics. The Diagnostics ITPA group can be a forum for open exchange of information between ITER representatives and the larger community, both public and private. Any barriers to openness must be reduced to the extent possible. Stronger partnerships between physicists and engineers in the development and testing of ITER diagnostics (both US and international contributed) would strengthen the basis for diagnostic designs, reduce the risks to in-service performance, and develop deeper understanding of each diagnostic's operation and capabilities to ensure well-prepared use in advancing the ITER research program as well as transfer of knowledge to US FPP diagnostic development.

Beyond successful delivery of US diagnostics, it is not a given a priori that the US will have responsibility for operating and exploiting those diagnostics. It is strongly recommended that the US develop a plan and a workforce for participation in the assembly, installation, commissioning, operation, and exploitation of the US-contributed ITER diagnostics and claim leadership responsibilities as appropriate. Long-term continuity is needed, commensurate with the timescales of ITER operations. Programs should begin now, for example with opportunities for graduate students who could then be experts when ITER begins operating. Developments of diagnostic hardware, data analysis schemes, and PCS tools should be supported using domestic and international facilities now, to be available for whenever IO contract opportunities are released. A dedicated US diagnostics effort, both on-site and off-site from ITER, allows direct experience with operation and maintenance of diagnostics at reactor scale, as well as a first glimpse into ITER physics results as the data become available.

For ITER success, the US should ensure reliable operation and appropriate/timely data analysis, resolving issues as they arise not only for US-credited diagnostics but for the entirety of the ITER diagnostic suite. Participation in the ITPA Diagnostics working group should be supported and expanded, with mechanisms to contribute to the diagnostics of other parties and the IO in order to gain broader experience. Near term opportunities may include hosting an ITPA Diagnostics meeting, utilization of direct IO contract and ITER on-site work through frameworks like ITER Project Associates (IPA) or ITER Science Fellows networks, and facilitation of US proposals for direct contracts with relevant parties, or other funding mechanisms, to enable researchers to address specific R&D requests from ITER IO and ITPA Diagnostics.

Even in non-nuclear phases, diagnostics and control schemes for critical elements such as H-mode transition, large heat dissipation by radiation, disruption mitigation, etc. can accelerate ITER's path to success and simultaneously inform FPP design. The US should develop synthetic diagnostics as well as data interpretation workflows for inclusion in predictive simulations for ITER and for plasma control demonstrations using the ITER diagnostic set. Data analysis techniques (e.g. Bayesian analysis) that combine information from several diagnostics can be a powerful advantage to leverage both US and partner diagnostics. Additionally, these tools should be complemented with experimental testing either on relevant test stands and/or tokamaks.

In FPO, ITER will provide real-world experience with the control of a reactor scale, long-pulse burning plasma – experience that is essential for the safe and reliable operation of an FPP. While ITER's diagnostic set will be broader than that of an FPP and will include more direct measurements of the properties of interest, ITER can be used to inform on a minimum set of measurements needed to control a reactor and better inform the operation of an FPP through control tests using a representative reduced diagnostic set. To complement these tests, control algorithms that include real-time synthetic diagnostics, integrated data analysis of multiple measurements using machine learning, and other approaches can be developed that are tailored to the needs of an FPP. ITER will provide invaluable information on issues and diagnostic lifetimes in a harsh reactor-like environment that can only be approximated on present devices. Reliability and utility for an FPP should be evaluated using quantitative risk assessment based on ITER data, including the impact of 14 MeV neutron/gamma-ray irradiated materials on diagnostic performance. This knowledge is also essential for the licensing of a US FPP, and leveraging nuclear experience on ITER can lead to development of robust, radiation hardened diagnostics and materials for an FPP.

**TABLE DIAG -1: DIAGNOSTICS BEFORE AND DURING EACH IRP PHASE**

| IRP Phase | Configuration | ITER Needs | FPP Needs |
|---|---|---|---|
| First Plasma | Basic main parameter diagnostics (magnetics, breakdown, X-rays, density) | Address remaining design needs, diagnostic hardware development, data analysis schemes, and PCS tools | ITER diagnostic design concepts and detailed hardware solutions. |
| PFPO-I | Operation at low power levels (20-30 MW EC) and low density, expanded diagnostics for plasma parameters, control, neutrons | Ensure operation and data from US-led ECE, TIP, XRCS-Core, UWAV, DRGA. | Knowledge transfer from commissioning, maintenance, operational experience. |
| PFPO-II | All heating power available in H (or He) plasmas, 15 MA/5.3 T, high performing, routine systems for core, edge, divertor measurements | Ensure operation and data from US-led MSE, LFSR. Demonstrate advanced control and data analysis. | Evaluate utility and reliability of FPP relevant diagnostic set. |
| Fusion Power | Full power, integrated DT operation and measurement capability for fusion power production and burning plasma physics | Demonstrate burn control algorithms that include real-time synthetic diagnostics, integrated data analysis | Utilize real-world experience with control of burning plasma to validate FPP diagnostics and control algorithms. Use information on diagnostic issues/lifetimes in reactor-like environment. |

***Question 1: How can US research most effectively contribute to the success of ITER?***

**Initiative DIAG-1A: The US should develop a plan and a workforce for participation in the commissioning, operation, and exploitation of the US-contributed ITER diagnostics**

- Ensure reliable operation, appropriate/timely data analysis, and issue resolution
- Ensure long-term knowledge transfer of data analysis techniques, engineering designs, performance statistics, maintenance procedures and logs, operational experience, troubleshooting information, etc.

**Initiative DIAG-1B: Develop synthetic diagnostics and data interpretation workflows for inclusion in predictive simulations for ITER and for plasma control demonstrations using ITER diagnostic set**

- Leverage the entire ITER diagnostic set (both US and from other partners) and prepare with experimental testing on relevant facilities.

***Question 2: What essential ITER research products are needed to strengthen the domestic program to address strategic objectives aimed at the development of a fusion pilot plant?***

**Initiative DIAG-2A: Utilize ITER to gain experience with the control of a reactor scale, long-pulse burning plasma**

- Inform the operation of an FPP through control tests using reduced diagnostic set more representative of that in an FPP
- Gather data on diagnostic reliability and lessons learned
- Place priority on diagnostics used for boundary plasma, divertor heat flux control, and fluctuations

## E.10. Technology and Integration

ITER's technology has immediate and long term value to US fusion – many ITER systems have final or preliminary designs backed by robust analysis that would be of immediate value. Information sharing enables US efforts to further validate advanced fusion technology hardware and develop operational practices for ITER and future burning plasmas.

ITER will be a vital test of key enabling technologies for burning plasma physics in its operation. Among these are long pulse heating and current drive sources (ECH, neutral beams and ICH); fuel cycle systems (pellet injection fueling and ELM pacing, D, T and He exhaust pumping and processing), and magnet and blanket cooling (cryogenic plants and heat removal systems.) All of these technologies are vital to the FPP mission, in particular providing experience with long pulse DT operation. ITER will demonstrate cryogenics and tritium handling for a long pulse burning device, previously untested on this scale. ITER technology development and experience with Low-Z actively cooled PFCs has direct application to US DOE program objectives such as MPEX and FPP. System level radiation transport code development and data from ITER DT operation will have substantial value to development of long pulse devices.

Substantive US contributions to ITER technology, magnets, fueling, tritium exhaust pumping and processing, ICH and ECH transmission line systems, heat removal and diagnostics have already been designed and tested, and many are now in fabrication and delivery. In the present and near term, the US needs to deliver our ITER technology contributions and also pivot efforts to ensure robust knowledge transfer to US institutions. This can begin now through participation in the installation, commissioning and startup operations of both the US and systems from other ITER parties. Programs for undergraduates at US labs, graduate participation in technology development, positions for US scientists to visit facilities such as the ITER Neutral Beam test facility and the FALCON ECH test facilities should be complemented by the development of US facilities and a dedicated organization that will help insure the broadest and most equitable possible US access to ITER research product.

Technology consists of not just hardware, but also the software and analysis needed to design and operate it. Wherever possible, the US should participate in the development of software and analysis and call on ITER to make it available to the larger community. US private and public development within an open-source framework enables quality science. While recognizing some developments might be restricted by issues of intellectual property, ITER has not just an opportunity but an obligation to sort through IP concerns [4.2. Working With the Broader Community] to address these issues with US stakeholders. This is a near term opportunity contributing to an FPP program well before ITER's final research products are available.

Decades of investment in ITER have enhanced the technical readiness of fusion technology, but there are limited mechanisms to transmit lessons learned – primarily publications and a limited number of public ITER reports. There are currently limited infrastructure and limited US/IO collaborative agreements for effective transfer of research and development output. US DOE and US ITER should expand near term ITER technology collaborations that can extend wider to future scientific and operational efforts, enabling the future of not just private or publicly funded science, but net-energy US fusion [see also General Initiative - Knowledge].

ITER's mission will also benefit from these near and long term engagements of US technology experts, matched to US strengths and US needs for an FPP. A substantial onsite presence will clearly be needed for effective participation but it can be enhanced through remote participation [3.2.2] and via structures for broad data access [Initiative MODSIM-2B], thereby providing the greatest scientific return on our efforts and investment throughout the life of the ITER program.

**TABLE TECH-1: TECHNOLOGY NEEDS FOR EACH IRP PHASE**

| IRP Phase | Configuration | ITER Needs | FPP Needs |
|---|---|---|---|
| First Plasma | Full data system ready, remote participation ready, 8 MW EC, glow discharge cleaning | Tests of remote participation and data sharing technologies, validation of ITER software tools | Prompt efforts to transfer technology from ITER to us research across the full range of technologies |
| PFPO-I | 20-30 MW EC, initial validation of models, some understanding for hardware challenges and reliability questions | Validation of high power ECH performance and efficiency, Prompt technical solutions to any issues uncovered during first plasmas | Operational and engineering understanding of technology performance and reliability statistics |
| PFPO-II | All heating power available in H (or He) plasmas - long pulse | Testing and development, work finalizing NBI and ICRF systems, studying their efficient and reliable operation | Understanding of the full range of H&CD sources and behavior at high power |
| Fusion Power | Full power IC, NBI, ECH, integrated DT operation with NEUTRONS | Testing of neutron resilience of materials and system involved in ITER | Comparison of heating and current drive systems in DT plasma, data on neutron damage |

***Question 1: How can US research most effectively contribute to the success of ITER?***

**Initiative TECH-1A: Become actively involved in commissioning and startup**

- Position US experts (on-site and remotely) to lead commissioning and operation of systems they helped design and deliver

**Initiative TECH-1B: Engage US technology experts in addressing near-term ITER needs**

- Conduct a comprehensive review of domestic capabilities and map these to near-term ITER needs. The US has strengths in a wide breadth of technology areas including tritium fuel cycle, beryllium handling, magnets, disruption mitigation, nuclear design and integration, plasma heating and current drive, superconductor fatigue, diagnostics, and safety and licensing.

***Question 2: What essential ITER research products are needed to strengthen the domestic program to address strategic objectives aimed at the development of a fusion pilot plant?***

**Initiative TECH-2A: Develop relevant technology and integration tools**

- Develop and validate an ITER digital twin
- Develop and validate simulation tools for radiation transport, tritium breeding and transport, power conversion, and safety analysis. ITER will be the largest and most relevant neutronics and tritium transport validation experiment in fusion, and making full use of this data is critical to the safety and licensing of any future plant.

- Capitalize on experience gained in plasma facing component installation and operation, particle control, and heating and current drive

**Initiative TECH-2B: Gain experience with a tritium fuel cycle**

- Execute an international export control agreement between partners to enable a more collaborative tritium research effort
- Make full use ITER data to validate computational transport models of the facility and its subsystems to predict tritium flows, inventories, and loss terms
- Develop a tritium and fuel cycle systems simulator.
- Consider options to  gain access to TBM data from ITER or ITER members

**Initiative TECH-2C: Transfer knowledge of fusion reactor engineering and operations**

- Ensure full access to technology design, testing, reliability information, commissioning, operation, repair and maintenance procedures - as well as the failure rates of all hardware during all phases of ITER assembly and operation  have extensive value to US efforts.
- Support the robust on-site and remote participation needed to ensure US technology researchers obtain the “tacit knowledge” that can only be transmitted through human interaction in operations.

## Appendix F. Workshop Survey Results

As part of the final workshop meeting, all registered participants were requested to complete an anonymous survey based on the draft report. The survey gave respondents the option to respond to the following questions:

a) Rank the top 5 topical initiatives that the US should focus on for ITER's success.
*Consider factors such as : leveraging US program strengths, impact on achieving both timeline and technical goals, and providing possibly unique support to ITER's success.*

b) Rank the top 5 topical initiatives where ITER research would help contribute to a US FPP.
*Consider factors such areas where ITER may provide unique experience and data that impacts the design of an FPP. Here, an FPP is defined by the 2020 Community Planning Process report to have the deliverables: 1) produce net electricity from fusion, 2) establish the capability of high average power output, and 3) demonstrate the safe production and handling of tritium, as well as the feasibility of a closed fuel cycle.*

There were 415 total workshop registrants and 98 total survey responses, with 88 people responding to (a) and 79 people responding to (b). The distribution of topical area expertise of the survey respondents was similar to the distribution of the total registered workshop participants. Respondents could pick one or several closely related initiatives per priority level. Each ranking level was issued a value (i.e. top priority = 5 points, second priority = 4 points, etc.) and point values were normalized by the total number of initiatives selected (i.e. point value divided by 5 if a total of 5 initiatives were ranked, or divided by 8 if 8 total initiatives were ranked).

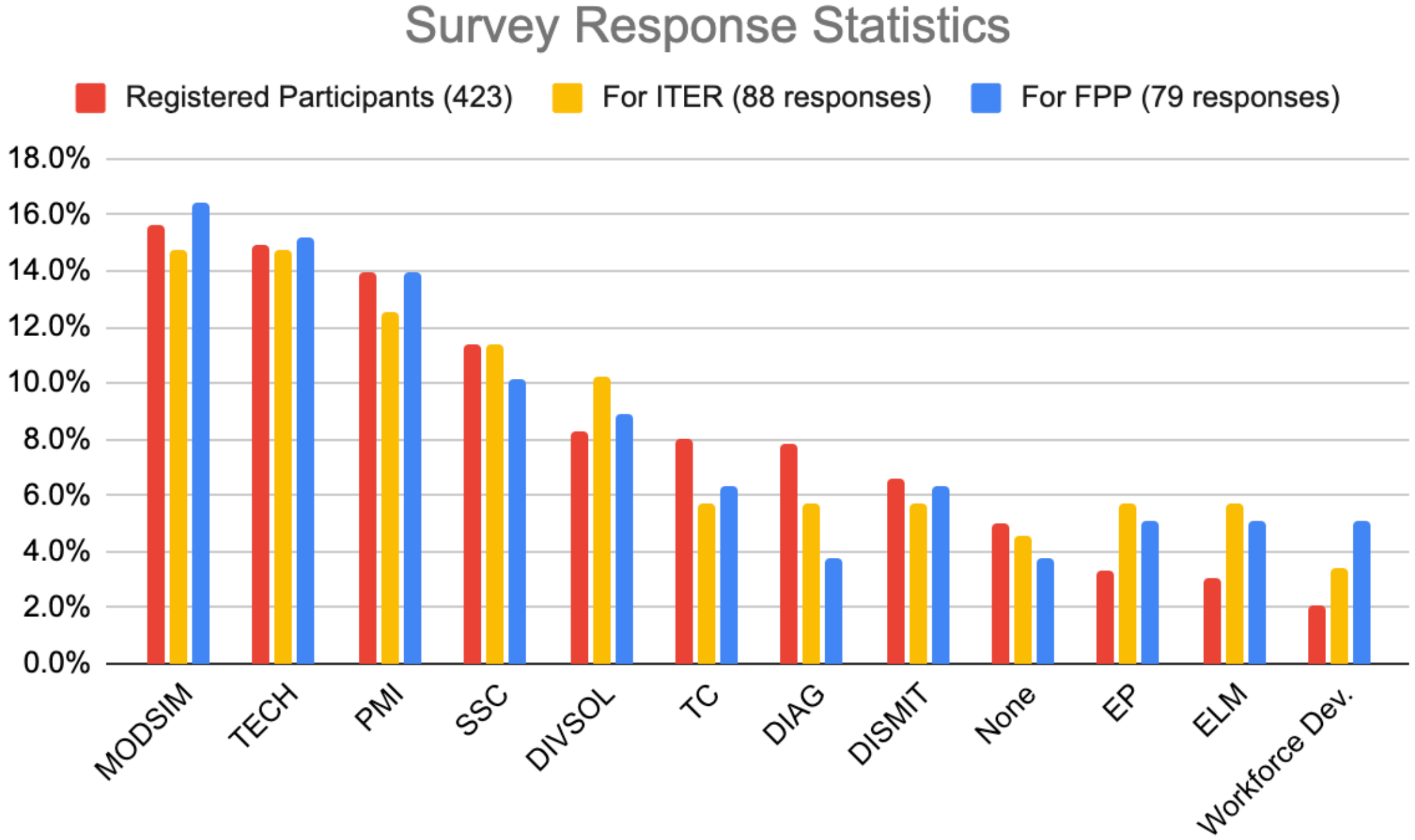

Figure F.1. Comparison of topical area distribution of total registered workshop participants vs. survey respondents.

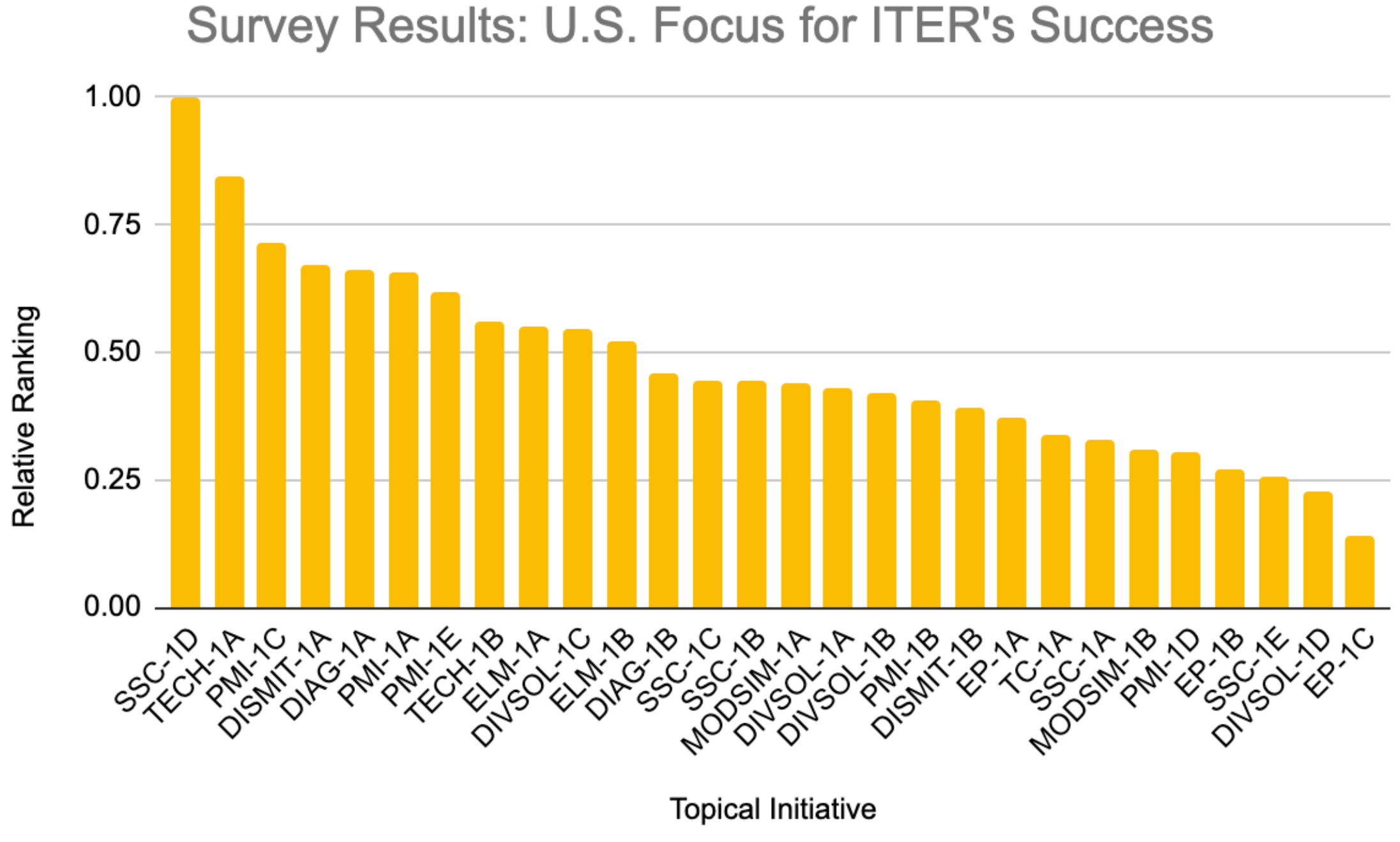

Figure F.2. Initiative ranking results in response to the top 5 topical initiatives that the US should focus on for ITER's success. Higher value corresponds to higher ranking.

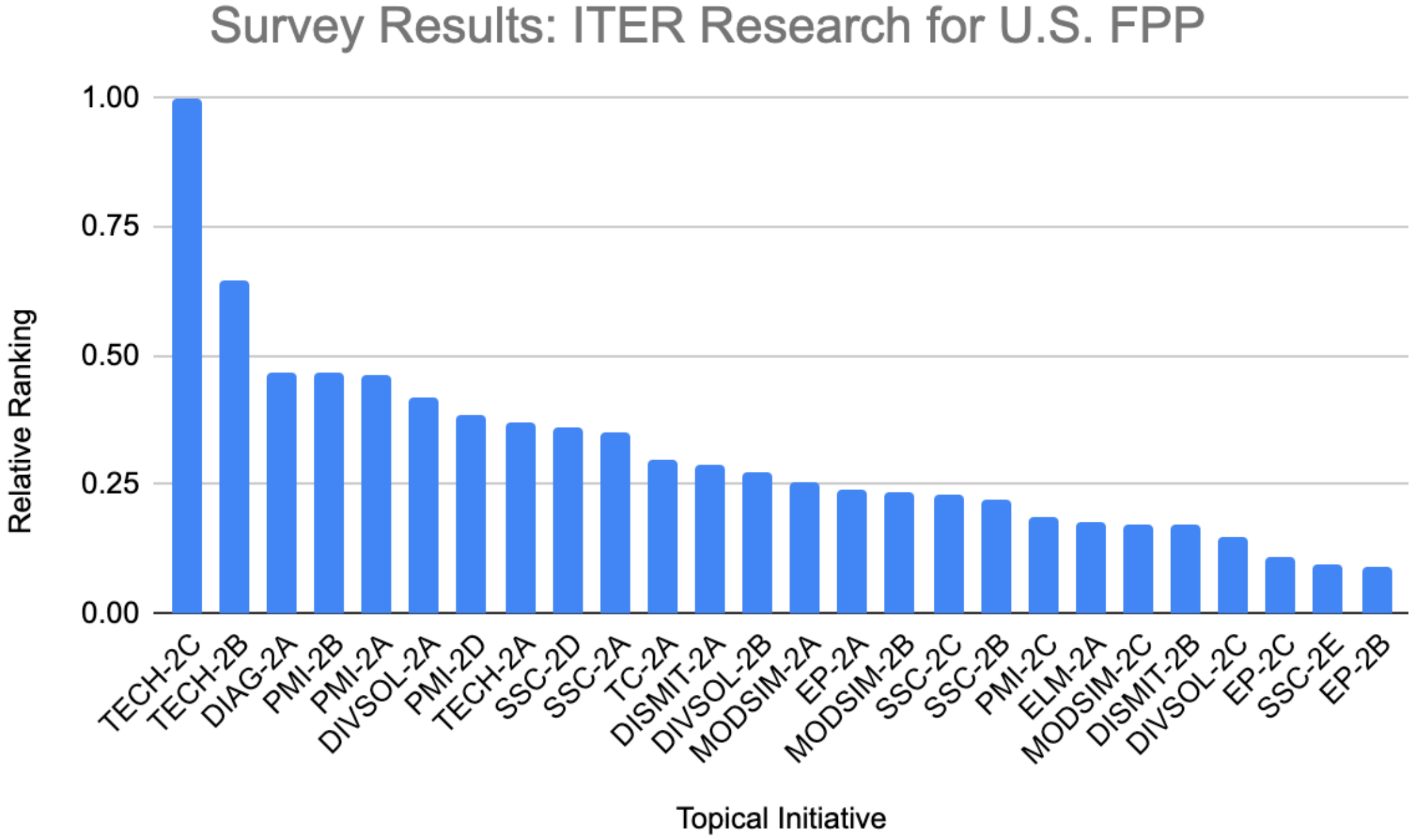

Figure F.3. Initiative ranking results in response to the top 5 topical initiatives where ITER research would help contribute to a US FPP.

# Appendix G. US Diagnostics for ITER

**TABLE G.1 – US DIAGNOSTICS SUMMARY**

| US Diagnostic System | Brief Purpose | Scheduled ITER Phase Availability |
|---|---|---|
| Electron Cyclotron Emission (ECE) | Electron temperature profile | PFPO-I |
| Toroidal Interferometer Polarimeter (TIP) | Chord-averaged electron density | PFPO-I |
| Core X-ray Crystal Spectrometer (XRCS-Core) | Ion temperature profile | PFPO-I |
| Upper Wide Angle Viewing (UWAV) | Visible and IR imaging of divertor | PFPO-I |
| Diagnostic Residual Gas Analyzers (DRGA) | Exhaust gas composition | PFPO-I |
| Motional Stark Effect (MSE) | Current density profile | PFPO-II |
| Low Field Side Reflectometer (LFSR) | Edge electron density & fluctuation profiles | PFPO-II |

# Appendix H. Leadership Selection and Term Lengths

In Chapter 3, we discussed various leadership roles within the US ITER Research Team. In this Appendix we propose selection processes and term lengths for each of these roles. We anticipate some of these processes will need to be adjusted for conformance to governmental requirements.

## H.1. US ITER Research Team Missions and Projects Leadership

Leaders of USIRT Missions or Projects should be selected by USIRCO from nominations made by USIRT members, and with concurrence from USIRAB and DOE-FES to ensure diverse and representative leadership. In case selected candidates are not adequately funded to perform the proposed duties, augmented funding should be made available.

Term lengths of two years are suggested, assuring some long-term continuity while providing leadership opportunities to a larger group. These terms can be extended once.

## H.2. US ITER Research Coordination Office Appointments

USIRCO appointments should be funded full-time, diverse, and rotating. Appointments should include an overall Director and Deputy Directors for Research Coordination in major technical areas such as Engineering & Technology, Theory & Modeling, and Experiment and Control (Fig. 3.1). An additional appointment, at the same level, should be made of a Deputy Director for Workforce Development. Individuals holding leadership roles or other roles within USIRCO should remain employees of their original institution during their USIRCO term, as with a secondment, and funding from DOE should be made available to support these arrangements.

These roles are outward-facing, interfacing at a high level with the ITER Organization and other ITER Member research organizations, and benefitting from longer-term continuity. Hence, we propose longer terms, usually four years, renewable once. In some cases, USIRCO personnel may be employed by institutions that limit outside appointments to shorter terms. This should be accommodated to keep the candidate pool as broad and diverse as possible.

The Deputy Director positions should be filled through open searches by the USIRAB with concurrence from the Director and DOE-FES. The Director should be selected by the USIRAB with concurrence from DOE-FES and input solicited from the outgoing Director (if appropriate). In each case, the Advisory Board's deliberations should start by soliciting nominations from all USIRT members.

Early in the formation of the USIRT we anticipate there may not be sufficient resources to fully staff the USIRCO as described here. There may be a need to combine some of the roles shown in Figure 3.1 to keep the office lean. By the time ITER begins research operations during PFPO-I we anticipate the full complement of Deputy Directors will be in place.

## H.3. US ITER Research Advisory Board membership

The initial number and membership of the Advisory Board should be chosen by FES and its members should be fully supported for their activities in this role. Membership should include representation from all major US stakeholder groups, including national laboratories, universities, and private industry. Subject matter experts in workforce development, DEI, and organizational effectiveness should also be included. Importantly, the membership of the Advisory Board should reflect the diversity of the workforce to which we aspire, both in terms of demographic diversity and diversity of background and expertise.

Replacement of outgoing USIRAB members should be done using a hybrid process. Half of the members should be chosen via open election by all USIRT members. Elections of this sort tend to favor well-known candidates. In order to further broaden the USIRT membership, half of its members should be appointed by the Director with concurrence of the USIRAB and DOE. This selection should be guided by the need to broaden service opportunities, carefully considering both demographic and institutional diversity.

To keep the USIRAB's perspective broad and fresh, its members should serve three-year non-renewable terms. The initial membership should be divided into three groups, with one-third having a two-year term, one-third three, and one-third four. This will assure that each year,starting at the end of the second year, one-third of the USIRAB membership will be replaced via the process described above.

## Appendix I. State of the ITER Organization Workforce

As part of the exercise of coming up with a strategy to include workforce development in the effort to create deeper engagement of the US with the ITER project, it is necessary to establish a baseline as to the current participation status of US personnel in the ITER workforce. To that end, we reached out to Hyunejune Choe, HR Project Responsible at ITER, and Sophie Gourod, Talent Management Section Leader at ITER. They kindly agreed to share some relevant statistics regarding US participation on responding to recruiting calls and participating in other programs such as student internships. What follows is a summary of the HR intelligence shared with us by Mr. Choe and Mrs. Gourod. To better understand how the talent and recruitment information is broken down, the most current ITER organizational chart is shown here for reference. US presence on each ITER domain is indicated in Fig. I.1.

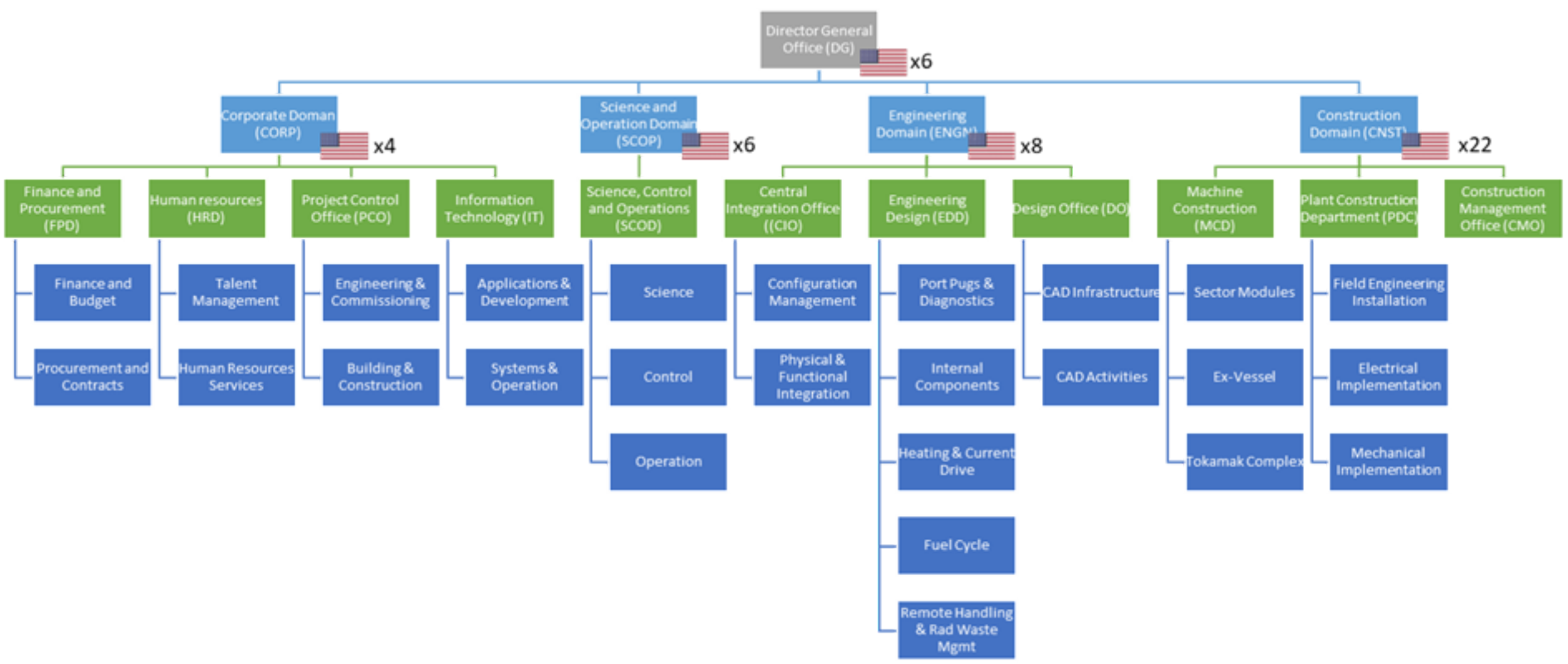

*Figure I.1. ITER organization chart, indicating US personnel on each domain.*

As of December 2021, ITER employs 1035 persons. 70% of this staff comes from the EU, and the US participates with 4.4% (46 people), behind China, Korea and Russia. Of those 46 people, 37 are direct professional employees, 3 are contractors for the tokamak cooling water system and 9 are direct administrative employees. Fig. I.2 presents a breakout of American IO staff by declared gender and ITER domain; the largest involvement is in construction, and the lowest presence is in commissioning and operations and scientific coordination. 26% of the US participants in ITER identify themselves as women.

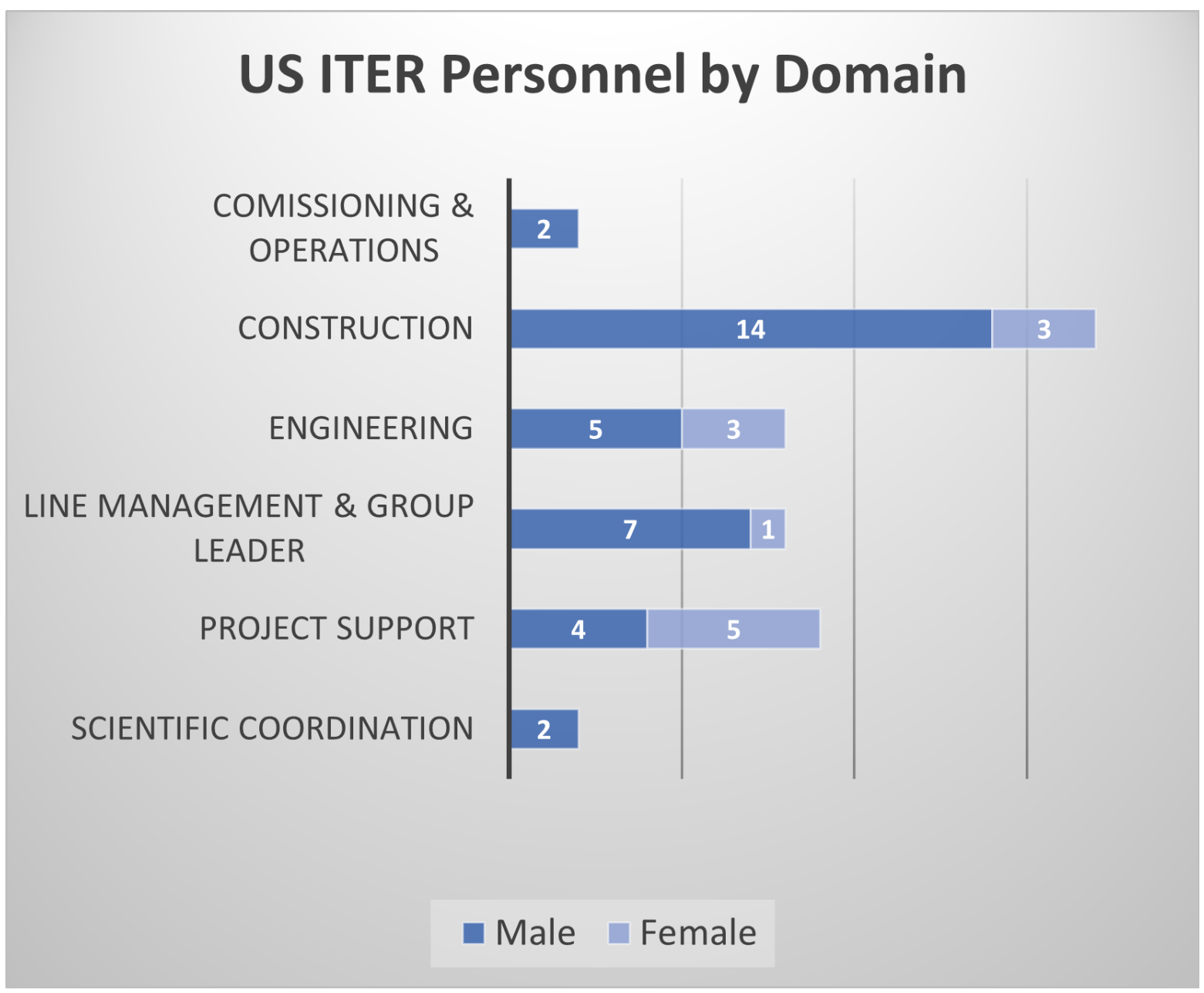

*Figure I.2. IO staff from the US, broken down by domain and gender*

Current ITER staffing represents an increase of 5% with respect to December 2020, and the current staff is at 95% of the project staffing target; this means that only 5% of the planned staff positions remain to be filled by new personnel. The implication is that if the US wants to increase its presence on the ITER staff, it will need to aggressively go after the remaining 5% positions, and also look to fill out positions that may become vacant during the life of the project. There have been continuous efforts to increase non-EU appointments, but despite these, Non-EU staff departures reached 52% (33 out of 64 in total) during 2021 compared to 62% during 2020. A high percentage of departures was observed for staff from China, India and the US (26 by head count or 40.6%). The arrival rate from those 3 members has continuously been decreasing for the last 5 years.

During 2021, 101 positions within CNST, ENGN and SCOP domains were fulfilled, as well as 14 managerial vacancies. The following chart presents the historic participation of US citizens on ITER recruiting calls, both in terms of number of nominations received and as a percentage of the total nominations received. There is a very clear downward trend in the participation of the US in the ITER recruitment process. During 2021, of the 157 applications received from US citizens, 94% were nominated from the US ITER domestic agency. Of this number, 26 applications were shortlisted (16%, compared to the global average of 11%, which is a good indicator of the quality of US applicants) and in the end the US contributed 5.7% to the ITER appointments during 2021. The Figure I.3 presents the evolution of US participation in ITER recruiting.

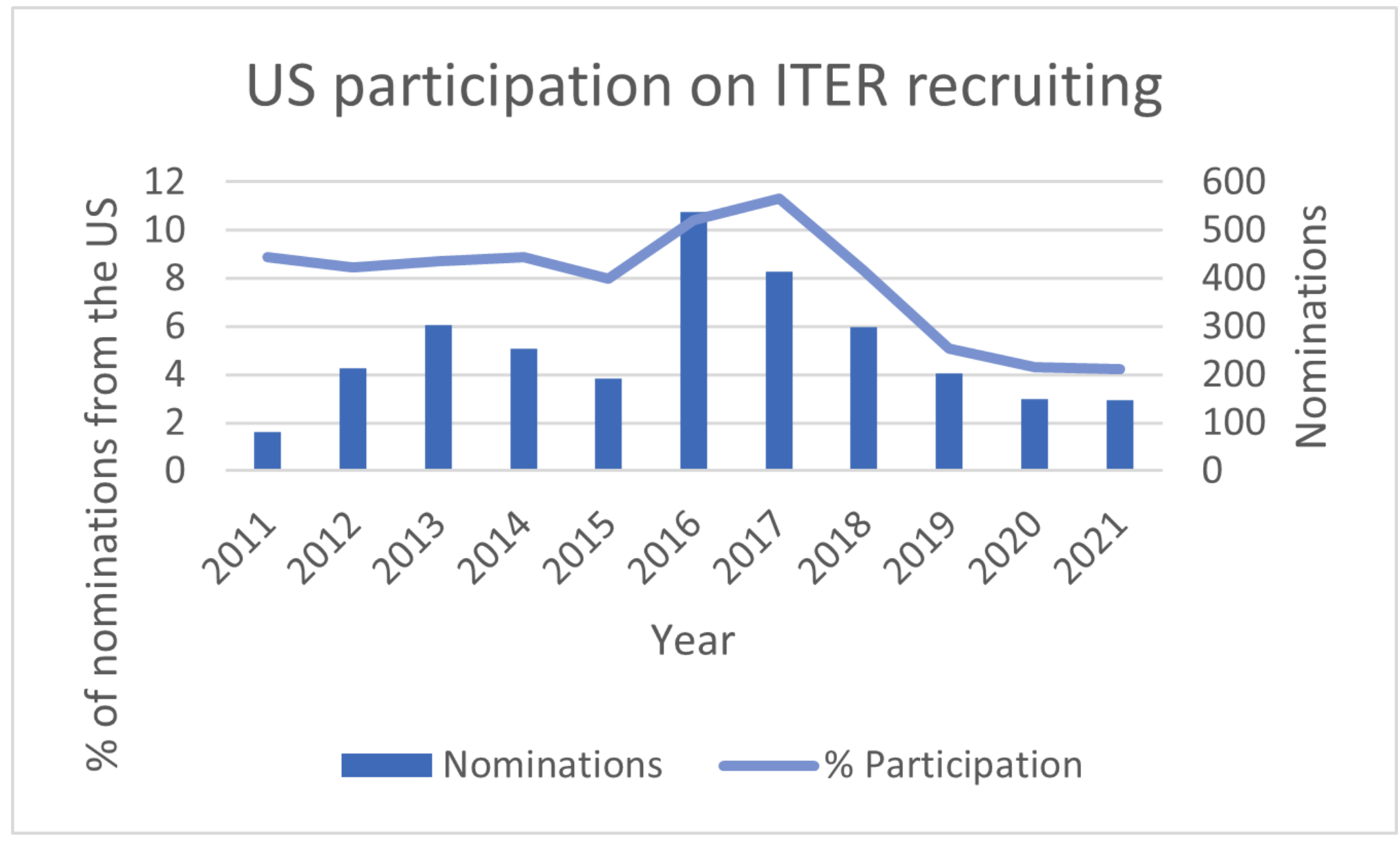

*Figure I.3. Total nominations for IO staff positions and the fraction of those nominations coming from the US*

The global number of nominated non-EU applications is 28.2%, and the ITER Organization continuously takes a proactive approach in pre-selecting Non-EU candidates during the hiring process; however, non-EU appointments were slightly decreased (38 Non-EU appointments among 105 in total, excluding 16 IO staff). The pre-selection rate of non-EU candidates decreased to 38.7% in 2021 with respect to the 40.1% in 2020 due to a lack of diversity in candidacies. A global decrease of female applications triggered a decrease of female appointments accordingly: female candidates' application rate dropped to 12.2% in 2021 vs. the 16.6% registered in 2020. Just to give a very concrete example, 22 published vacancies had to be extended up to two months on average due to the limited numbers of applications, which were mainly in IT, Control System and Diagnostics areas. ITER HR pointed out that it has not received any requests from the US ITER Domestic Agency to coordinate recruitment workshops and/or information sessions, which they have found extremely effective at attracting talent to the ITER Organization. The ITER Global Career Event, which had more than 1500 online attendees, also had a very small participation on the part of US citizens.

In terms of diversity, ITER HR has found that DG and CORP represent the domains where most of the female staff is recruited, since about 1/3 of the applications to those domains are from applicants identifying as women. In the case of ENGN, CNST and SCOP, the percentage drops to about 8%. Overall, the percent of staff identified as women has dropped from 25.2% in 2020 to 9.9% in 2021, since most of the recent hirings have occurred in ENGN, CNST and SCOP domains.

Another important aspect to gauge US talent involvement on ITER is the participation of students from US institutions on the ITER internship programs. There are three tiers of internship on ITER:

- Scientific or technical internships (Category A). The selected Interns are highly involved in IO activities and undertake a specific project under the supervision of an ITER staff member. This type of internship is open to students enrolled in their last year of a postgraduate program at a university (e.g. last year of Master or last year of Engineering School). The internship can last up to six months, with the possibility of extending it up to a year. Interns are paid a monthly allowance of 1300 Euros if the internship duration is at least five months and if it is performed in person at ITER Headquarters.
- Technical Internships (Category B). The selected Interns contribute to projects or research in their field of study under the supervision of an IO staff member. This tier of internship is open for candidates with at least one year of studies post-high school. The internship can last up to six months, with the possibility of extending it up to a year. Interns are paid a monthly allowance of 650 Euros if the internship duration is at least two months and if it is performed in person at ITER Headquarters (i.e. not fully remotely).
- Job-shadowing internships (Category C). Interns observe working conditions and may assist their supervisors in various tasks. This category is for students enrolled at an international Secondary School, international High School or international section where internships are mandatory (e.g. “3ème” and “2nde” students in France); or English speaking students from a school outside France where internships are mandatory; or upon approval by the Office of the Director General (e.g. for English speaking children registered in national schools). These internships last between 1 and 4 weeks. This internship category has no allowance or travel expenses.
- Specific internship (Category S). These are scientific or technical internships subject to the IO having a particular agreement (e.g. Memorandum of Understanding) in place with a laboratory, industry, university or government. Interns are highly involved in IO activities and undertake a specific project under the supervision of an IO staff. Students shall either be pursuing a PhD at a university or an equivalent institution, or be participating in a program in a scientific or technical field, which has a special agreement with the IO. The duration of the internship can be up to 4 years, and the IO does not provide monetary support unless specific provisions are established in the MoU indicating otherwise. The amount of the allowance shall be defined on an individual basis in the Internship Agreement.

The ITER internship program is very competitive: in 2021, the IO received around 1800 applications for internships, and accepted 71 (19 A, 20 B and 32 C). More than half of these were from EU students, and 9 were from US students. Most of the interns accepted in 2021 come from France (45), so money associated with interns mobility is sub-utilized. Figures I.4 and I.5 show the evolution of the participation in ITER internships between 2018 and 2021 on each of the Internship categories, and the breakout of the 2021 internships by organizational domain.

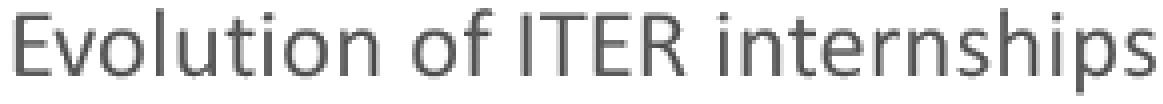

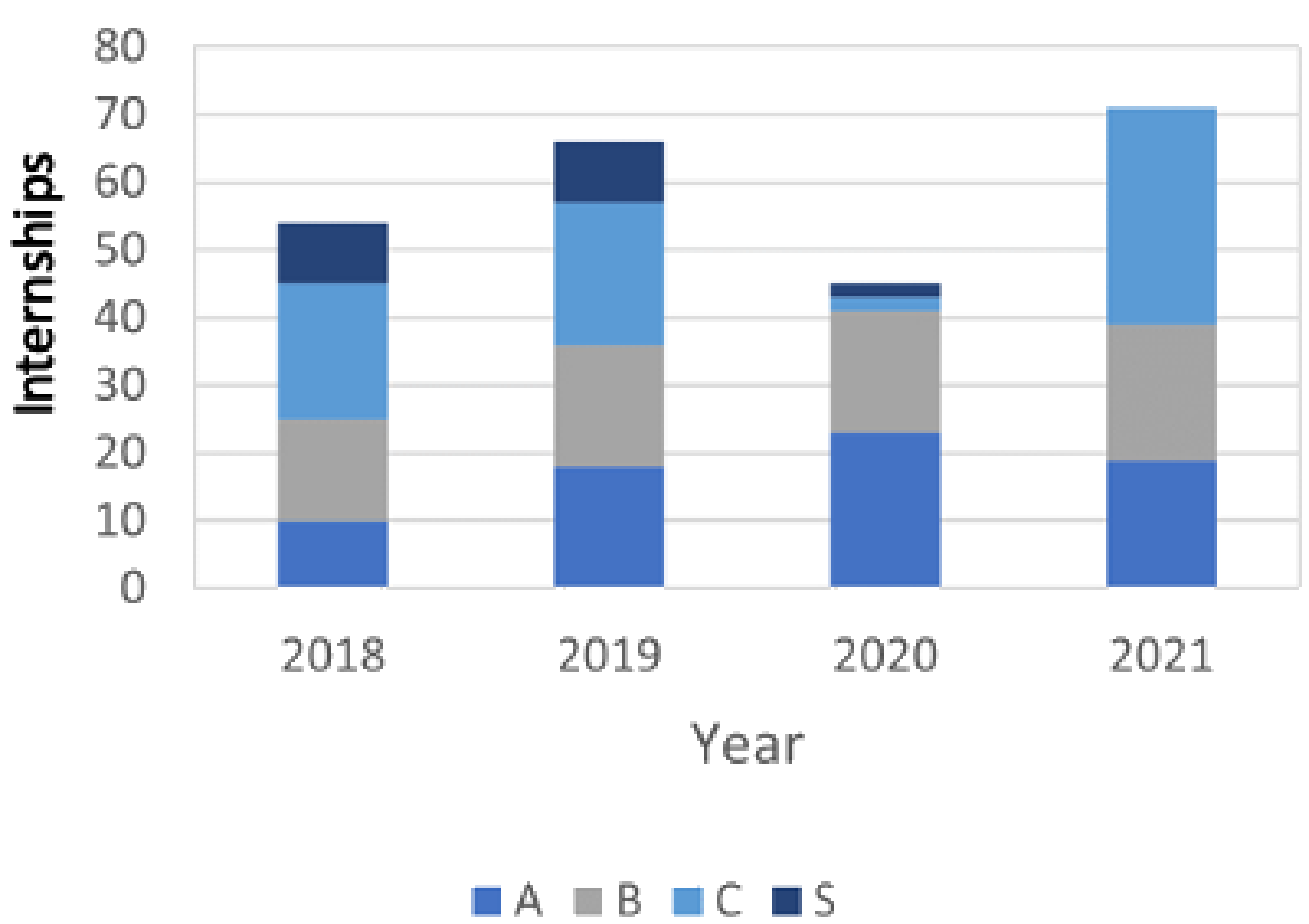

*Figure I.4. IO internships during 2018-2021*

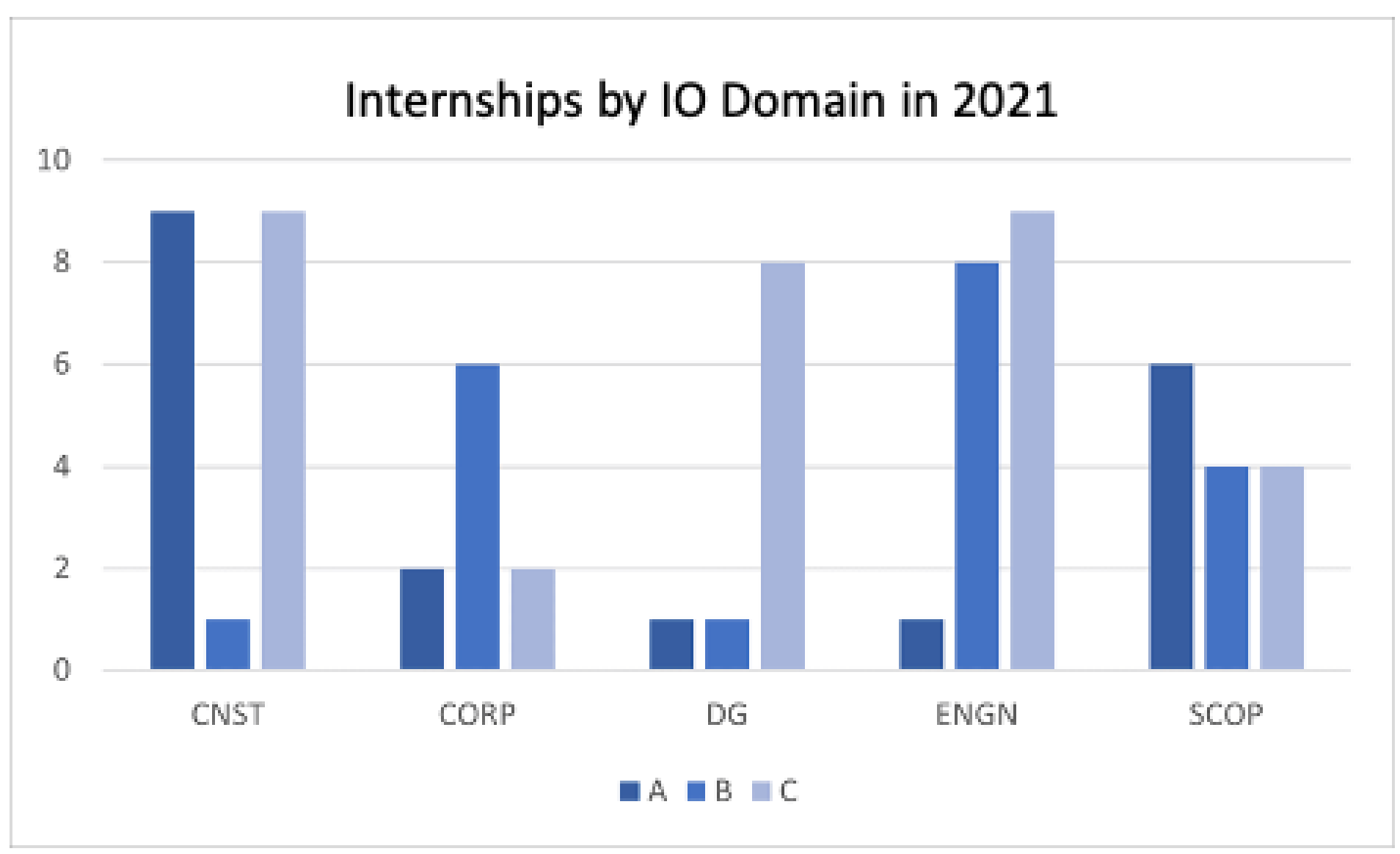

*Figure I.5. Breakdown of internship by domain and tier in 2021*

Participation of students in internships requires a Memorandum of Understanding to be signed between the University and the ITER Organization. Although a specific number was not provided, it was mentioned that the number of MoUs signed between US Universities and the ITER organization is below 10 (details on this statistic can be made available upon request by ITER HR, but individual MoUs cannot be made available). Among the universities with signed MoU with the IO are the University of Utah, San Diego State University, the University of Michigan, University of Illinois Urbana-Champaign, and Columbia University. Conflicts between ITER's and US Universities' policies on Intellectual Property have been a sticking point preventing wider university participation; indeed the MOUs currently in place have tended toward avoiding rather than resolving IP issues.

For the 2022 Internship campaign, the IO has limited the number of topics that can be submitted by each Domain: CSNT and ENGN submitted 20 topics, while SCOP submitted 15. 29 students from the US submitted applications, compared to 140 from the EU. The decision to cap the number of topics was taken after analysis of the efficiency of previous internship campaigns, and taking into account feedback from hiring managers. Also, the number of applications per student shall be limited to five. Q1 internship topics starting the first quarter of 2022 have been organized with a maximum of 2 internship topics per Domain that met this requirement was advertised with specific deadlines. The internship team with the support of the supervisors and the assistants have been working on it on priority. The following numbers have been approved by the DG office:

- Category A: 35
- Category B: 20
- Category C: 30
- Category S: 5

It is suggested that in order to take full advantage of the internship program, an ongoing scientific or technical collaboration with ITER staff is highly desirable and can lead to much better success in obtaining internships, since then the scope of the internship can be agreed upon beforehand. Careful planning is also desirable, since the internship program has specific deadlines.

## Appendix J. Acronyms Appearing In This Report

| | |
|---|---|
| CPP | See DPP-CPP, below |
| DEI | Diversity, Equity, and Inclusion |
| DIAG | Diagnostics Topical Area |
| DISMIT | Disruption Mitigation Topical Area |
| DIVSOL | Divertor and Scrape-Off Layer Topical Area |
| DMS | Disruption Mitigation System |
| DOE | US Department of Energy |
| DPP-CPP | APS Division of Plasma Physics Community Planning Process |
| ECH | Electron Cyclotron Heating |
| EF | Error Field |
| ELM | Edge Localized Mode (also associated Topical Area) |
| EP | Energetic Particles (also associated Topical Area) |
| FES | DOE Office of Fusion Energy Sciences |
| FESAC | Fusion Energy Sciences Advisory Committee |
| FP | First Plasma |
| FPO | Fusion Power Operation campaign |
| FPP | Fusion Pilot Plant |
| H&CD | Heating and Current Drive |
| H-mode | High confinement operational regime |
| ICRH | Ion Cyclotron Resonance Heating |
| IMAS | Integrated Modeling and Analysis Suite |
| IO | ITER Organization |
| ITER | Tokamak facility currently under assembly in France. ITER is Latin for "The Way." |
| ITPA | International Tokamak Physics Activity |
| MHD | Magnetohydrodynamics |
| MODSIM | Modeling and Simulation Topical Area |
| NASEM | National Academies of Science, Engineering, and Medicine |
| NBI | Neutral Beam Injection |
| NTM | Neoclassical Tearing Mode |
| PDS | Pulse Design Simulator |
| PFC | Plasma-facing component |
| PFPO-I | First Pre-Fusion Power Operation campaign |
| PFPO-II | Second Pre-Fusion Power Operation campaign |
| PFURO | Plasma and Fusion Undergraduate Research Opportunities Program |
| PMI | Plasma-Material Interactions (also associated Topical Area) |
| RE | Runaway Electron |
| RMP | Resonant Magnetic Perturbation |
| RWM | Resistive Wall Mode |
| SciDAC | Scientific Discovery through Advanced Computing (DOE program) |
| SOL | Scrape-off Layer |
| SPI | Shattered Pellet Injection |
| SSC | Scenarios, Stability, and Control Topical Area |
| TC | Transport and Confinement Topical Area |
| TECH | Technology and Integrations Topical Area |
| USIPO | US ITER Project Office, responsible for contributing US hardware to ITER |
| USIRAB | US ITER Research Advisory Board |
| USIRCO | US ITER Research Coordination Office |
| USIRP | US ITER Research Program |
| USIRT | US ITER Research Team |